\documentclass[twocolumn]{aastex61}
\usepackage{graphicx}
\usepackage{multirow}
\usepackage{amsmath} 
\usepackage{color}

\usepackage{amssymb}
\usepackage{natbib}
\newcommand{\tsys}{$T_\text{sys}$}

\shorttitle{phased-ALMA Calibration in VLBI observations}
\shortauthors{Goddi, Mart\'i-Vidal, Messias, et al.}

\begin{document}

\title{{\small Calibration of ALMA as a phased array} \\
{\scriptsize ALMA observations during the 2017 VLBI campaign}
}

\author{
C. Goddi}
\affil{ALLEGRO/Leiden Observatory, Leiden University, PO Box 9513, NL-2300 RA Leiden, the Netherlands }
\affil{Department of Astrophysics/IMAPP, Radboud University, PO Box 9010, NL-6500 GL Nijmegen, the Netherlands}

\author{
I. Mart\'i-Vidal
}

\affil{
Centro Astron\'omico de Yebes (Instituto Geogr\'afico Nacional), Apartado 148, 19180 Yebes, Spain
}

\affil{
Onsala Space Observatory (Chalmers University of Technology), 43992 Onsala, Sweden
}

 \author{
H. Messias 
}
\affil{
Joint ALMA Observatory, Alonso de Cordova 3107, Vitacura 763-0355, Santiago de Chile, Chile
}

\author{
 G. B. Crew 
}
\affil{
Massachusetts Institute of Technology Haystack Observatory, 99 Millstone Road, Westford, MA 01886, USA
}

  \author{
  R. Herrero-Illana 
 }
\affil{
European Southern Observatory, Alonso de C\'ordova 3107, Vitacura, Casilla 19001, Santiago de Chile
}

\author{V. Impellizzeri
}
\affil{
Joint ALMA Observatory, Alonso de Cordova 3107, Vitacura 763-0355, Santiago de Chile, Chile
}

\author{H. Rottmann
}
\affil{Max-Planck-Institut  f\"{u}r Radioastronomie, Auf dem H\"{u}gel 69, D-53121 Bonn, Germany
}

\author{J. Wagner
}
\affil{Max-Planck-Institut  f\"{u}r Radioastronomie, Auf dem H\"{u}gel 69, D-53121 Bonn, Germany
}

\author{E. Fomalont
}
\affil{
Joint ALMA Observatory, Alonso de Cordova 3107, Vitacura 763-0355, Santiago de Chile, Chile
}

\author{L. D. Matthews
}
\affil{
Massachusetts Institute of Technology Haystack Observatory, 99 Millstone Road, Westford, MA 01886, USA
}

\author{D. Petry
}
\affil{European Southern Observatory, Karl-Schwarzschild-Strasse 2,
D-85748 Garching bei M$\ddot{u}$nchen
}

\author{
N. Phillips
}
\affil{
Joint ALMA Observatory, Alonso de Cordova 3107, Vitacura 763-0355, Santiago de Chile, Chile
}

\author{R. Tilanus}
\affil{ALLEGRO/Leiden Observatory, Leiden University, PO Box 9513, NL-2300 RA Leiden, the Netherlands }
\affil{Department of Astrophysics/IMAPP, Radboud University, PO Box 9010, NL-6500 GL Nijmegen, the Netherlands}

\author{E. Villard
}
\affil{
Joint ALMA Observatory, Alonso de Cordova 3107, Vitacura 763-0355, Santiago de Chile, Chile
}

\author{L. Blackburn}
\affil{Harvard-Smithsonian Center for Astrophysics, 60 Garden St., Cambridge, MA 02138, USA
}

\author{M. Janssen}
\affil{Department of Astrophysics/IMAPP, Radboud University, PO Box 9010, NL-6500 GL Nijmegen, the Netherlands}

\author{M. Wielgus}
\affil{Harvard-Smithsonian Center for Astrophysics, 60 Garden St., Cambridge, MA 02138, USA
}
\affil{Black Hole Initiative at Harvard University, 20 Garden St., Cambridge, MA 02138, USA}

\begin{abstract}
We present a detailed description of the special procedures for calibration and quality assurance  of Atacama Large Millimeter/submillimeter Array (ALMA) observations in Very Long Baseline Interferometry (VLBI) mode. 
These procedures are required to turn the phased ALMA array into a fully calibrated VLBI station.
As an illustration of these methodologies, we present full-polarization observations carried out  with ALMA as a phased array at 3mm (Band 3) and 1.3mm  (Band 6)  as part of Cycle-4. 
These are the first VLBI science observations conducted with ALMA and were obtained during a 2017 VLBI campaign in concert with other telescopes worldwide as part of 
the Global mm-VLBI Array (GMVA, April 1-3) and the Event Horizon Telescope (EHT, April 5-11) in ALMA Bands 3 and 6, respectively.  
\end{abstract}


\correspondingauthor{Ciriaco Goddi, Ivan Mart\'i-Vidal}
\email{C.Goddi@astro.ru.nl, I.Marti-Vidal@uv.es}

%________________________________________________________________

%==============================================================================
\section{Introduction}
%==============================================================================

Very Long Baseline Interferometry (VLBI) is an astronomical technique to  make images of cosmic sources with the highest angular resolution presently achievable in astronomy. 
VLBI uses a global network of radio telescopes spread across different continents as an interferometer to form a virtual Earth-sized telescope. 
By recording radio wave signals at individual antennas and afterwards cross-correlating the signals between all pairs of antennas  using time stamps of atomic clocks for synchronization, one obtains the interferometric 
visibilities that can be used to reconstruct an image of the source using Fourier transform algorithms as normally done in standard connected-element interferometers \citep{ThompsonMoranSwenson2017}.  

At centimeter wavelengths, VLBI has been used for many decades to measure
the size and structure of radio sources on angular scales as small as one
milliarcsecond \citep[e.g.,][]{Pearson1988,Kellermann1998,Jorstad2001,Goddi2006}.
Since the achievable angular image resolution of an interferometer can be expressed as $\theta \sim \lambda/B$, where $\lambda$ is the observed wavelength  and $B$ is the maximum distance between pairs of telescopes (or "baseline"),  
the higher frequencies (shorter wavelengths) provide the higher resolving power. 
While extension of VLBI techniques to the millimeter (mm) regime (hereafter mm-VLBI)
provides the highest angular resolution (as fine as a few tens of microarcseconds for a typical Earth-size baseline of $\sim$ 10000~km),  
it  faces significant observational and technical challenges: 
higher surface accuracy needed for telescopes operating at mm-wavelengths, 
higher stability required for   atomic clocks and receiver chains, and, above all, stronger  distortion effects on the radio-wave fronts by the troposphere which decreases the coherence timescales to only a few seconds.  
Therefore and because of the limited sensitivity of existing
networks of VLBI antennas, the use of mm-VLBI  has been restricted to the study of a relatively small number of bright sources \citep{Krichbaum1998,Doeleman2008,Doeleman2012}. 
 
As a critical step toward overcoming these limitations, an  international  consortium has  built  a  beamformer  for  the  Atacama  Large  Millimeter/submillimeter Array (ALMA)  within the ALMA Phasing Project (APP) \citep{APPPaper}. 
 ALMA is the most sensitive (sub)mm-wave telescope ever built and consists of two main components: 
 50 individual antennas of 12-m diameter  comprise the so-called "12-m Array", 
which  is used in conjunction with the 64-antenna Baseline (BL) correlator
 and an additional sixteen antennas (twelve
7-m antennas and four 12-m antennas) which comprise the ALMA "Compact Array" (ACA) and which can be operated independently with a separate ACA correlator.

The beamformer can  aggregate the entire collecting area of ALMA (usually limited to the 12-m Array) into a single, very large aperture by aligning in phase and summing up the signals from individual antennas. 
This turns ALMA into a virtual single-dish telescope (equivalent to a telescope of 84-m diameter if one could phase all the 12-m antennas in the array)  
 where all antennas act jointly as one giant element in a VLBI experiment, boosting the achievable signal-to-noise ratio
(SNR) of VLBI baselines to the site.
The extraordinary sensitivity of ALMA as a phased array (hereafter phased-ALMA), combined with the extremely high angular resolution available on North-South baselines, enable transformational science on a variety of scientific topics, including  tests of Einstein's general theory of relativity near black holes \citep{Doeleman2008,Doeleman2012,Goddi2017,EHT2019I}, accretion and outflow processes around black holes in active galactic nuclei or AGNs \citep{,Boccardi2017}, jet launch and collimation from AGN \cite{AsadaNakamura2012}, pulsar and magnetar emission processes,  maser science \citep{Issaoun2017}, and astrometry \citep[see][for detailed descriptions of the science case for phased-ALMA]{Fish2013,Tilanus2014}. 

Joint VLBI observations that include phased-ALMA with other telescopes worldwide were conducted for the first time  in  2017 April as part of ALMA Cycle-4.
This paper describes the entire analysis processing chain for ALMA data acquired during the 2017 VLBI campaign, 
with particular  focus  on the calibration of interferometric visibilities recorded while ALMA observes in VLBI mode (VOM). 

The current paper is structured as follows.
\S~\ref{APP} summarizes the main properties of phased-ALMA as a VLBI station. 
\S~\ref{2017campaign} gives an overview of the Cycle 4 observations during the April 2017 VLBI  campaign, focusing on the ALMA observational setup. 
\S~\ref{calibration} describes in detail the  calibration procedures, and \S~\ref{polcal}   focuses on the polarization calibration of  interferometric ALMA data in VOM.
\S~\ref{vlbi} describes the  procedures adopted to apply the ALMA data calibration tables to the VLBI visibilities. 
 Finally,   \S~\ref{summary} provides a summary. 

%--------------------------------------------------------------------------------------
\begin{figure*}
\centering
\includegraphics[width=\textwidth]{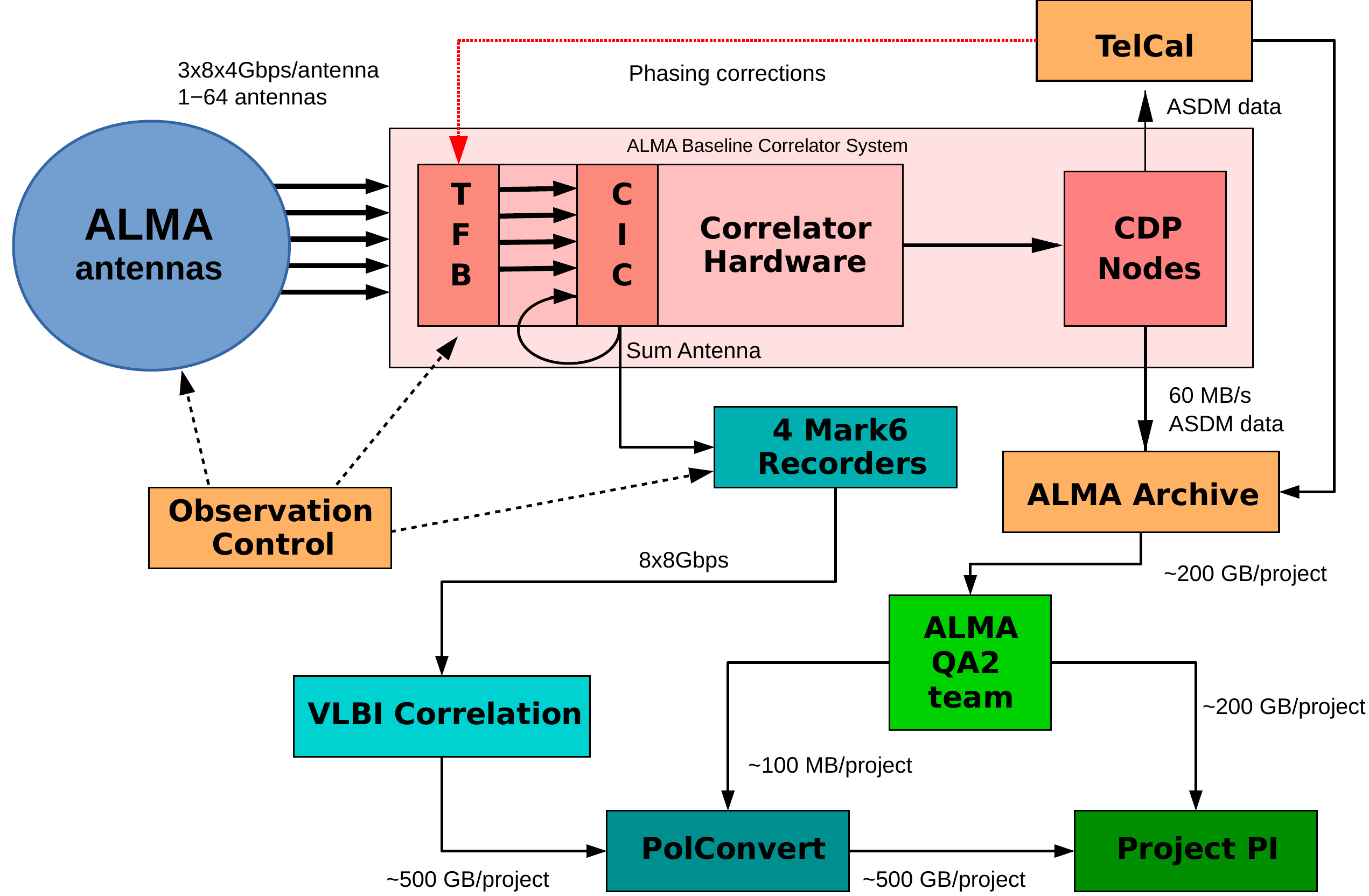}
\vspace{0.5cm}
\caption{
General data-flow diagram for VLBI observations with phased-ALMA. 
The ALMA correlator receives data from up to 64 antennas whose dual-pol
receivers are sampled in 3~bits at 4~Gbps (for a total data rate of $3 \ [bits] \times 4 \ [baseband] \times 2 \ [polarization] \times 4 \ \rm{ Gbps} = 3 \times 8 \times 4 \ \rm{  Gbps}$).  At the front-end of the correlation, the
tunable filter bank (TFB) cards convert these streams to 2-bit signals for correlation (with a total data rate of $2 \times 8 \times 4 \ \rm{  Gbps}$) which are fed to the
Correlation Interface Cards (CIC) which manufacture the sum antenna signal.
The latter replaces one of the correlator antenna inputs (in Cycle 4, antenna "DV03" was overridden and used to store the  phased signal).
 At the back-end of the correlation process, the Correlator
Data Processor (CDP) Nodes provide correlated data  to TelCal for calculating
the phase adjustments, which are then applied  in the TFBs. 
The sum signal is also sent to the VLBI Mark6 recorders (each recorder takes both polarizations of one baseband at a recording rate of 16 Gbps, resulting in a total rate of 64 of Gbps for four recorders) and the ALMA correlated data is sent to the ALMA Archive.  
The APS and VLBI activities
are managed by Observation Control processes which also orchestrate the normal interferometric  observations.  
The archived data is subsequently analyzed by the "Level 2 Quality Assurance" (QA2) Team  and then  delivered to the ALMA project PI along with the QA2 calibration tables for ALMA-interferometric data analysis.  
Meanwhile, the recorded VLBI baseband data is shipped to the VLBI correlators for the correlation of the full VLBI  experiment. This process can only be completed once
ALMA is converted into circular polarization basis through the use of \texttt{PolConvert} using calibration products (deliverables) from the QA2 process.  Lastly, the VLBI dataset is finally delivered to the PI for the full VLBI data analysis.
Note that the flow is continuous in time, and in the direction of
the arrows (the red line for the phasing corrections  
indicates that changes made in one sub-scan affect the next).
}
\vspace{0.5cm}
\label{aps_workflow}
\end{figure*}
%%--------------------------------------------------------------------------------------

%==============================================================================
\section{Observing with ALMA as a phased array}
\label{APP}
%==============================================================================
A full  overview of the ALMA Phasing System (APS) (and its hardware and software components)
as well as a description of the operation of ALMA as a VLBI station,  is provided in \cite{APPPaper}. 
In this section, we summarize some of the APS elements relevant
to understanding interferometric  data taken with the APS  (\S~\ref{APS}),
the specialized procedures required for the VLBI correlation (\S~\ref{vlbicorr}), and the conversion of linearly polarized data to a circular basis (\S~\ref{intro:polconvert}). 
A general data-flow diagram for VLBI observations with phased-ALMA  is shown in Fig.~\ref{aps_workflow}.

%________________________________________________________________
\subsection{The ALMA Phasing System (APS)}
\label{APS}
%________________________________________________________________

The APS performs phase adjustments to the individual ALMA antenna signals  to create a {\it phased array} from a designated subset of the full observing array. 
The phasing corrections are computed relative to a specified {\it reference antenna}.
The phased signals are then summed  within the ALMA BL correlator cards to create a virtual
antenna (the {\it sum antenna}) whose signals are fed back into the ALMA BL correlator and correlated as if it were a real, much larger antenna co-located with the reference antenna.
This allows the sum antenna to be cross-correlated with other non-phased antennas (within the {\it comparison array}) to provide feedback on the efficiency of the phasing process.
 The phasing adjustments are calculated within the ALMA Telescope Calibration ({\it TelCal}; \citealt{Broguiere2011}) subsystem of the software and then applied by the BL correlator. 
The information is therefore processed  in a closed "phasing loop" between the BL
correlator (where measurements are made and applied) and
TelCal (where corrections are calculated).
We discuss some important details on the spectral specifications and the timing of the phasing loop in the next two sub-sections.

%%--------------------------------------------------------------------------------------
\subsubsection{Spectral specifications of the phasing loop}
\label{APS_spectral}

 In the ALMA BL correlator, each 2~GHz input band (one out of four quadrants) is  subdivided into thirty-two 62.5~MHz channels by sets of tunable digital
filter bank (TFB) cards.  
Data are delivered from the correlator to TelCal in the form of "channel averages" (spectral sub-regions), which correspond to sets of TFBs.
Therefore, phasing corrections are calculated by TelCal on portions of the full spectrum.  
Averaging visibilities over frequency ranges within the full band allows increasing the SNR of the phase solutions while reducing the volume of the data used for the phasing calculations.

The number of channel averages can be defined in input and may be tuned (in principle) according to the source strength. In Cycle-4 (and Cycle-5)  APS observations have adopted averages over 4 TFB channels, resulting in eight frequency chunks\footnote{The number used for the channel averages   may change in future cycles.}, each spanning 250~MHz\footnote{The TFB channels are actually overlapped
slightly in frequency  yielding an effective bandwidth of 1.875~GHz per quadrant, therefore each frequency chunk  spans actually 234,375~MHz.}, per baseband.
This specific choice  
was  a compromise between having a set of channel averages
with sufficient signal to robustly calculate phases, while still providing an effective correction to  the 
static baseband delays\footnote{The static baseband delay is the sum of all of the stable signal path delays from the receivers to the correlator.}.   
Normally, the Correlator Data Processing (CDP) cluster makes corrections for such baseband delays.  When the APS is active, however, TelCal  solves for phase adjustments in channel averages, and in order to  apply the calculated values  within the TFBs, these delay corrections must be disabled \cite[see][]{APPPaper}.
The solution implemented during ALMA Cycle-4 (and Cycle-5) is to compute and
apply the needed delay correction as part of the phasing corrections.
Specifically, TelCal splits each baseband into eight contiguous frequency chunks and
fits the X and Y phase gains at each chunk and for each antenna (using one of the phased antennas as the reference). 
The set of phase adjustments across the channels provides effectively a delay-like correction, 
which mostly removes the generally large baseband delays in the phased signals. 
However, the correction
is imperfect, as it is not identical to subtraction of a single linear phase slope as a function of frequency across the full band. 
This results in a small correlation loss caused by the small residual delay within each channel average chunk.
It also adds an additional frequency-dependent X-Y offset that produces small
discontinuities at the edges of the frequency  chunks. Such phase offsets and
jumps must be determined off-line (\S~\ref{cross-phase}), using
observations of the polarization calibrator. 
The proper handling of the baseband delay correction may be addressed in a future ALMA software release, by  enabling TelCal to take the baseband delays into account in the
calculation of the phasing solutions.

%%--------------------------------------------------------------------------------------
\subsubsection{Timing of the phasing loop}
\label{APS_timing}

Each "VLBI scan" is partitioned into "sub-scans" for correlation and for processing in TelCal.
In order to choose a suitable integration period to calculate the phasing corrections  in the channel averages, 
one should consider that longer integrations result
in better (less noisy) phase corrections in stable atmospheric conditions, whereas 
shorter integrations are required for acceptable efficiency during sub-optimal 
observing conditions.  
The operational compromise adopted for Cycle-4 was to
program the correlator for four 4-s integrations (4.032 s) per sub-scan with
2~s (2.016~s) channel averages across four adjacent TFBs\footnote{ 
The high time resolution on
the channel averages is provided to allow data from the start
of the correlator subscan to be excluded without significant loss of
accuracy in the phasing calculation. The shorter cadence allows some portion of the first integration
to be used for the next phasing correction.}. 
These 16~s (16.128~s) correlator sub-scans require a setup and dump gap between them of 2.064~s.
Thus the total loop time (between phasing updates) is 18~s (18.192~s).
The phasing corrections are calculated at the end of each 16-s sub-scan and
applied at the beginning of the next. 

\begin{figure*}
\centering
\includegraphics[width=\textwidth]{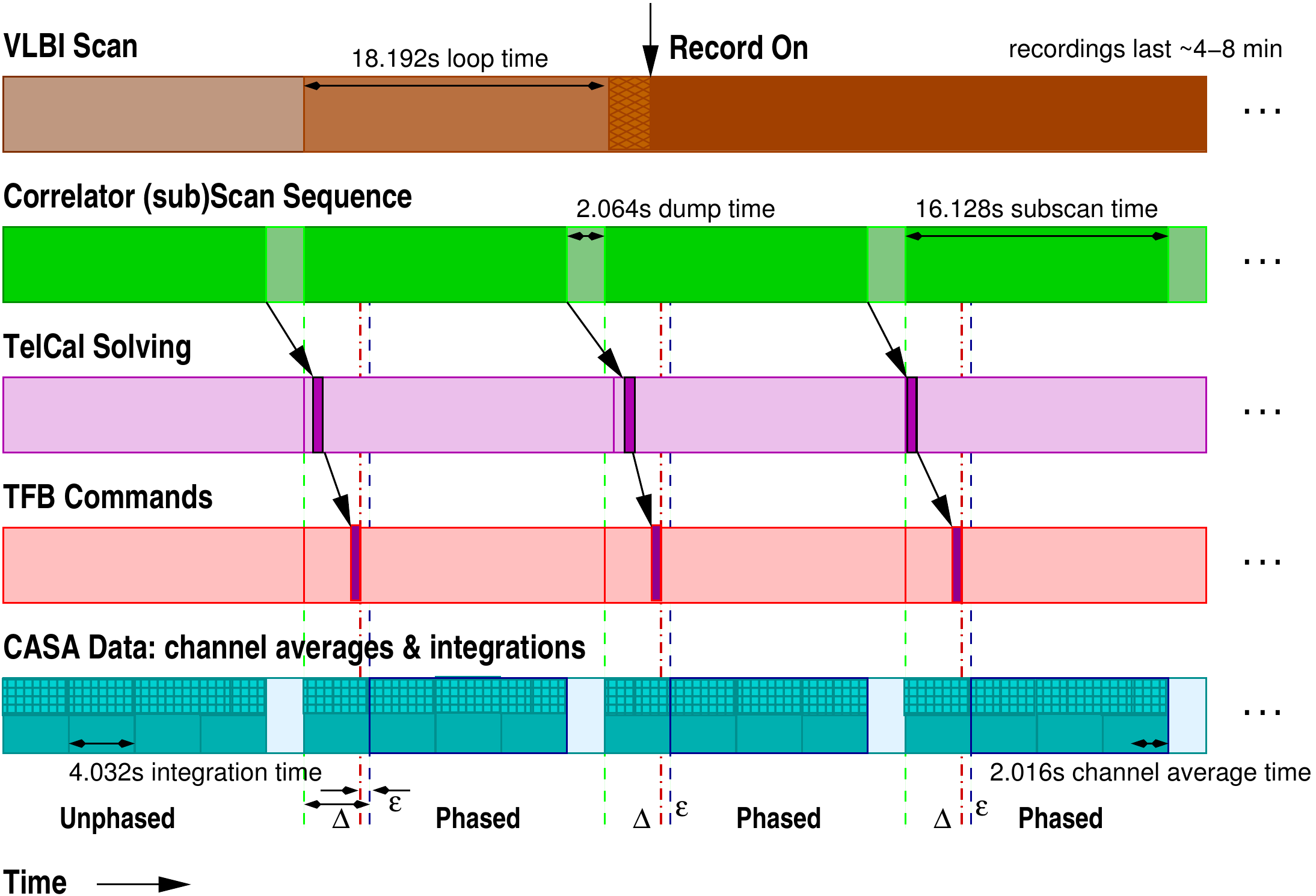}
%\vspace{0.5cm}
\caption{
General APS timing diagram adapted from Fig.~11 of \cite{APPPaper}. This diagram shows the timings associated with the different components of  the system as discussed in the text.
The top bar (brown) reflects the scheduled VLBI scan: the ALMA BL correlator
is started prior to the planned recording start in order to allow phase-up to occur.
The correlator (green bar) performs its work in so-called sub-scans
separated by short "dump'' periods when the hardware is read out to the CDP nodes for generating the integrations and channel averages (bottom bar, teal).
After every sub-scan, the correlator sub-scan data is passed to TelCal
to calculate (purple bar) the phasing corrections which are applied in
the TFBs (at the input to the BL correlator).  As these corrections are
made, the integrations and channel averages (available in the measurement sets)
become phased.  The timing is such that the  first portion of every
block of integrations corresponding to a sub-scan (marked $\Delta$) is
either unphased (first sub-scan) or the least-well phased of each group.}
\label{casa-scans}
\end{figure*}

Note that the full 16-s sub-scan cannot be used to determine the phasing solution.
In practice, the transfer of data to TelCal is not instantaneous, so a few  seconds are needed to obtain the data to process (Telcal does the data processing in $<1$~seconds); the arrival of the phasing adjustment from TelCal to the correlator is usually within the first 4-s integration. Therefore,  12 or 14 s of channel average data are typically considered valid for the next TelCal computation.
 Note also that the correlator continues to process data through the gaps for the 
VLBI recordings, but that these 2~s intervals are not represented in the
ALMA ASDM files (the raw data in the ALMA archive).
\emph{An operational consequence of this is that the ALMA data only cover 90\% (16.128/18.192) of the VLBI data and some interpolation through the gap is required.}
A timing diagram sketching   the timings associated with the different components of the system is shown in Fig.~\ref{casa-scans}.

A final consideration is that the phasing system is imperfect: there are $\sim$10~deg (RMS) phasing errors under typical conditions.  This is  captured within the ASDM data sets and calculations may be made after the observations to calculate the phasing
efficiency and correct the amplitude of the summed VLBI signal for decorrelation losses (see \S~\ref{pheff} for details).

%--------------------------------------------------------------------------------------
%%--------------------------------------------------------------------------------------

%________________________________________________________________
\subsection{VLBI correlation and polarization basis}
\label{vlbicorr}
%________________________________________________________________

Conventionally VLBI is performed using  circularly polarized feeds (with quarter wave plates) to avoid parallactic angle issues in the correlation.  
The ALMA antennas however have linearly polarized feeds, which provide a high polarization purity, i.e. a low polarization leakage between polarizers  (e.g., \citealt{Rudolf2007}; see also \S~\ref{pol-basics}). 
Several options were possible for the
adaptation of the ALMA linear polarizations into a circular basis for
VLBI. 
These included applying the conversion to the raw data streams either at the antenna frontends or computing it at the correlation stage at the VLBI backend.
This option however, has two major drawbacks: 
first,  the additional hardware  required  can potentially increase the instrumental polarization effects, and, second, it is an irreversible process. 
Therefore,  the final chosen strategy was to apply a {\it post-correlation} conversion for
the ALMA signal polarization. 
The VLBI correlation is executed with the DiFX software \citep{Deller2011},
which correlates data streams and has no intrinsic understanding of polarization other than as labels. 
Since ALMA provides X and Y polarization recordings, while the rest of the VLBI stations provide R and L circular polarization signals,   DiFX  reports XL, XR, YL and YR correlation products in its native binary (so-called SWIN) output.
The VLBI fringes in this  {\it mixed-polarization} basis,
can then be converted into a pure {\it
circular-polarization} basis using an algorithm
based on hybrid matrices in the frame of the Radio Interferometer
Measurement Equation \citep{Hamaker1996,Sault1996}. 
 
 %________________________________________________________________
\subsection{ALMA data QA2 and Polarization conversion} 
\label{intro:polconvert}
%________________________________________________________________

The process of polarization conversion can be divided into two main parts.
In the first part, the visibilities among the ALMA antennas (computed by the ALMA
correlator, simultaneous with the VLBI observations) are calibrated
using ALMA-specific algorithms for full-polarization data reduction (see \S~\ref{calibration} and \S~\ref{polcal}).
Within the ALMA organization, this process is known as "Level 2 Quality Assurance'' (QA2). 
In the second part, 
the calibration tables derived in the QA2 stage are sent to the VLBI correlators, 
where a software tool known as   \texttt{PolConvert}  \citep{PCPaper} (run at the correlator computers) applies these tables directly to the VLBI visibilities produced by the DiFX software. 
It is the \texttt{PolConvert} program that transforms the linear-polarization ALMA data streams into a circular basis for VLBI, and generates the calibration information for phased-ALMA.
One of the main advantages of this "off-line" conversion is that it is "reversible",
in the sense that one can perform the full QA2 analysis of the ALMA data multiple times, in order to find the best estimates of the pre-conversion correction gains prior to the
polarization conversion. 
Details about this process are given in \S~\ref{vlbi}  \citep[see also][for a full description of the \texttt{PolConvert} algorithm]{PCPaper}.

\vspace{0.3cm}
In summary, VLBI observations with phased-ALMA is a four-part process: (i) observe at ALMA using the APS, (ii) correlate the VLBI data from ALMA and the other participating stations with DiFX in a mixed polarization basis, (iii) calibrate the ALMA data for polarization and other calibration products, and finally (iv) apply these products to the DiFX output using \texttt{PolConvert}  in a postprocessing step prior to VLBI data calibration.

%=============================================================
\section{VLBI Observations with phased-ALMA during Cycle 4}
\label{2017campaign}
%=============================================================

To date, phased-ALMA has been commissioned and approved for
science observations in ALMA Band 3 (3~mm) and
Band~6 (1.3~mm), where the
Global mm-VLBI Array (GMVA) and the Event Horizon Telescope (EHT) are available to serve as the respective partner networks. 
The GMVA
sites include eight stations of the Very Long Baseline Array (VLBA),
along with
the Robert C. Byrd Green Bank Telescope (GBT), the Effelsberg 100-m
Radio Telescope, the Yebes Observatory 40-m telescope, the
IRAM 30-m telescope, the Mets\"ahovi 14-m telescope, and the Onsala
Space Observatory 20-m telescope.
The GMVA data is correlated at the Max-Planck-Institut  f\"{u}r Radioastronomie (MPIfR) in Bonn (Germany). 
The EHT (as per 2017) includes  the  Arizona Radio Observatory's Submillimeter Telescope (SMT),
the Submillimeter Array (SMA), 
the James Clerk Maxwell Telescope (JCMT),
the Large Millimeter Telescope Alfonso Serrano (LMT),  
the Atacama Pathfinder EXperiment (APEX), 
the IRAM 30-m telescope, 
and the South Pole Telescope (SPT). 
The EHT data are cooperatively correlated at the MIT Haystack Observatory and at MPIfR. 
A full description of the EHT array is presented in \citet{EHT2019II}.

The current APS has been commissioned for use for continuum observations of (non-thermal) compact  sources bright enough to allow on-source phasing of the array (with correlated flux densities of $>$0.5 Jy on intra-ALMA baselines).

In Cycle 4, nine Principal Investigator (PI) projects were approved, three  in Band 3 with the GMVA  (Table~\ref{table:exp_b3}) and six in Band 6 with the EHT  (Table~\ref{table:exp_b6}). 
The projects were executed as part of  the global VLBI observing campaign  from April 2 to April 11 2017. 
\begin{table}
\caption{Projects observed in Band 3.}
\centering  
\small
\begin{tabular}{cccc}
\hline\hline                  
\noalign{\smallskip}
 Project & Target &Date  & UT range \\
\noalign{\smallskip}
\hline
\noalign{\smallskip}  
%1\,mm      &   213.1       & 215.1        &  227.1       &  229.1       &   7812.5  & 4.03    \\
2016.1.01116.V     & OJ287  & 2016 Apr 02  & 06:55:08.2  --  15:19:42.7     \\
2016.1.00413.V     & Sgr A*  & 2016 Apr 03  & 20:52:28.0  --  04:43:54.0    \\
2016.1.01216.V     & 3C273  & 2016 Apr 04  & 00:24:56.9  --  05:32:46.0     \\
\noalign{\smallskip}
\hline   
\end{tabular} 
\label{table:exp_b3}
\end{table}

\begin{table*}
\caption{Tracks and Projects observed in Band 6.}
\centering  
\small
\begin{tabular}{ccccc}
\hline\hline                  
\noalign{\smallskip}
Track & Date  & Project & Target & UT range \\
\noalign{\smallskip}
\hline
\noalign{\smallskip}  
D                  & 2017 Apr 05   &&& 04/22:12:24 --  05/09:12:39  \\
&& 2016.1.01114.V  & OJ287   &  04/22:12:24 --  05/03:22:12  \\
&& 2016.1.01154.V  & M87      &  05/03:24:01 --  05/07:17:28    \\
&& 2016.1.01176.V  & 3C279  &  05/07:19:31 --  05/09:12:39   \\
\noalign{\smallskip}
\hline   
\noalign{\smallskip}
B                  & 2017 Apr 06   &&& 06/00:18:36 --  06/16:18:34  \\
&& 2016.1.01154.V  & M87      &  06/00:18:36 --  06/08:01:34    \\
&& 2016.1.01404.V  & Sgr\,A*  &  06/08:03:18 --  06/14:39:32     \\
&& 2016.1.01290.V  & NGC1052  &  06/14:51:06 --  06/16:18:34    \\
\noalign{\smallskip}
\hline   
\noalign{\smallskip}
C                  & 2017 Apr 07   &&& 07/03:45:42  --  07/20:46:36  \\
&& 2016.1.01404.V  & Sgr\,A*  &  07/03:45:42 --  07/14:30:32     \\
&& 2016.1.01290.V  & NGC1052  &  07/19:23:51 --  07/20:46:36    \\
\noalign{\smallskip}
\hline   
\noalign{\smallskip}
A                  & 2017 Apr 10  &&&  09/23:02:48 --  10/10:01:39  \\
&& 2016.1.01114.V  & OJ287   &  09/23:02:48 --  10/03:48:34  \\
&& 2016.1.01176.V  & 3C279  &  10/03:51:14 --  10/06:20:33  \\
&& 2016.1.01198.V  & Cen\,A  &  10/06:23:07 --  10/10:01:39  \\
\noalign{\smallskip}
\hline   
\noalign{\smallskip}
E                 & 2017 Apr 11   &&& 10/21:44:54 --  11/10:31:04  \\
&& 2016.1.01114.V  & OJ287   &  10/21:44:54 --  11/00:21:54  \\
&& 2016.1.01154.V  & M87      &  11/00:23:20 --  11/05:02:44    \\
&& 2016.1.01176.V  & 3C279  &  11/05:05:06 --  11/08:44:34   \\
&& 2016.1.01404.V  & Sgr\,A* &  11/08:46:18 --  11/14:02:41     \\
\noalign{\smallskip}
\hline\hline  
\end{tabular} 
\label{table:exp_b6}
\end{table*} 
In Cycle 4 approximately 40 antennas were available for Science use. Since two or more are withheld from the phased array for online estimation of the phasing efficiency (the "comparison" antennas), about 37 antennas\footnote{The  APS cannot phase an even number of antennas \citep[see][]{APPPaper}.} within a radius of 180~m\footnote{Compact configurations  are  the most desirable for phased observations because they have smaller delays between antennas and thus delay adjustments are made less frequently.} 
 were normally phased together (which is equivalent to a telescope of $\sim$73~m diameter).

In both Band 3 and 6, 
the spectral setup includes four spectral windows (SPWs) of 1875~MHz,
two in the lower and two in the upper sideband, 
correlated with 240 channels per SPW (corresponding to a spectral resolution of 7.8125 MHz).  The full-resolution spectral data are
available in the 4.032-s integrations; the channel average data (\S~\ref{APS_spectral})
are not used for calibration.

%________________________________________________________________
\subsection{Band 3 with the GMVA}
The GMVA observed the three approved programs on three consecutive nights spanning 2017 April 2-4 (Table~\ref{table:exp_b3}). 
The spectral setup includes a total of four SPWs, centered at 86.26, 88.2, 98.26, and 100.26 GHz (Table~\ref{table:freq_b3}). 
\begin{table*}
\caption{ALMA frequency setting in Band 3.}
\centering  
\small
\begin{tabular}{ccccccc} 
\hline\hline                  
\noalign{\smallskip}
 Band & \multicolumn{4}{c}{  Central Freq. (GHz)} &   Chan. Width &   Integ. time \\
   ($\lambda$)        &   SPW\,0   &  SPW\,1 &   SPW\,2 &  SPW\,3 &   (kHz) & (s) \\
\noalign{\smallskip}
\hline
\noalign{\smallskip}  
3 (3\,mm)      &    86.268     & 88.268       &  98.328      & 100.268      &   7812.5  & 4.03   \\
\noalign{\smallskip}
\hline   
\end{tabular} 
\label{table:freq_b3}
\end{table*}
Note that only the lowest frequency ($\sim$86 GHz) SPW (0)
was recorded on VLBI disks due to the limited recording rates available at the other GMVA stations.
The list of observed sources and their calibration
intent is given in Table~\ref{table:sources_b3}. 
%-----------------------------------------------------------------------------
 % 
\begin{table*}
\caption{Observed sources (and their calibration intent) in the Band 3 projects with the GMVA.}              
\label{table:sources_b3}       
\centering  
\small
\begin{tabular}{ccccccc} 
\hline\hline                  
\noalign{\smallskip}
Project & Flux Calib. & Gain Calib. &  Bandpass Calib.   &Polarization Calib. & VLBI Calib. & Target\\
\noalign{\smallskip}
\hline
\noalign{\smallskip}  
2016.1.00413.V   & Callisto & J1744-3116 & 4C 09.57 & B1730-130 & B1921-293 & Sgr A* \\
2016.1.01116.V & J0510+1800 &  J0830+2410  & 4C 01.28 & 4C 01.28 & --- & OJ287 \\
2016.1.01216.V & Callisto & J1224+0330 & 3C279 & 3C279 & J1058+0133 & 3C273 \\
&&&&& (4C 01.28) &  \\
\noalign{\smallskip}
\hline    
\end{tabular}
\end{table*}
%--------------------------------------------------------------------------------------

%________________________________________________________________
\subsection{Band 6 with the EHT}

The EHT observed the six approved projects  in five nightly schedules or {\em tracks}, labelled A through E  (see Table~\ref{table:exp_b6}), during an observing window of ten days (where the observations are triggered based on the weather).
As a consequence, subsets of different projects shared common tracks. While this strategy was adopted to optimise the efficiency of the VLBI campaign, the arrangement of different projects within the same observing block is not a standard scheduling mode for ALMA. 

The EHT spectral setup includes a total of four SPWs,  centered at 213.1, 215.1, 227.1, and 229.1 GHz (Table~\ref{table:freq_b6}). Note that only the SPWs in the upper sideband (SPW=2, 3) 
were recorded on VLBI disks as done by the other EHT stations.
\begin{table*}
\caption{{ALMA frequency setting in Band 6.}} 
\centering  
\small
\begin{tabular}{ccccccc} 
\hline\hline                  
\noalign{\smallskip}
 Band & \multicolumn{4}{c}{ Central Freq. (GHz)} &   Chan. Width &  Integ. time \\
  ($\lambda$)            &    SPW\,0  &  SPW\,1 &   SPW\,2 &  SPW\,3 &    (kHz) &  (s) \\
\noalign{\smallskip}
\hline
\noalign{\smallskip}  
6 (1.3\,mm)      &   213.1       & 215.1        &  227.1       &  229.1       &   7812.5  & 4.03    \\
%3\,mm      &    86.268     & 88.268       &  98.328      & 100.268      &   7812.5  & 4.03   \\
\noalign{\smallskip}
\hline    
\end{tabular} 
\label{table:freq_b6}
\end{table*}
The list of observed sources and their calibration
intent is given in Table~\ref{table:sources_b6}.
%-----------------------------------------------------------------------------
 % 
\begin{table*}
\caption{Observed sources (and their calibration intent) in the Band 6 projects with the EHT.}              
\label{table:sources_b6}       
\centering  
\footnotesize
\begin{tabular}{ccccccc} 
\hline\hline                  
\noalign{\smallskip}
Experiment & Flux Calib. & Gain Calib. &  Bandpass Calib.   &Polarization Calib. & VLBI Calib. & Target\\
\noalign{\smallskip}
\hline
\noalign{\smallskip}  
Track A   & 3C279 & [ J0837+2454,J1246-0730, & 4C 01.28 & 3C279 & ---  & M87, OJ287 \\
&& J1321-4342 ] &&&& Cen A\\
&&&&&& \\
Track B& Ganymede & [ J1243+1622,J0243-0550, &  B1730-130 & 3C279 & [ 3C273, B1921-293 & M87, Sgr~A* \\
& J1058+0133$^{a}$ & J1225+1253,J1744-3116 ] &&&  B0003-066,B0130-17 ] & ngc1052 \\
&&&&&& \\
Track C& Ganymede& [ J1744-3116, &  B1730-130 &  B1921-293 &  [ B0003-066, 3c84, & Sgr~A*  \\
&&J0243-0550 ] & && B0130-17 ] & ngc1052 \\
&&&&&& \\
Track D & 3C279 & [ J1246-0730,J1243+1622,  & 4C 01.28 & 3C279 && M87, OJ287\\
& J0750+1231$^{a}$ & J0837+2454 ] &&&& \\
&&&&&& \\
Track E & 3C279 & [ J0837+2454,J1243+1622, & 4C 01.28 & 3C279 & [ B1921-293, & Sgr~A* \\
& [ J0750+1231,  & J1246-0730,J1744-3116 ] &&& B1730-130 ] & M87 \\
& J1229+0203 ]$^{a}$  &&&&& OJ287 \\
\noalign{\smallskip}
\hline    
\end{tabular}
\tablenotetext{a}{J1058+0133, J0750+1231, and J1229+0203 were  observed with the \texttt{CALIBRATE\_FLUX} intent, but were not used as flux calibrators. }
\end{table*}
%--------------------------------------------------------------------------------------

%________________________________________________________________
\subsection{Observing schedules and data structure}

The VLBI schedule is governed by the VLBI EXperiment (VEX) file. ALMA observes the VLBI targets  
specified in the  VEX file with the APS actively phasing the array. 
To enable calibration of the ALMA array,  a block of 15-min duration before the start of the VLBI schedule is devoted to observations of flux density, bandpass, and polarization calibrators in ordinary interferometric mode (i.e. with the APS off). Scans on the phase and polarization calibrators (also in ordinary interferometric mode) are then cycled through the schedule in the gaps between VLBI scans.  
Therefore, the ALMA scheduling blocks (SBs) include scans when the phasing is activated (\texttt{APSscans})  and scans during ordinary ALMA observations (\texttt{ALMAscans}). 
These two modes of operation are usually referred to as {\it ALMA-mode} and {\it APS-mode}.   
This operational scheme enables full calibration of the ALMA visibilities.   

In principle, the ALMA calibrations within each project on any given
track would be sufficient to properly calibrate the project.
However, in practice, some calibration scans were not completed,
and it became necessary to extend the calibration across the full
observing night (track).  This is not normally done with ALMA observations. 
Instead, ALMA would normally re-observe such SBs. 
For VLBI, however, this is not an option given the participation of the other global sites. 
As a result, {\em ad-hoc} calibration procedures were developed to
handle the QA2 of VLBI experiments
(see Sections~\ref{calibration} and \ref{polcal}).

\subsubsection{Choice of reference antenna}
Since the antennas in the phased array are in phase with the reference antenna, it follows that the calibration needed for the VLBI correlation is in fact essentially that of the
reference antenna.  Specifically, this calibration effort is dominated by the X-Y phase difference as well as the delay between these two signals (\S~\ref{polcal}).  
 If the antenna is shadowed, however, this will seriously   compromise the calibration. It is therefore imperative to design the ALMA SB to insure that the reference antenna will not be in shadow for any part of the observations\footnote{This task is much harder with more compact arrays normally used for phased observations.}.

%==============================================================================
\section{Ordinary data calibration}
\label{calibration}
%==============================================================================

As described in \S~\ref{APS}, during phased-array operations, the data path from the antennas to the ALMA correlator is different with respect to standard interferometric operations and some corrections (e.g. the baseband delays) are turned off while the APS is active. 
This makes the  calibration of \texttt{APSscans}  and \texttt{ALMAscans} during VLBI observations  intrinsically different.  
As a consequence, the two types of scans need to be processed independently within the Common Astronomy Software Applications package (CASA)\footnote{The "observation intents" of each scan stored in the measurement set (MS) metadata indicate which scans are observed in ordinary ALMA mode and which ones were observed in APS mode: the latter contain the string \texttt{CALIBRATE\_APPPHASE\_ACTIVE}.}.  
ALMA calibrators may be processed in CASA with standard procedures, whereas VLBI targets  may still be processed by the same CASA analysis tasks but with some essential modification in the procedures. 

The special steps added to the standard ALMA QA2 calibration are described in the following subsections.
Details on polarization calibration will be presented in \S~\ref{polcal}. 
The data reduction presented here was done using CASA version 5.1.1 (but also versions 4.7.2 and 5.3.0 were successfully  tested). 
The work-flow diagram for  calibration of APP interferometry data in CASA is sketched in Figs.~\ref{casacal_workflowa} and \ref{casacal_workflowb}.

%--------------------------------------------------------------------------------------
\begin{figure*}
\centering
   \includegraphics[width=0.75\textwidth]{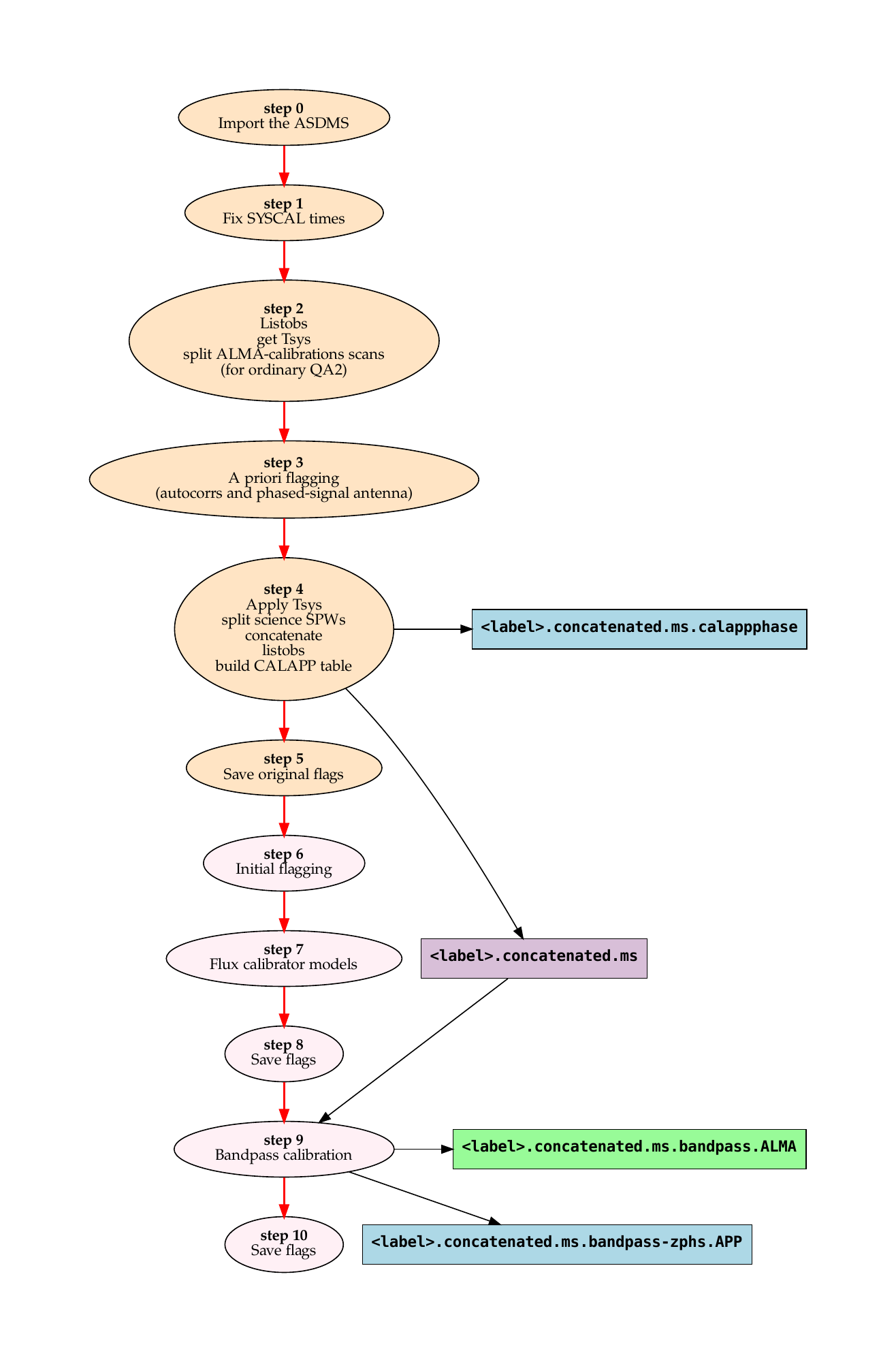}
\caption{
Part 1 of the workflow diagram for CASA calibration of APS interferometry data (continued in Fig. \ref{casacal_workflowb}).
The script is partitioned into 21 steps which are represented by the
corresponding ovals in the flow.  
The colors of the ovals gather the various steps into four main parts of the data processing path: 
(a)  initial data import (light orange); 
(b) ordinary calibration similar to normal ALMA interferometry modified for APS-mode; 
(c) polarization calibration specific to APS-mode; 
(d)  imaging and QA2 products packaging.
The blue and green boxes refer to calibration tables generated and
delivered to the VLBI correlators for running \texttt{PolConvert};
the purple boxes refer to measurement sets uv-data files generated at various stages;
the orange "artifacts" box refers to a directory full of diagnostics and 
other script output included in the deliverables to the correlators;
and a few text objects are shown in grey.
For clarity, not all of the arrows
(e.g., those to the measurement sets) are shown,
or arrows are shown to flow through products (e.g., step 15).
}
\label{casacal_workflowa}
\end{figure*}

\begin{figure*}[h!]
\centering
\vskip -2cm
\includegraphics[width=0.95\textwidth]{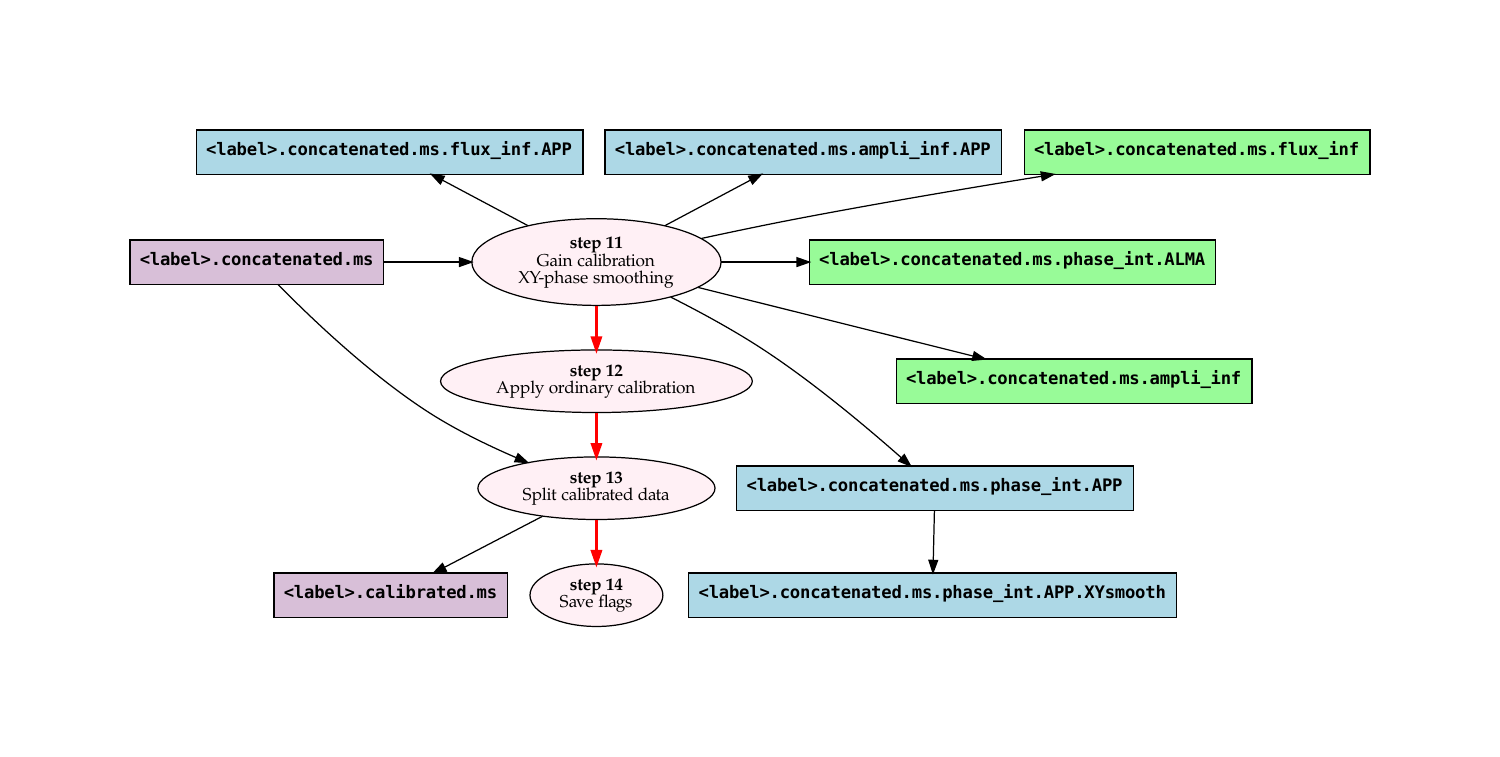}
\vskip -1cm
%\hline
\vskip -1cm
\includegraphics[width=0.95\textwidth]{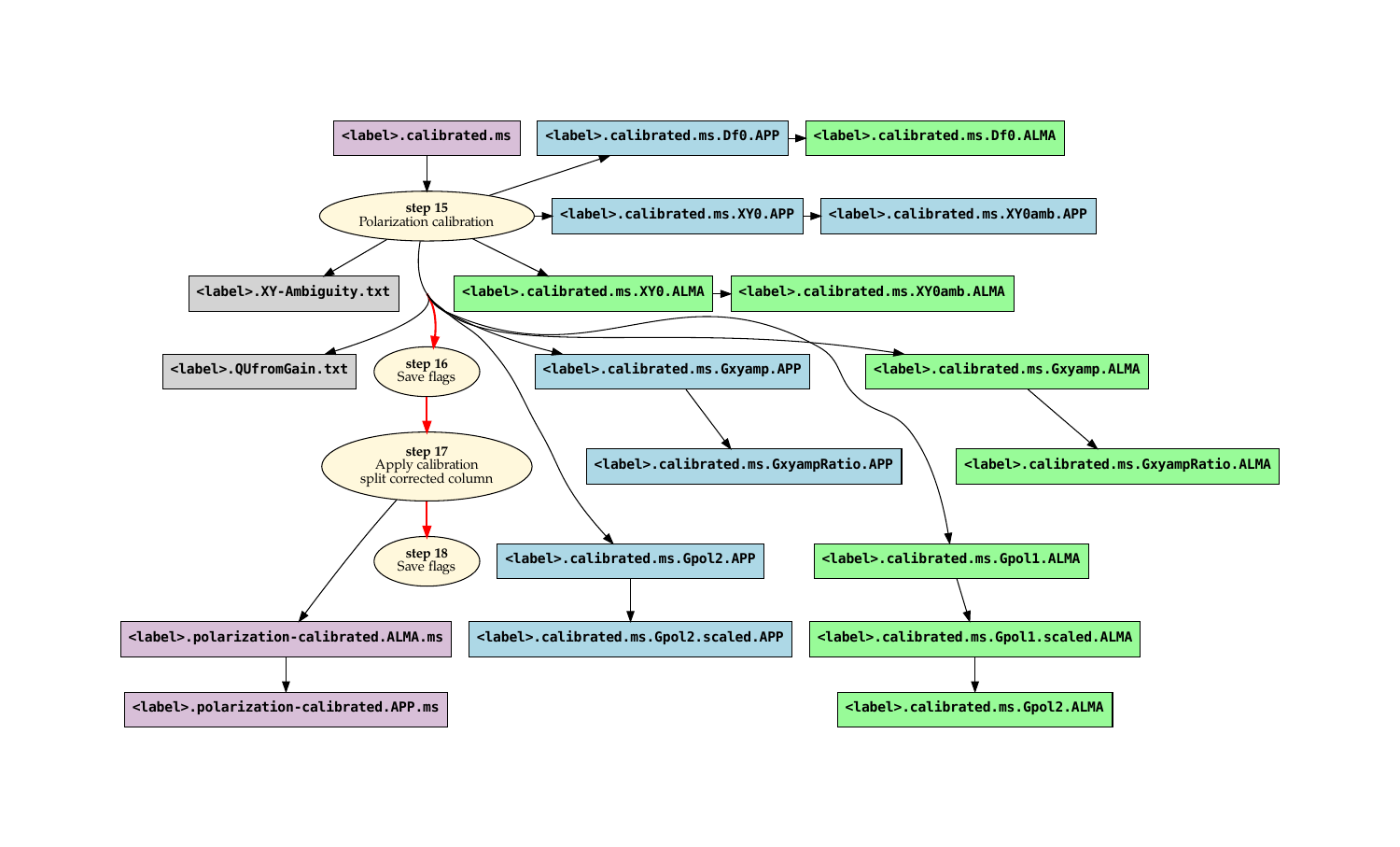}  
\vskip -1cm
%\hline
\vskip -1cm
\includegraphics[width=0.95\textwidth]{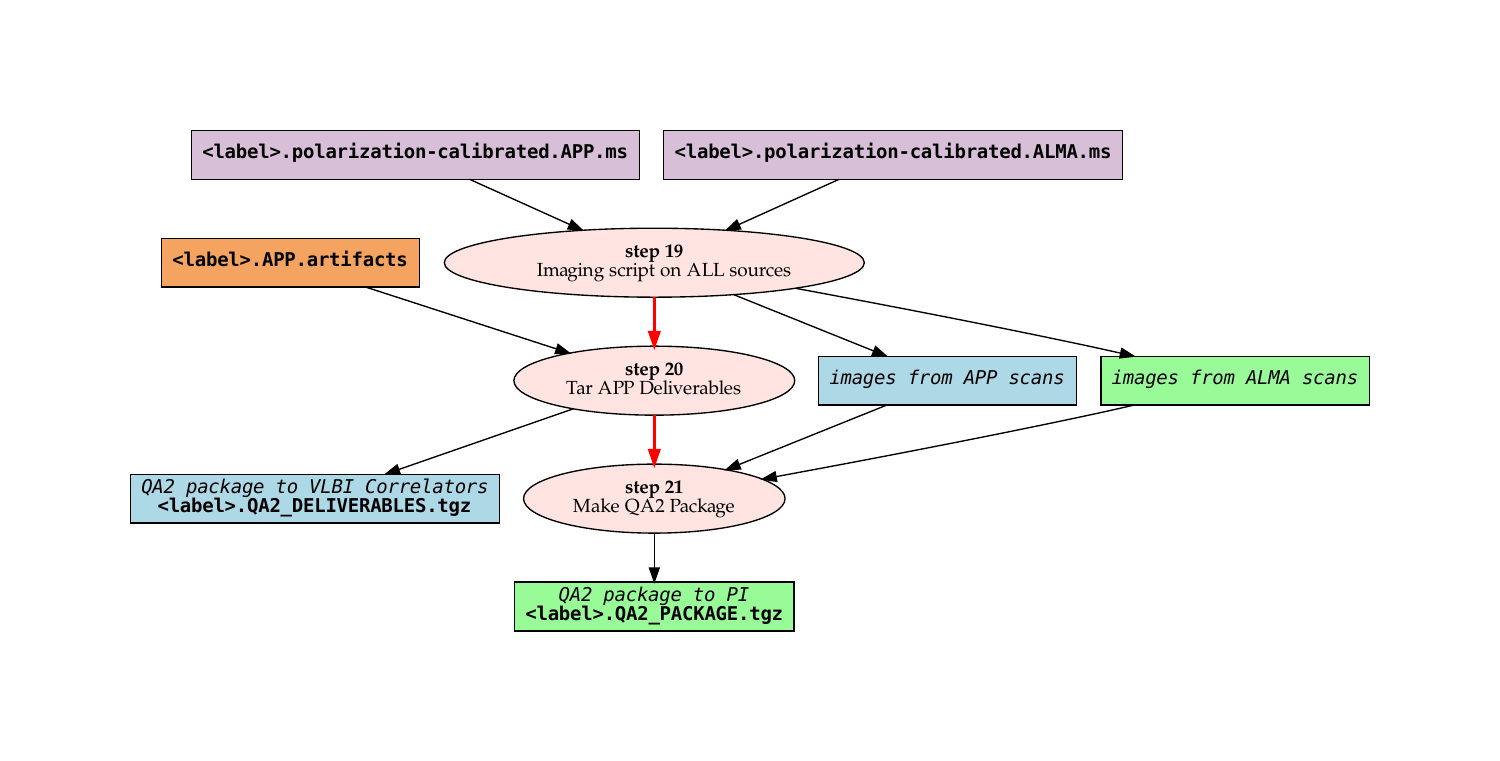}
\vskip -1cm
\caption{
Continuation of the workflow diagram for CASA calibration of APS interferometry data from Fig. \ref{casacal_workflowa}.
}
\label{casacal_workflowb}
\end{figure*}
%%--------------------------------------------------------------------------------------

%________________________________________________________________
\subsection{Pre-calibration stage} 
%________________________________________________________________

%%----------------------------------------------------------------------
\subsubsection{Data import} 
\label{import}
In VOM, ALMA still produces ordinary ASDM data files.
However, these ASDM files contain additional metadata specific to the APS, 
including information on \texttt{APSscans}  and \texttt{ALMAscans}
and a \texttt{CalAppPhase} table which captures the performance of the phasing system.
The table reports mainly phase values (one entry per subscan per channel average per polarization per antenna) and the
time range within the scan of stable phase. 
The table also contains entries defining the category of individual antennas (phased antennas, reference antenna, and comparison antennas) and an indication of whether or not this represents a change from the previous scan. 
 This table may be used in conjunction with the ALMA data for a particular scan to calculate the phasing efficiency (see \S~\ref{pheff}). 
 
As opposed to standard ALMA calibration procedures, the ASDM files for a given VOM project or track are concatenated before the calibration. 
This step is necessary for  \texttt{PolConvert},  since there is no CASA task for the concatenation of already-existing calibration tables, and  \texttt{PolConvert} interprets different calibration tables as  incremental calibration (i.e., not as tables to be appended). 
The data on different \texttt{CalAppPhase} tables from  different ASDM files are collected into a single \texttt{CalAppPhase} table.

Another important difference to the standard procedure is that water vapor radiometer (WVR),   system-temperature (\tsys), and 
antenna-position corrections are not applied to the data before concatenation. 
The reason for this is that TelCal solves for the antenna phases by
self-calibrating the intra-ALMA cross-correlations with no a-priori \tsys\ and WVR
corrections.
Applying these corrections before the antennas gain calibration (see next subsections), 
 implies that the phase gains would not be derived exactly on the same data used by TelCal,
and the incremental phase gains (with respect to TelCal solutions) could be biased. 
The effect of this potential bias is likely more important for the WVR   than the \tsys\ corrections. 
However, an additional good reason to avoid  using  \tsys\ corrections is that in the case some  antennas have failed \tsys\  measurements, 
 opacity corrections would be applied for some phased antennas and not in others\footnote{Note that in ordinary ALMA observations,  
  antennas (and/or scans) with failed $T_{sys}$ measurements are usually removed from the analysis, but this is not an option in APS observations, because those antennas have already been added to the phased sum.},  biasing the phased-array calibration  with \texttt{PolConvert}  (\S~\ref{polconvert}). 
In short, by not applying these a-priori corrections,  all phased antennas are treated equally in the calibration (the phased signal for VLBI is an {\it unweighted} sum of the signals from all the phased antennas).

%%----------------------------------------------------------------------
\subsubsection{Data flagging} 
\label{flagging}
Standard a-priori flagging of autocorrelated data (pointing and atmosphere measurements, times when the antennas were slewing, etc.) is applied to individual MSs (one per execution block) before concatenation. 
One main difference with respect to standard procedures, is that in the pre-calibration stage no data  flagging is applied for the shadowing among the antennas. This is because the APS software does not flag  phased antennas based on shadowing, so  flagging them offline during calibration would again bias the phased-array calibration  with \texttt{PolConvert}  (\S~\ref{polconvert}). 
While this limitation may affect the ALMA visibilities (e.g., introduce cross-talk between the antennas involved in the shadowing or degrade the polarization purity of the signal), shadowing flags can still be applied after the phase calibration and before the polarization calibration, since the latter is done using all the observations of the polarization calibrator together (i.e., any visibility flagged due to the shadowing would not affect the calibration process).

In APS observations, the sum antenna, which stores the phased signal, is  a virtual\footnote{Technically, it does not exist to the ALMA control system and thus the normal metadata related to an antenna do not exist either.} antenna
and therefore must be flagged before calibration. 
  In principle, the visibilities of any baseline related to the sum antenna will be equal to the auto-correlation (with some delay) of the signal of the other antenna in the baseline, plus a small contribution from the cross-correlations with all the other phased antennas in the array.
Including these baselines would  introduce a bias in the antenna gains and thus must be excluded.
In Cycle 4, the sum antenna appears to CASA as "DV03" and was automatically flagged.

In addition to the standard a-priori flagging, the first few integrations of each source observing block,  during which the phases are still being adjusted, are flagged as well. 
In particular, the APS scans are started two
sub-scans (ca. 18~s) prior to the start of the VLBI recording to allow the APS to calculate and apply the phase adjustments. The "phase-up" occurs after the first sub-scan; the second sub-scan receives a phase update in the first 4-s integration (Fig.~\ref{casa-scans}).  Thus, the first $\sim$~22~sec of each $\sim$~4-8~min block (which is the typical length of a VLBI scan) are routinely flagged to prevent using poorly phased data.

%________________________________________________________________
\subsection{Bandpass Calibration} 
\label{bandpass}
\begin{figure*}[ht!]
\includegraphics[width=10cm]{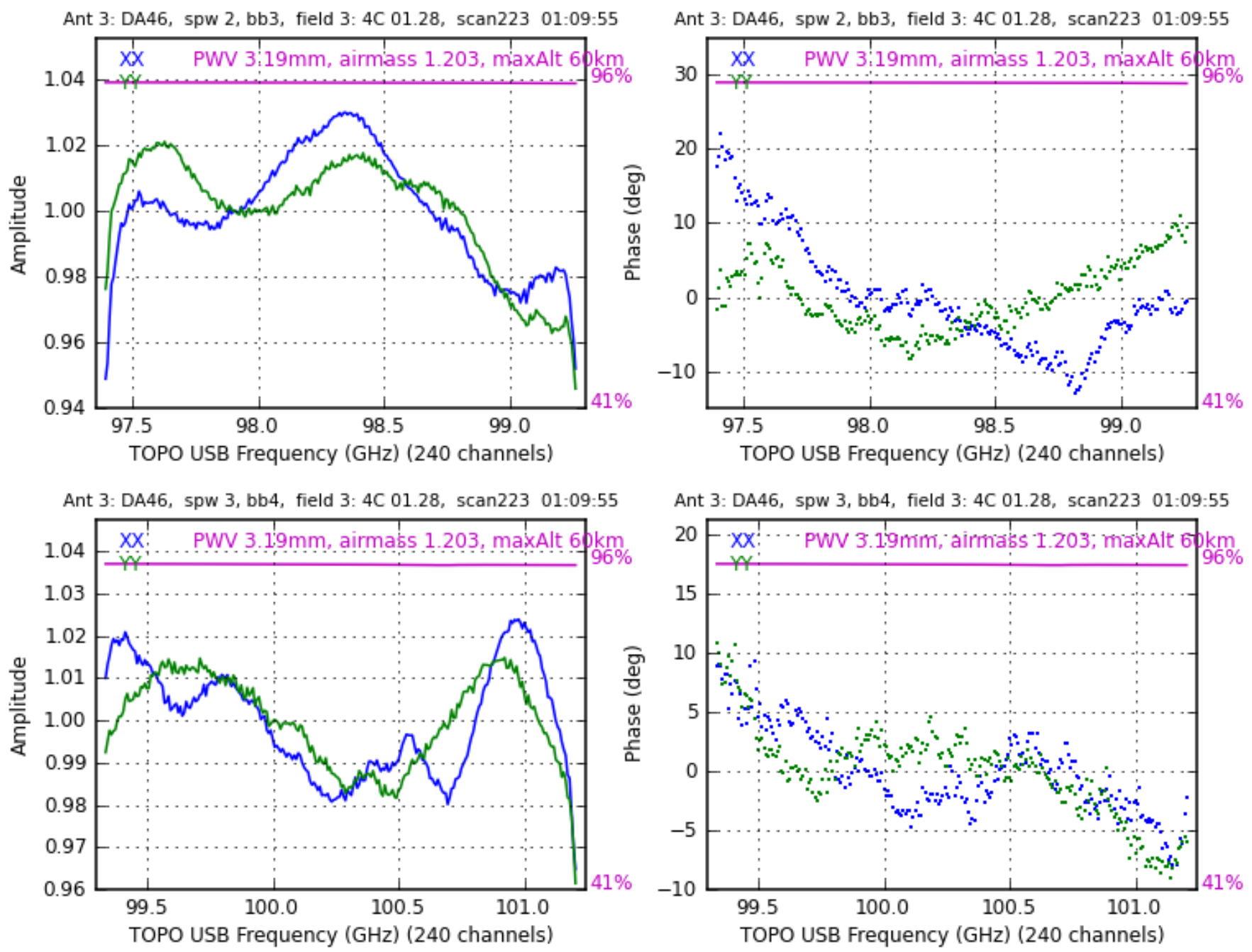} 
\hspace{-10mm}
\vspace{0.5cm}
\includegraphics[width=10cm]{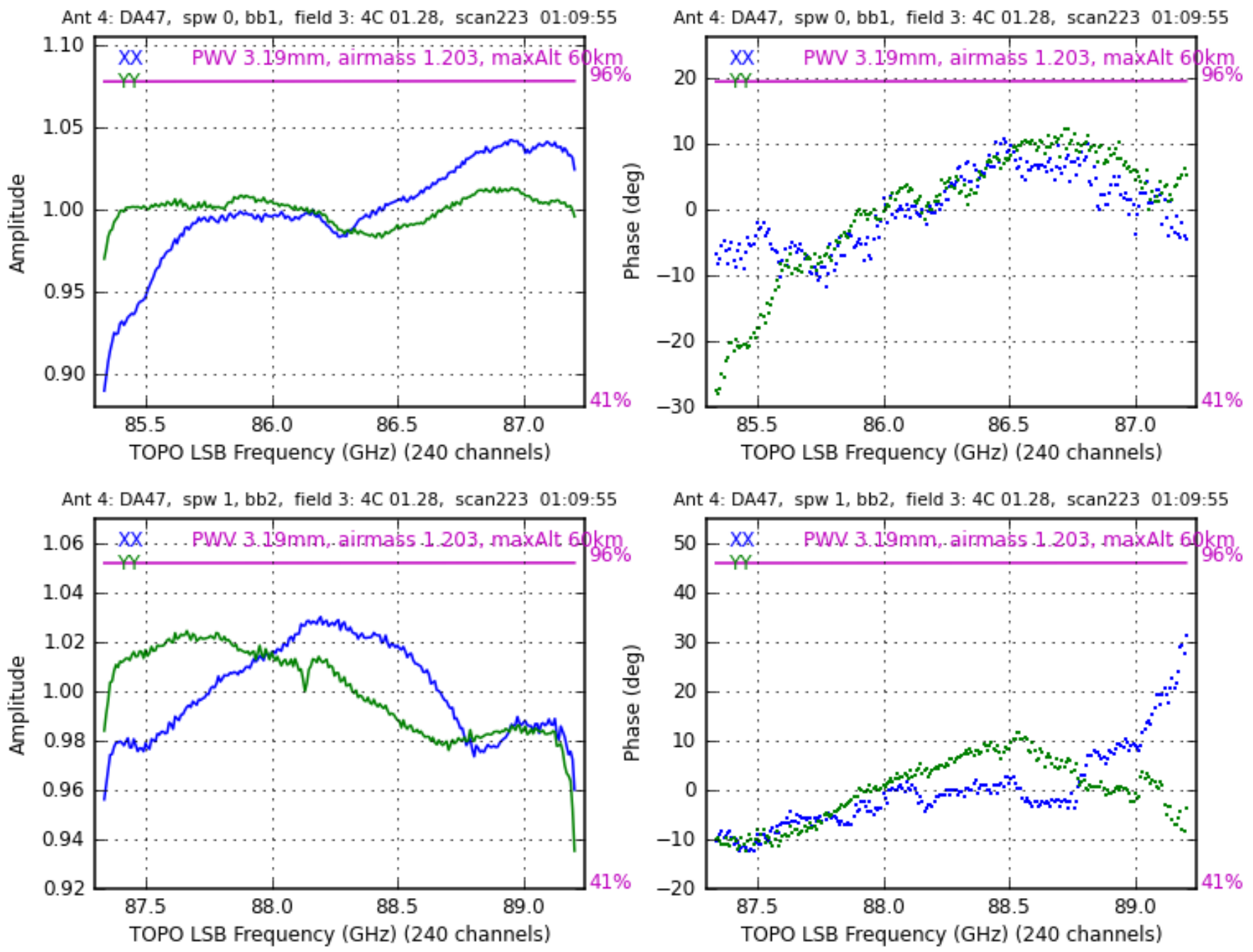}
\includegraphics[width=10cm]{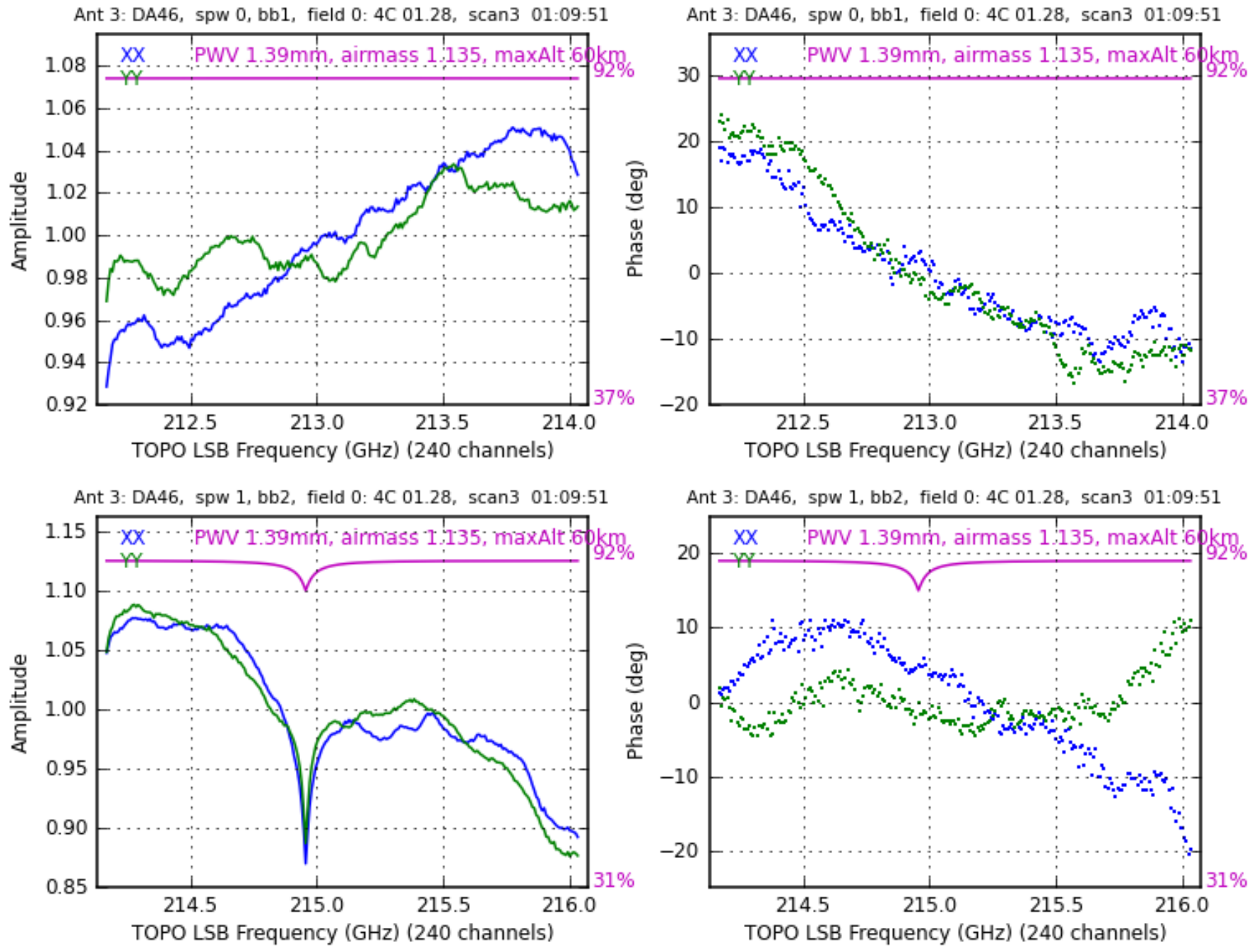} 
\hspace{-10mm}
\includegraphics[width=10cm]{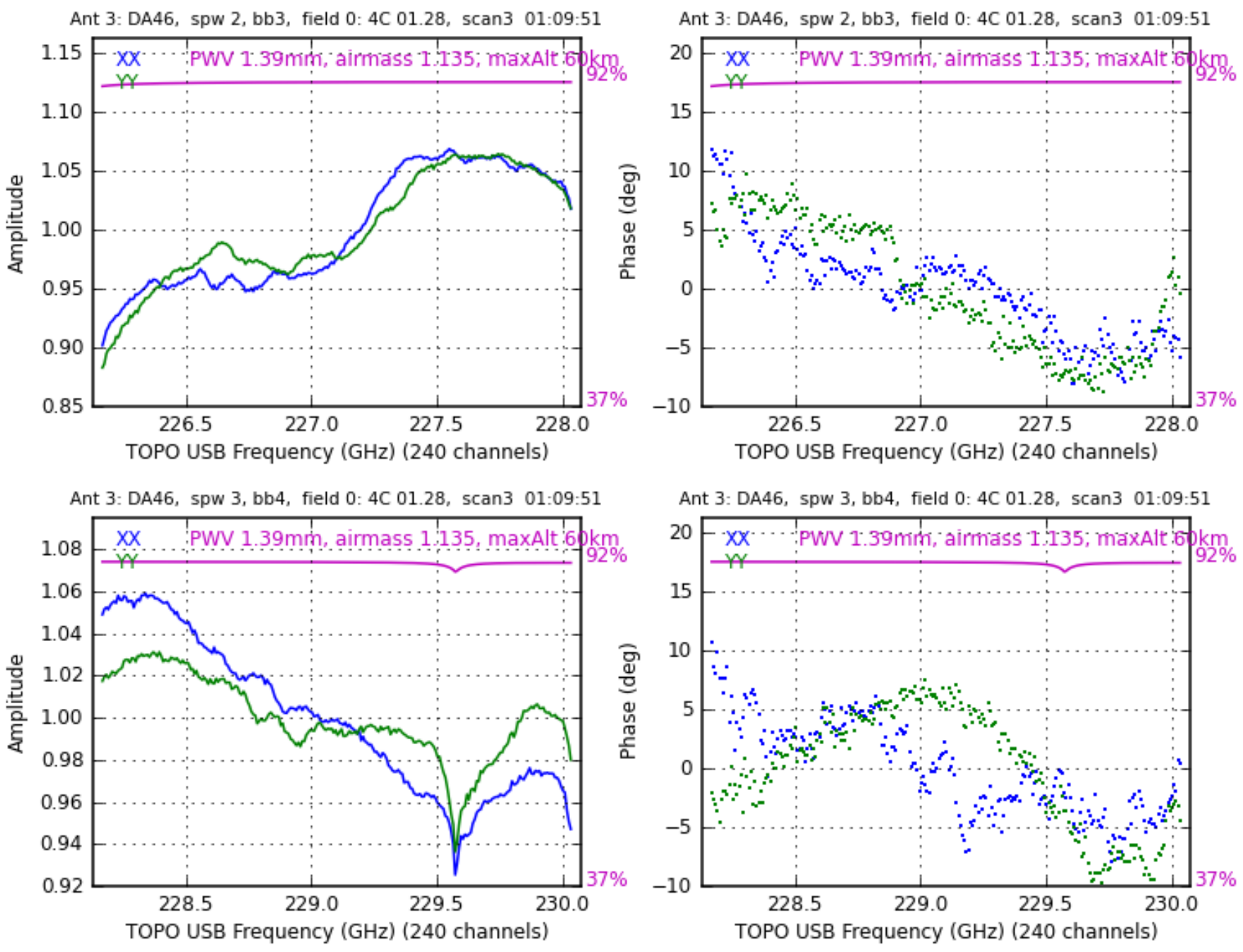}
\caption{Bandpass (amplitude and phase) of the phased antenna DA46  for SPW=0,1 (left panels) and SPW=2,3 (right panels) in Band 3 (project 2016.1.01116.V; top panels) and Band 6 (Track A; bottom panels), respectively. The bandpass calibrator is 4C~01.28 in both bands. 
Note the prominent atmospheric (ozone) absorption lines (at $\sim$214.9 GHz and $\sim$229.6 GHz).
}
\label{fig:BP}
\end{figure*}

The bandpass calibration tables are derived from observations of the bandpass calibrators
in ALMA mode.  The same targets are usually also observed in the VLBI schedule for the VLBI bandpass calibration. 
Figure~\ref{fig:BP} shows the amplitudes and phases of the bandpass solutions for one of the phased antennas in both Band 3 and Band 6. 
Two different bandpass tables are used in the calibration. 
One is obtained from an ordinary calibration using the \texttt{ALMAscans} 
and is applied to the \texttt{ALMAscans}. 
The other table is a copy of the  first one, but with all phases zeroed;  
this is applied to the \texttt{APSscans}. 
This scheme is necessary because of the intrinsic difference between \texttt{ALMAscans} and  \texttt{APSscans}. 
In ALMA-mode, any difference between the X and Y bandpass phases reflects residual cross-delays from the ALMA correlator model, which is not used in APS-mode. 
In APS-mode, the TelCal phase adjustments introduce additional X-Y cross phases (in frequency chunks) which are, by construction, zero for the reference antenna. The phases of the chunks must be solved for using the polarization calibrator, but with no bandpass phases applied to the data.

Since APP observations may be done with a 
"flexible'' array\footnote{%
Since VLBI observations must carry on with the VLBI schedule in the event of antenna failures, the APP
design allows for array antennas to be removed and restored as
necessary.} in which
 different antennas can be present at different times during the execution of a given project or observing track, 
 there could be cases where some antennas are not in the array during observations of the bandpass calibrator (but are either added later or dropped earlier). Such  antennas would miss bandpass calibration and would be flagged following standard procedures. 
   Two solutions are implemented in the APP calibration scheme to address this issue: 
   a)  more than one source can be listed as  bandpass calibrator (in \texttt{ALMAscans}); 
   b) in  cases where some (well-functioning) antennas did not observe any suitable bandpass calibrator during an entire track, 
   a flat bandpass is assumed by setting a unity-gain for the bandpass amplitude solutions (this avoids these good antennas to be  flagged). 
    The latter option was never used in the Cycle-4 data processing. 

%________________________________________________________________
\subsection{Phase Gains Calibration}
\label{phase-gains}
\begin{figure*}[ht!]
\hspace{-5mm}
\includegraphics[width=10cm]{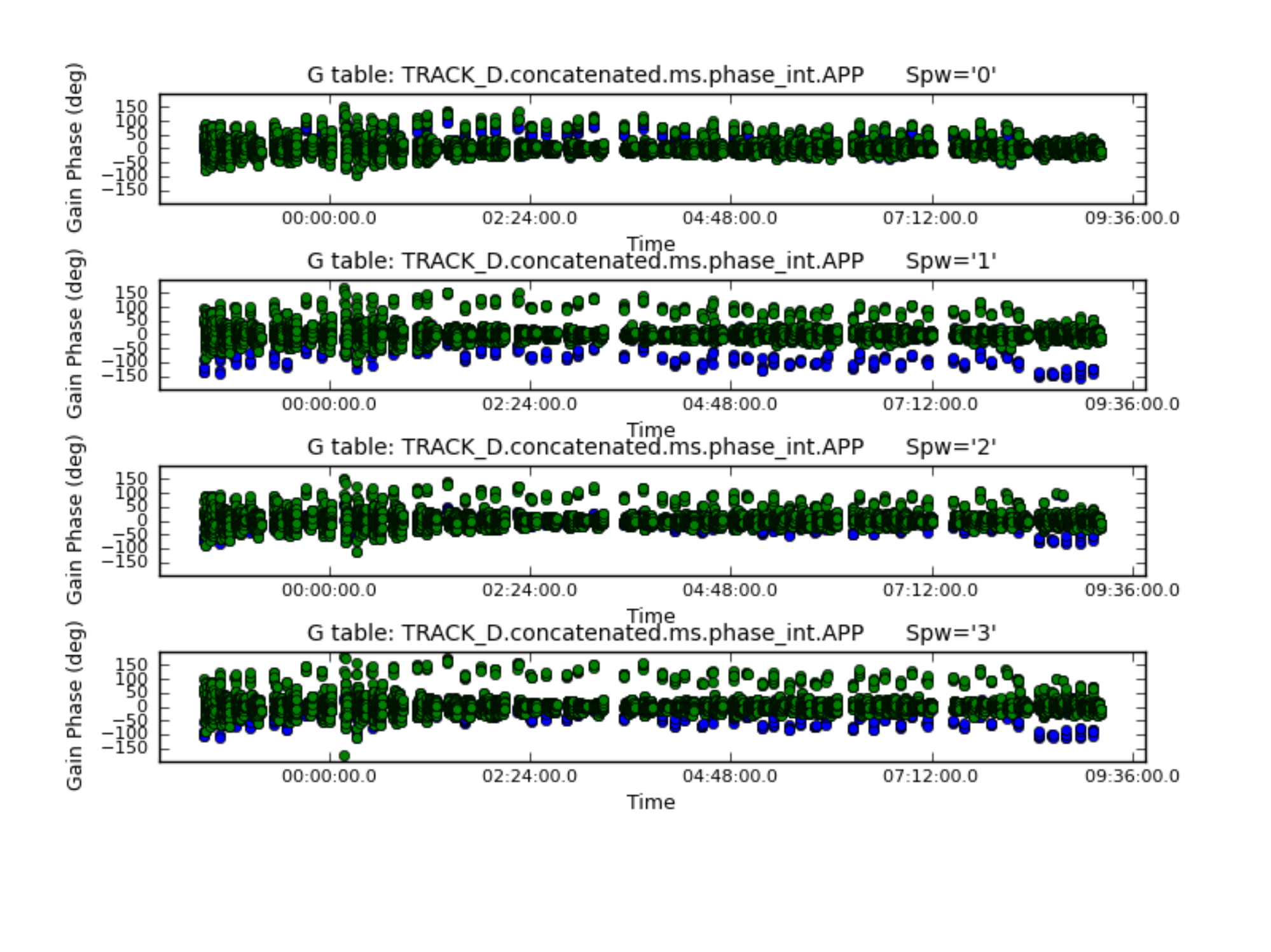} \hspace{-5mm}
\includegraphics[width=10cm]{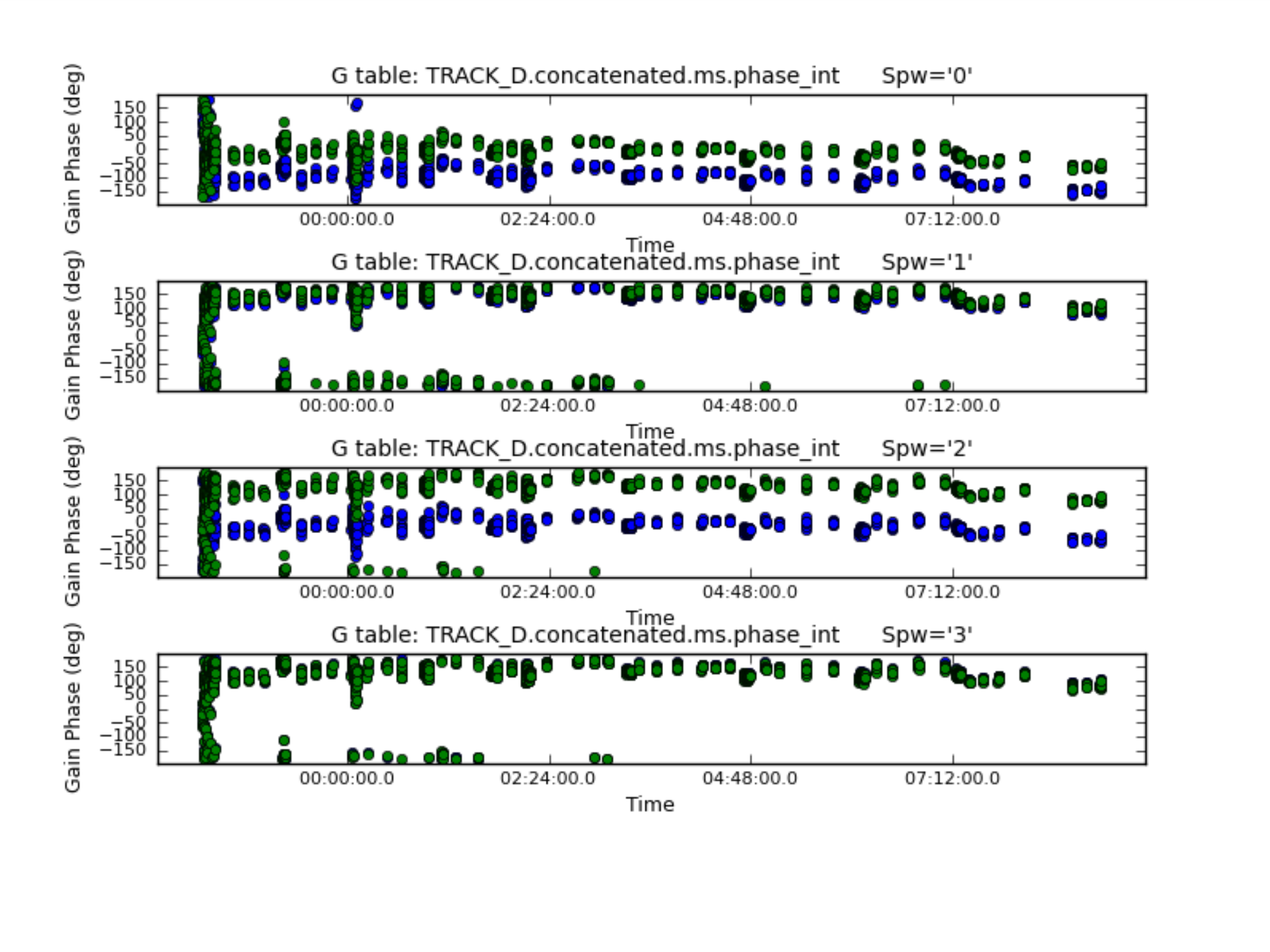}
\vspace{-15mm}
\caption{Phase gains of the phased antenna DA41 in Band 6 (Track D) in \texttt{APSscans} (left) and \texttt{ALMAscans} (right). 
Blue and green show XX and YY polarizations. 
The points with phases far from zero 
correspond to the  first integrations of every VLBI scans  where the antenna are not yet properly phased (see \S~\ref{APS_timing} and Fig.~\ref{casa-scans}).
}
\label{fig:phase}
\end{figure*}

The phase gains are also intrinsically different between the \texttt{ALMAscans} and \texttt{APSscans} (the bandpass tables to be pre-applied are different), and are therefore found separately for each set of scans. 
The \texttt{gaincal} task derives phase gains tables  
with a solution interval equal to the integration time of the ALMA correlator (\texttt{solint = 'int'}). 
Figure~\ref{fig:phase} shows the time evolution of phase gains for a typical phased antenna both in \texttt{APSscans}   and \texttt{ALMAscans}.  
For the \texttt{APSscans} (Figure~\ref{fig:phase}, left), the phases are around zero with no offset between the polarizations.  
During ordinary ALMA observations (Figure~\ref{fig:phase}, right), there are clear phase offsets between polarizations (X/Y cross-phase) and the phases are offset from zero. 

\begin{figure*}[ht!]
%\centering
\vspace{-5mm}
\hspace{-5mm}
\includegraphics[width=10cm]{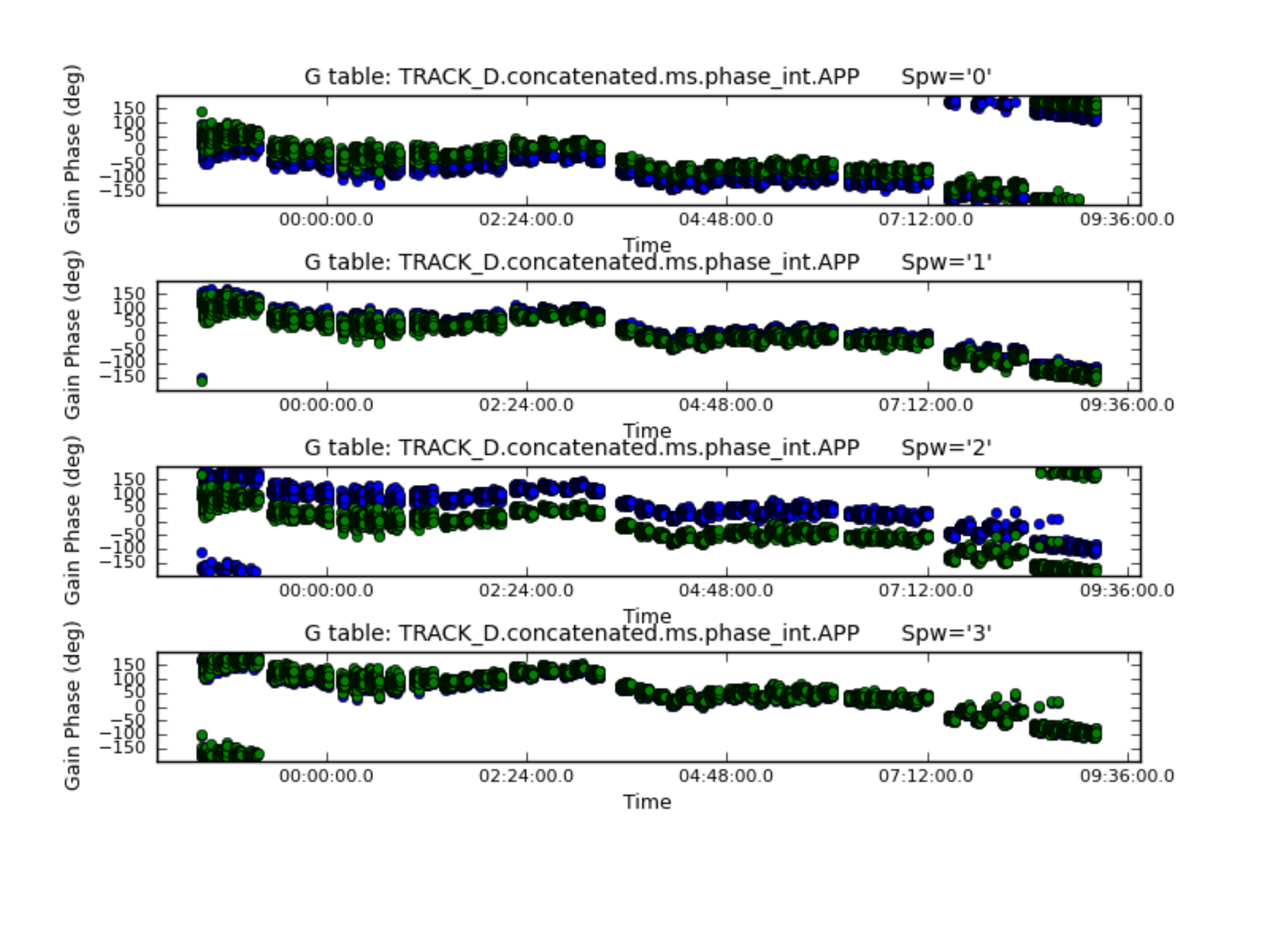} \hspace{-5mm}
\includegraphics[width=10cm]{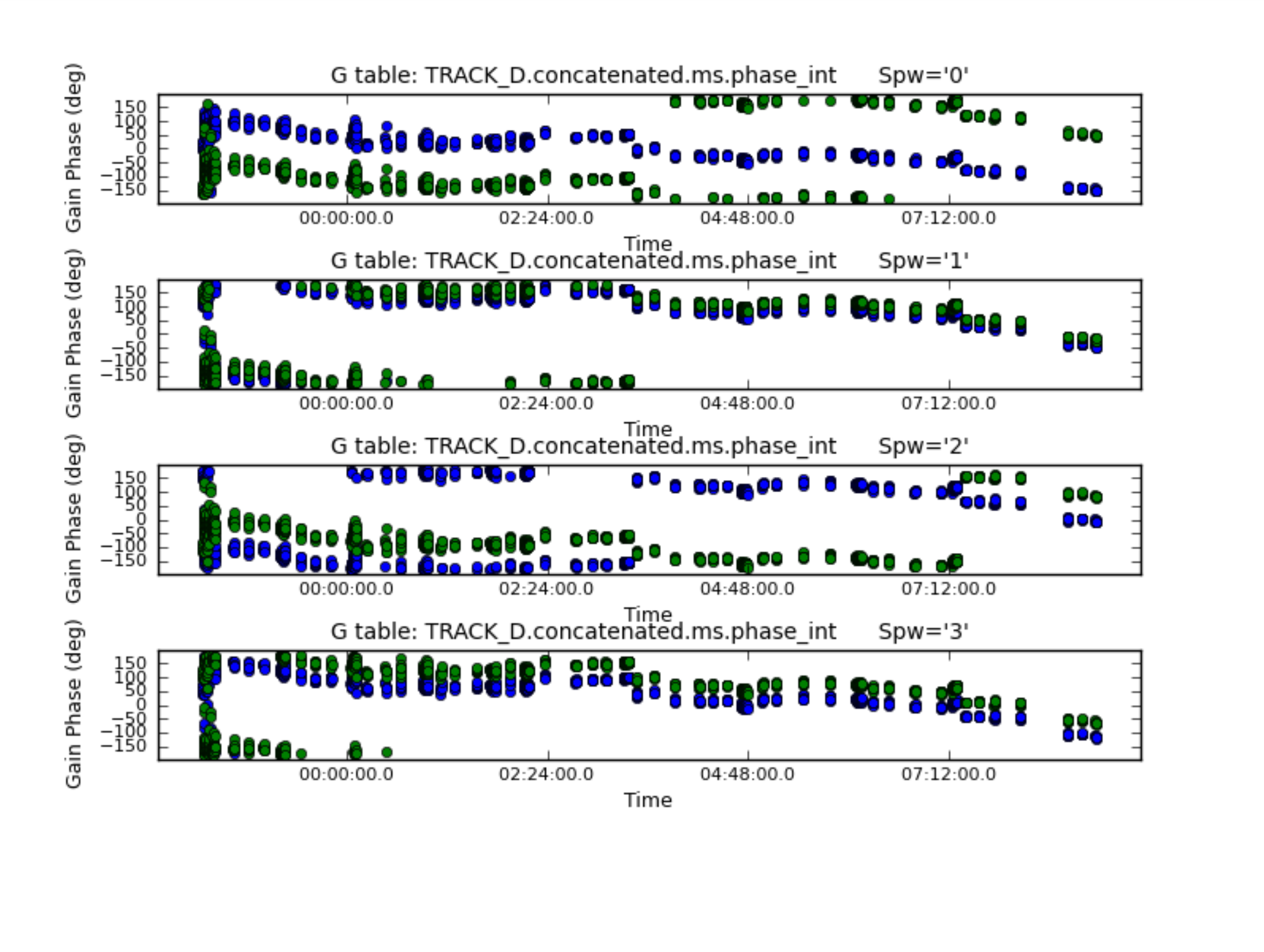}
\vspace{-15mm}
\caption{Phase gains of the comparison (i.e. non-phased) antenna DA64 in Band 6 (in Track D) in \texttt{APSscans} (left) and \texttt{ALMAscans} (right).
Note that also in this case,  there is a clear difference between \texttt{APSscans} and \texttt{ALMAscans}, owing to the fact that the  corrections from the ALMA correlator
model are not applied in the APS-mode.  Blue and green points show XX and YY polarizations, respectively.}
\label{fig:phase_ctrl}
\end{figure*}

Figure~\ref{fig:phase_ctrl} shows the time evolution of phase gains for a "comparison" antenna (i.e., an
antenna participating in the observations, but not being phased). 
Also in the case of a comparison antenna, there is a difference  (in the phases of each polarization channel)  between \texttt{APSscans} and \texttt{ALMAscans}, owing to the fact that the  corrections from the ALMA correlator model are not applied in the APS-mode.

%________________________________________________________________

\subsection{Amplitude Calibration and Absolute Flux-density Scale}
\label{ampcal}
The amplitude calibration and the absolute flux-density scaling are mostly performed  as in standard ALMA observations. 
This process uses a primary flux-density calibrator and is based on self-calibration  of the observed sources, and consists of three steps:

\begin{enumerate}
\item The CASA task \texttt{setjy} is used to scale the model of the flux-density calibrator to its correct value. The models of all the other sources are set to a flux density of 1\,Jy.

\item The CASA task \texttt{gaincal} is then used to calibrate the amplitudes (with a solution interval  equal to the  scan length) of the antenna gains for all sources, after applying on the fly the bandpass corrections (\S~\ref{bandpass}) and the phase gains corrections (\S~\ref{phase-gains}). 

\item The CASA task \texttt{fluxscale} uses the amplitude gains (generated in the previous step) to bootstrap the flux-density from the primary flux-density  calibrator into all the observed sources  
which will also be scaled to Jy units.

\end{enumerate}

We note that the {\em changes} in atmospheric opacity (for each source) are encoded in the gain tables generated in step \#2. For instance, a higher opacity for a given VLBI scan will automatically result in a higher amplitude correction for all the affected antennas in that scan.

\begin{figure*}[ht!]
\hspace{-5mm}
\includegraphics[width=10cm]{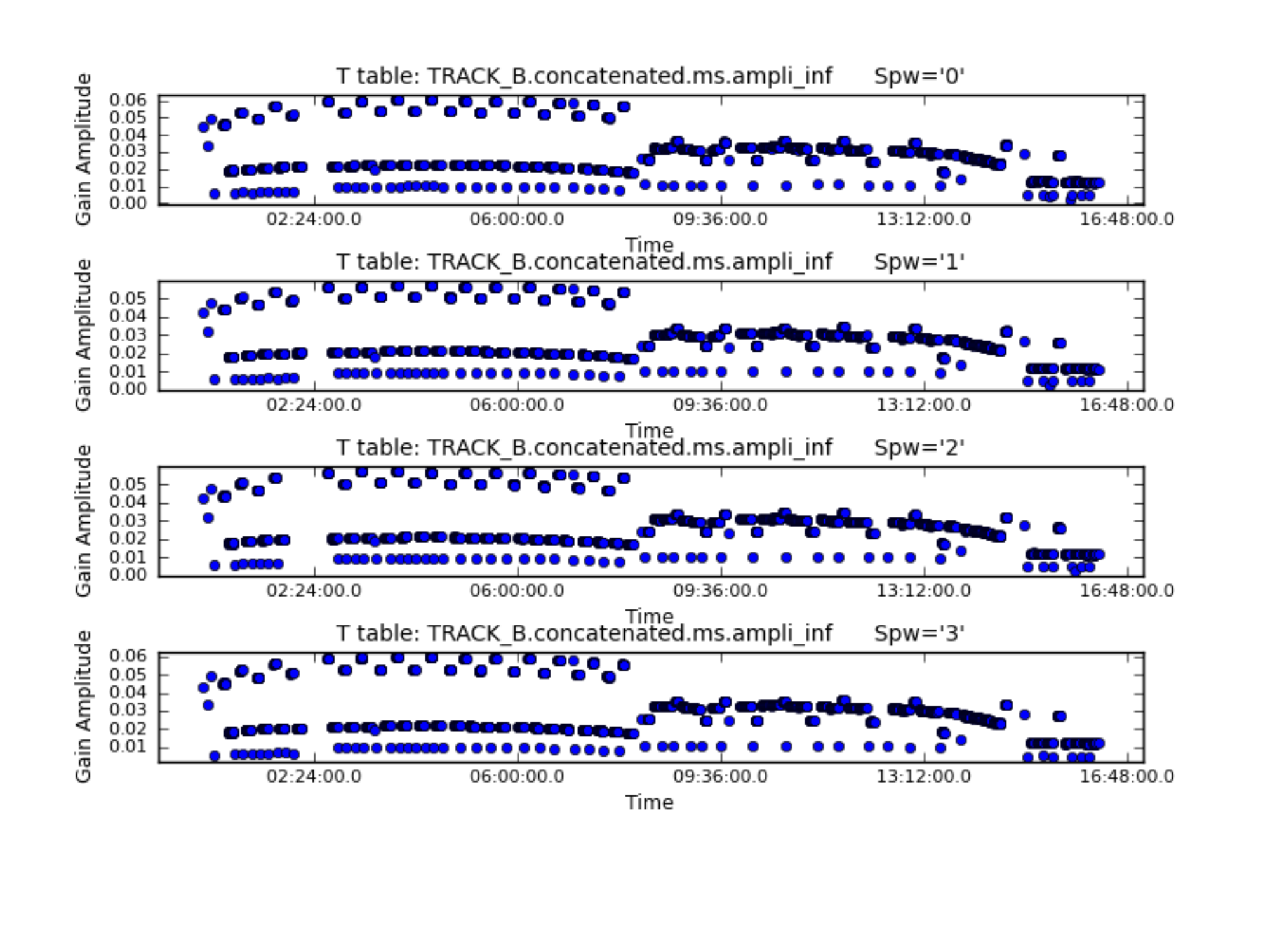} \hspace{-10mm}
\includegraphics[width=10cm]{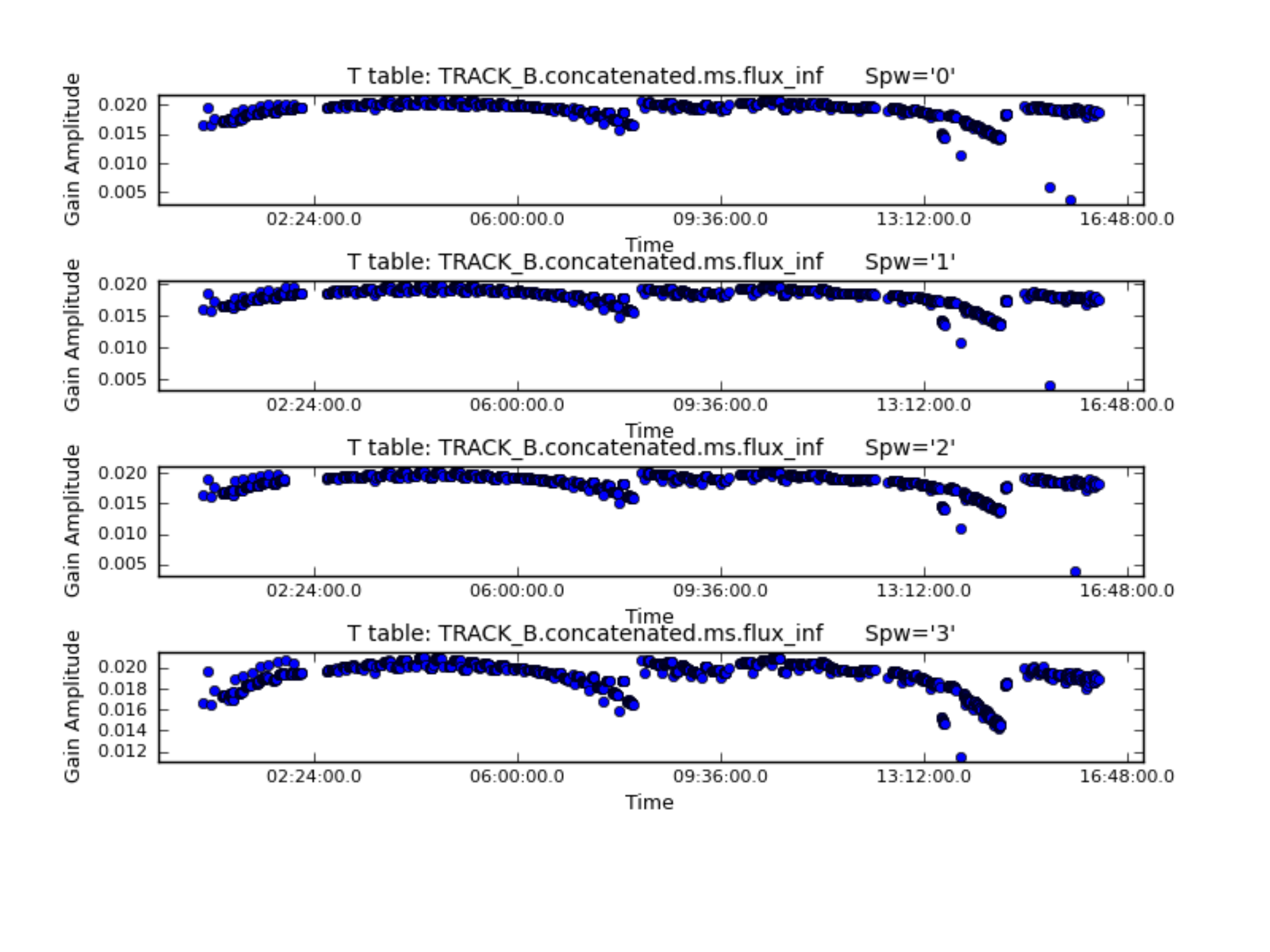}
\vspace{-15mm}
\caption{Amplitude gains for antenna DA\,41 during track B  in Band 6. In the left panel, unity flux-densities are used for all sources but the primary calibrator, Ganymede, whereas the right panel shows the gains after bootstrapping the flux density from the primary calibrator to all sources. 
}
\label{fig:amp}
\end{figure*}

Figure~\ref{fig:amp} shows the amplitude gains for the phased antenna DA41 in Band 6, including  \texttt{APSscans} and \texttt{ALMAscans}. 
The left panel shows the  gains  assuming a normalized  flux density for all sources (calculated in step \#2). 
The right panel shows  the amplitude gains  after bootstrapping the flux density from the primary calibrator to all sources  (calculated in step \#3). 
Note that most of the spread observed in the gains is removed after the absolute flux-density scaling\footnote{The  amplitude gains  should be independent of the observed source and thus  reflect the true antenna gains.} (although  some
dependence on the elevation remains, especially near the end of the track when sources are typically setting).

At this stage there are a few subtle differences with respect to standard ALMA amplitude calibration procedures. 
First, since each mode needs a different bandpass and phase gain tables, the amplitude gains must be found separately for \texttt{APSscans} and \texttt{ALMAscans}. 
Secondly,  the gain calibration is performed in "T" mode (i.e., one common gain for the two polarizations), in order to avoid altering the X/Y amplitude ratios for the polarization calibrator (this would  affect the estimate of the QU Stokes parameters from the XX and YY visibilities vs. parallactic angle; see \S~\ref{qu}). 
Thirdly, the  \tsys\ measurements  at the individual ALMA antennas, normally used to track  the atmospheric opacity, are not used in the calibration. 
Instead, in the APS data calibration, any effect from the time-variable atmospheric opacity during the observation of a given source is tracked by the amplitude self-calibration.  
While this scheme effectively removes the bulk of the opacity effect, 
it leaves a global scaling factor in the amplitude gains that is related to the difference between the opacity correction in the observation of the primary flux calibrator and the (average) opacity in the observation of a given source.
Such a difference should be of the order of a few \%, for high antenna elevations, but it could be much higher if the air mass difference between the primary flux calibrator and the target sources is higher (i.e., at low elevations). 
In summary, not accounting for the  \tsys\ of the individual ALMA antennas introduces an amplitude offset  which is source-dependent and constant during the observing epoch.
Appendix~\ref{tsys} provides an estimate of this amplitude scaling factor for each target (see values in Tables~\ref{tab:fluxes_b3} and \ref{tab:fluxes_b6}). 

%________________________________________________________________
\subsubsection{Primary flux-density calibrators}
\label{flux_calibrators}
As in ordinary ALMA projects, VLBI projects  include observations of primary flux-density calibrators (see Tables~\ref{table:sources_b3} and \ref{table:sources_b6}). In some cases, the ALMA system chooses a solar-system object (SSO), which provides the most accurate flux-density measurement, based on ephemeris estimates of the sub-solar illumination. 
In other cases, quasars (QSOs) are chosen from the ALMA flux-density monitoring database (`grid' sources) which includes measurements mostly in Band 3 and Band 7: flux values for Bands 4/5/6 are obtained by estimating the spectral index from a power-law fit from the Band 3 to Band 7. 
When the primary flux calibrator is an SSO, only baselines shorter than 100\,m are considered to determine the fluxes of the remainder fields, while for QSOs no $uv$-range cut is applied. 
Flux estimates from QSOs are affected by two systematics. First, a constant spectral index (from a power-law fit from the Band 3 to the Band 7) is not a valid assumption for many QSOs. Second, since the QSOs are variable, the accuracy of the flux-density calibration may depend on the time lag between the monitoring entries and the observing date. 
During the the 2017 VLBI campaign, there were also times when neither SSOs nor grid sources were observed. 

In order to provide the most self-consistent flux-density calibration among the different tracks, 
we bootstrapped the flux-density estimates from the tracks with SSOs (i.e., B, C for the EHT and 
2016.1.00413.V, 2016.1.01216.V for the GMVA)
into tracks with no SSO, using a commonly-appearing strong source as the primary flux-density calibrator.  
For the GMVA, we used J0510+1800 and for EHT tracks we used 3C279. 
While this method is not affected by the spectral index uncertainty, it is based on the assumption that the QSO is not variable across several days. 
In Appendix~\ref{tsys} we estimate flux scaling factors for both 3C279 and J0510+1800 on different days, by bootstrapping their flux-densities from Track B and 2016.1.01216.V into all other days, which effectively account for their intrinsic variability (these factors can be applied post-QA2 in the data analysis).  
 In Appendix~\ref{flux_comp_archive} we also quantify   systematics associated with the flux-density calibration and assess  that the overall uncertainty on the flux-density scale of targets observed with the GMVA and EHT is in most cases within 10\% (see Fig.~\ref{fig:fluxcomp_gs} and Tables~\ref{tab:fluxes_b3} and \ref{tab:fluxes_b6}).

%==============================================================================
\section{Polarization calibration}
\label{polcal}
%==============================================================================

VOM observations are always performed in full-polarization mode to supply the inputs to the polarization conversion process at the VLBI correlators (\S~\ref{polconvert}). 
This requires continuous monitoring of a polarized calibrator for calibration purposes.
Since the  delay corrections applied in the correlator to \texttt{APSscans} and \texttt{ALMAscans} are different (and 
it is non-trivial to transfer calibrations between  \texttt{ALMAscans} and \texttt{APSscans}), it is imperative that the polarization calibrator appears not only in the cyclic ALMA project calibration execution, but also in the VLBI scans.  

Here we first summarize some basic concepts of the standard procedure for polarization calibration of ALMA data (\S~\ref{pol-basics}) and then
we provide details on  individual steps in the data calibration procedure (\S~\ref{polcal-steps}).
In particular,  the gain calibration solution is first obtained without any source polarization model  (such a gain solution absorbs the source polarization; \S~\ref{polratio}).  
To extract the Stokes $Q$ and $U$ of the calibrator hidden in the gain solution, one can use the  \texttt{almapolhelpers} function \texttt{qufromgain}  in CASA (\S~\ref{qu}).  
The cross-hand phase differences relative to a reference antenna are calibrated using the CASA task \texttt{gaincal} with the mode \texttt{XYf+QU} (\S~\ref{cross-phase}).  
After the cross-hand phase calibration, the instrumental polarization calibration is performed using the CASA task \texttt{polcal}  (\S~\ref{dterms}).  
The selection of calibrators is justified in \S~\ref{polcals} and some special procedures with respect to the standard approach are listed in \S~\ref{special-polcal}. 

%________________________________________________________________
\subsection{Basics of polarization calibration}
\label{pol-basics}
%________________________________________________________________
In interferometers  having antennas with linearly polarized feeds (like ALMA) 
 both orthogonal linear polarizations (X and Y) are received  simultaneously  and  the data are correlated  to obtain XX, YY, XY, and YX correlations.
The polarization response can be described assuming that each feed is perfectly coupled to the polarization state to which it is sensitive, with the addition of a complex factor times the orthogonal polarization; this is called the "leakage" or "D-term" model. 
In the limit of negligible higher order terms in the instrumental polarization  and zero circular polarization\footnote{The  first-order $D$-term level is typically 2--3\%, therefore the second-order $D$-terms are assumed to be negligible.  Stokes V = 0 is  assumed for simplicity.}, 
the cross correlations for linear feeds on a baseline between antenna \textit{ i} and antenna \textit{ j} is given by \citep[e.g.,][]{Nagai2016}
\begin{eqnarray}
X_{i}X_{j}^{*} &=& (I+Q_{\psi})+U_{\psi}(D^{*}_{X_{j}}+ D_{X_{i}})  \\
X_{i}Y_{j}^{*} &=& U_{\psi}+I(D^{*}_{Y_{j}}+ D_{X_{i}})+Q_{\psi}( D^{*}_{Y_{j}}-D_{X_{i}}) \\
Y_{i}X_{j}^{*} &=& U_{\psi}+I(D_{Y_{i}}+ D^{*}_{X_{j}})+Q_{\psi}( D_{Y_{i}}-D^{*}_{X_{j}}) \\
Y_{i}Y_{j}^{*} &=& (I-Q_{\psi})+U_{\psi}(D^{*}_{X_{j}}+ D_{X_{i}})\,\,, 
\end{eqnarray}
where $Q_{\psi}=Q\cos 2\psi +U\sin 2\psi$, $U_{\psi}=-Q\cos 2\psi +U\sin 2\psi$, $\psi$ is the parallactic angle, and $D_{X}$ and $D_{Y}$ are the instrumental polarization $D$-terms. Therefore, a contribution from Stokes Q, U, and parallactic angle $\psi$ appears in the real part of all correlations. 
 Each Stokes parameter can be obtained from these four equations, and thus the calibration residuals of $D_{X_{i}}$, $D_{X_{j}}$, $D_{Y_{i}}$, $D_{Y_{j}}$ affect the $Q$ and $U$ visibilities.  
The instrumental contribution to the cross-hands visibilities (i.e., the effect due to leakage) is independent of parallactic angle (and thus is constant with time), whereas the contribution of linear polarization from the source rotates with parallactic angle for alt-az mount antennas and it is therefore time-dependent. This makes it possible to uniquely separate the source and instrumental contributions to the polarized interferometer response.
To that end, observations with an array using linear feeds need to include frequent measurements of an unresolved calibrator over a wide range of parallactic angle.

 %________________________________________________________________
\subsection{Detailed steps}
\label{polcal-steps}
%________________________________________________________________

%%--------------------------------------------------------------------------------------
\subsubsection{Gains for the polarization calibrator}
\label{polratio}
\begin{figure*}[ht!]
%\centering
\hspace{-5mm}
\includegraphics[width=10cm]{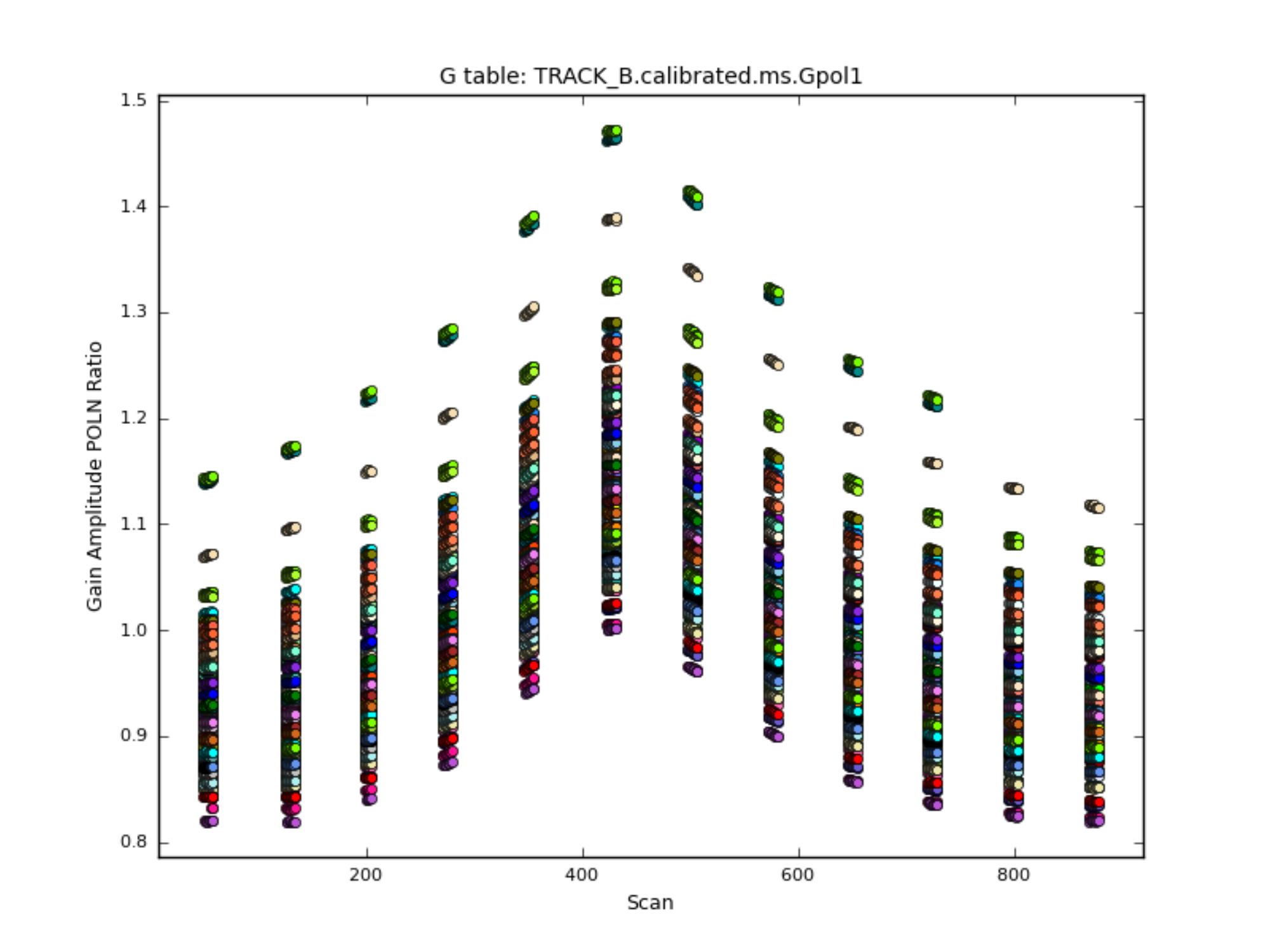} \hspace{-10mm}
\includegraphics[width=10cm]{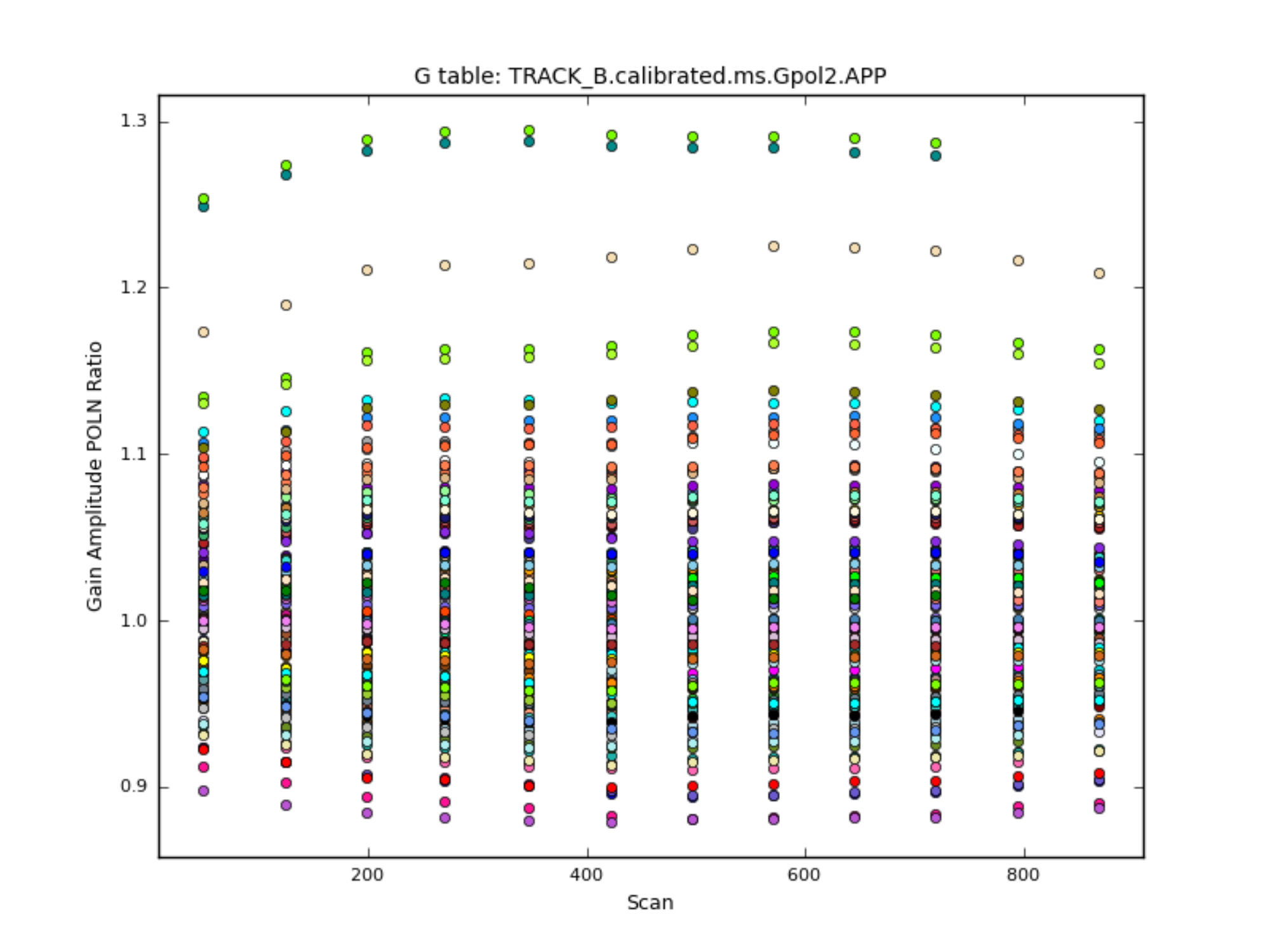} 
\caption{X/Y amplitude gain ratio versus scan number from the tables \texttt{'<label>.Gpol1'} (left) and \texttt{'<label>.Gpol2'} (right)  for Track B in Band 6. The polarization calibrator is  3C279. }
\label{fig:Gpol}
\end{figure*}

In order to   examine the polarization calibrator, the first gains are determined with \texttt{gaincal} using \texttt{gaintype='G'}  
(i.e., independent solutions for the XX and YY correlations),  providing in output the calibration table \texttt{`<label>.Gpol1'} : the  gain corrections in this  table absorb all of the polarization contributions. 
The (linear) source polarization can be displayed by plotting the (antenna-based) amplitude polarization ratio vs. time   (using \texttt{poln='/'} in \texttt{plotcal}), which reveals
a clear variation
as the linear polarization rotates with parallactic angle as a function of time (Fig.~\ref{fig:Gpol}, left panel). 
Once a full polarization source model is obtained for the polarization calibrator (\S~\ref{qu}), 
one can revise the gain calibration using such a model, 
yielding in output the calibration table \texttt{'<label>.Gpol2'}, where 
any signature of the source polarization is removed from the gains (Fig.~\ref{fig:Gpol}, right panel).

%%--------------------------------------------------------------------------------------
%%--------------------------------------------------------------------------------------

%%--------------------------------------------------------------------------------------
\subsubsection{X-Y Cross-phase}
\label{cross-phase}

\begin{figure}[ht!]
%\centering
\hspace{-5mm}
\includegraphics[width=0.5\textwidth]{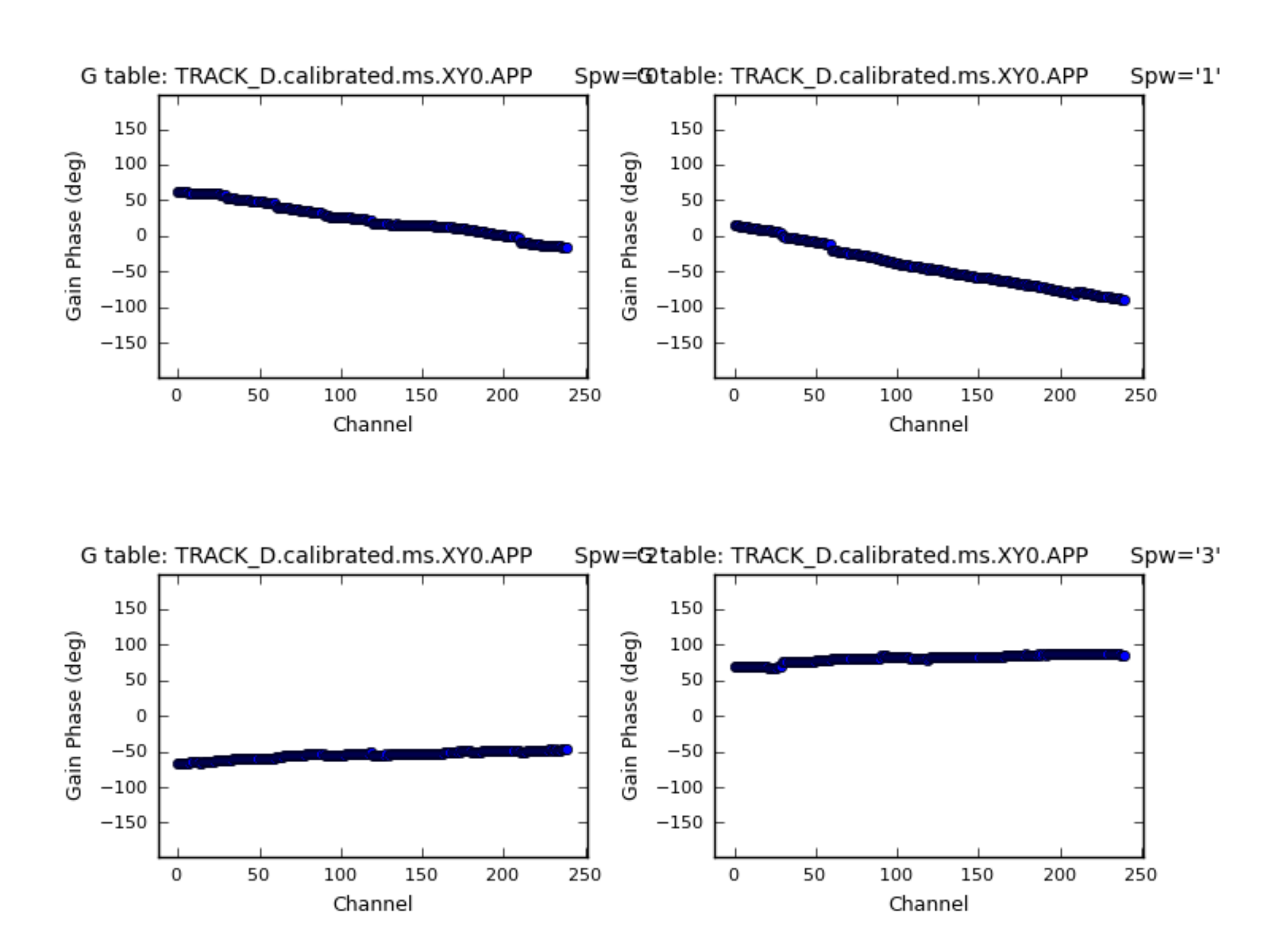} 
\caption{X-Y cross-phase of the reference antenna  in APS mode for Track D in Band 6  for each of the four SPWs.}
\label{fig:crossphase}
\end{figure}

The bandpass and gain tables computed in \S~\ref{calibration} are adequate for the parallel hands. 
Since the phase of the reference antenna is set to zero in both polarizations, yielding relative phases for all other antennas\footnote{In radio-interferometry absolute phase values are not measured.}, 
a single residual phase bandpass relating the  phase of the two hands of polarization in the reference antenna remains in the cross-hands of all baselines. 
The APS has already removed a bulk  cross-delay (linear phase slope) in this phase bandpass\footnote{As noted in \S~\ref{APS_spectral}, the APS makes phase corrections per channel average, and the aggregate of these corrections across a sub-band or SPW is equivalent to a delay correction.  Since this is done independently in X and Y,  the net effect is an X-Y cross-delay correction.}, but any residual  non-linear phase bandpass in the XY-phase needs also to be solved for,  so that both cross- and parallel-hands can be combined to extract correct Stokes parameters. 
This can be computed running \texttt{gaincal} in mode \texttt{XYf+QU}\footnote{\texttt{gaincal} in mode \texttt{XYf+QU} averages all baselines together and first solves for the XY-phase as a function of channel. It then solves for a channel-averaged source polarization (with the channel-dependent XY-phase corrected).}. 
Figure~\ref{fig:crossphase} shows the X-Y cross-phases of the reference antenna obtained for EHT track D in \texttt{APSscans}. 

There a number of considerations that apply specifically  to handling the X-Y  cross-phases for APS observations. 

\begin{enumerate}

\item 
Differences are expected between \texttt{ALMAscans} and \texttt{APSscans} due to the different corrections applied by the APS software.
For instance,  the cross-phases in individual SPWs of the \texttt{APSscans} show small jumps among the frequency chunks used by TelCal (\S~\ref{APS_spectral}).

\item 
The X-Y phase offset determined for the \texttt{APSscans} is independent of the antenna used as the reference in the QA2, since such an offset was applied to all
the antennas prior to the cross-correlation, while keeping all phases close to zero. 
 For the \texttt{ALMAscans}, using a different reference antenna in the QA2 calibration changes the derived X-Y offset.

\item 
 It is imperative for the success of the polarization conversion (\S~\ref{polconvert}), to flag "noisy" solutions in the X-Y phase difference in the 
  calibration tables obtained during ordinary calibration (see \S~\ref{phase-gains}), in order to minimize leakage-like noise in the VLBI visibilities (see Appendix~\ref{glitch}). 
 To this end, a cross-polarization running average of the
phase gains is applied 
before polarization calibration. 

\item
It is necessary to check and fix possible X/Y cross-phase jumps of 180 degrees within each SPW.  

\item
Only the X-Y phases are solved for, while the X-Y cross-delay is not computed: the bulk of the cross-delay is already removed by the APS.

\end{enumerate}

%%--------------------------------------------------------------------------------------
\subsubsection{Estimating QU for the polarization calibrator}
\label{qu}

The CASA  task \texttt{gaincal} used in mode \texttt{XYf+QU} determines not only the X-Y phase offset, but also
as a by-product the Q and U Stokes parameters for the polarization calibrator. 
Since \texttt{gaincal} only uses the XY and YX correlations to determine the X-Y phase offset, 
this estimate has a degeneracy of $\pi$ radians, which translates to an ambiguity in the signs of both Q and U Stokes parameters ($\pi/2$ in the polarization angle): (XY-phase, Q, U) $\rightarrow$ (XY-phase + $\pi$, -Q, -U)\footnote{Changing the sign of U and Q is equivalent to changing the sign of the cross-phase gain by adding $\pi$ radians.}.
To break this degeneracy, the actual signs of Q and U must be determined. 
This can be done with the \texttt{almapolhelpers} function \texttt{qufromgain}, 
which   "extracts" the source polarization information (Q and U values) encoded in the ratio between the X and Y gains  for the polarization calibrator (contained in the \texttt{'<label>.Gpol1'} table described in \S~\ref{polratio}). 
This estimate of Stokes Q and U is not as good as can be obtained from the cross-hands from the XY and YX correlations (\S~\ref{cross-phase}), since it relies on the gain  polarization ratio being stable, which is not necessarily true. Therefore it is mostly useful in removing the ambiguity that occurs in the cross-hand estimate.
Nevertheless, we assessed that both methods provide consistent values for the  polarization angles and fractional linear polarization, which is indicative of a self-consistent calibration. 

%--------------------------------------------------------------------------------------
\subsubsection{Polarization Leakage (D-terms)}
\label{dterms}
Once the polarization  model is obtained for the polarization calibrator and the X-Y phase offsets have been calibrated, 
one can solve for the instrumental polarization: 
the leakage terms or D-terms are estimated using the CASA task \texttt{polcal}. 
This task produces an absolute instrumental polarization solution on top of the source polarization and ordinary calibrations. 
Note that no reference antenna is used here because the polarized source provides sufficient constraints to solve for all instrumental polarization parameters on all antennas, relative to the specified source polarization. This is not possible in the case of an unpolarized calibrator (or in the circular basis, even if the calibrator is polarized), where only relative instrumental polarization factors among the antennas may be determined with respect to the reference antenna.  

The values fitted are of the order of a few percent (and generally $<$10\%; see Figures~\ref{fig:DT_IMAG}, \ref{fig:DT_REAL}).
The D-terms are then computed into the sky frame and arranged in a Jones matrix that can apply corrections (up to second order) to all four correlation products (XX, XY, YX and YY). 

We explicitly note that the problem of the leakage calibration of each individual ALMA antenna is not critical, since the effect of the D-terms on the final VLBI calibration is relatively small (below the thermal limit of the VLBI baselines).

%%--------------------------------------------------------------------------------------
\subsubsection{Amplitude gain ratios between X and Y}
Once the Stokes parameters of the polarization calibrator are
estimated,  they can be used as a model to estimate the ratio of
amplitude gains between the $X$ and $Y$ polarizers for each ALMA antenna.  
These X/Y amplitude gain ratios  are found by running \texttt{gaincal} (one solution, combining all scans of the polarization calibrator), using our estimate of Stokes parameters for the polarization calibrator and applying the X-Y phase and D-term calibration on the fly.  
%%--------------------------------------------------------------------------------------

\begin{table*}
\caption{Flux and polarization properties of the polarization calibrators employed in Band 3 across the GMVA campaign, as derived by \texttt{fluxscale} and \texttt{gaincal} in mode \texttt{XYf+QU} (and properly corrected for the $\pi$-radians ambiguities). 
A reference frequency of 93.084 GHz is assumed.}              
\label{table:polcal_b3}
\centering  
\small
\begin{tabular}{cccccccccc} 
\hline\hline                  
\noalign{\smallskip}
{Project} & { Source} & Flux (Jy) & Spectral & { p(\%)} & \multicolumn{4}{c}{ EVPA (deg.)} & { RM } \\
            &              &  & index & & { SPW\,0} & { SPW\,1} & { SPW\,2} & { SPW\,3} & { (rad/m$^2$)} \\
\noalign{\smallskip}
\hline
\noalign{\smallskip}  
%2016.1.01116.V &   4C\,01.28  & $7.6 \pm 0.4^{a}$ & $-0.45 \pm 0.04^{a}$ &  $4.3 \pm 0.4$ & -29.11 & -28.45 & -26.49 & -21.28 &  $-32000 \pm 10000$ \\
2016.1.01116.V &   4C\,01.28  & $7.6 \pm 0.4 ^{a}$ & $-0.45 \pm 0.04 ^{a}$ &  $4.5 \pm 0.1$ & -28.58 & -28.34 & -27.66 & -27.63 &  $-5000 \pm 500 ^{b}$ \\
2016.1.00413.V &  B1730$-$130 & $2.8 \pm 0.1$ & $-0.57 \pm 0.03$ &  $0.92 \pm 0.07$ &   35.94 & 36.78 & 40.67 & 41.52 & $-31000 \pm 6000$ \\
2016.1.01216.V  &  3C279 & $12.6 \pm 0.6$    &  $0.37 \pm 0.03$      & $12.2 \pm 0.2$  &  43.42 & 43.55 & 44.21 & 44.30 & $-5000 \pm 500$ \\
\noalign{\smallskip}
\hline    
\end{tabular} 
\tablenotetext{a}{The flux and spectral index for 4C\,01.28 are estimated after bootstrapping from Callisto observed on Apr 3 (the day before). }
\tablenotetext{b}{The EVPA  and RM values are computed using \texttt{ALMAscans} since \texttt{APPscans} yield a much lower EVPA value in SPW 3 ($\sim-21$deg); values for other SPW were comparable between the two modes. }
\end{table*}

% Previous calibration:
%\begin{table*}
%\caption{Polarization properties of the  polarization %calibrators used (other than 3C279) in Band 3 for the %GMVA. I TOOK THESE FROM THE IMAGES)}              
%\label{table:polcal_b3}
%\centering  
%\small
%\begin{tabular}{|c|c|c|c|c|c|c|c|}
%\hline 
%{\bf Experiment} & {\bf Source} & {\bf p(\%)} & %\multicolumn{4}{c|}{\bf EVPA (deg.)} & {\bf RM } \\
%            &              &  & {\bf spw\,0} & {\bf %spw\,1} & {\bf spw\,2} & {\bf spw\,3} & {\bf %(rad/m$^2$)} \\
%\hline
%2016.1.00413.V &  B1730$-$130 & 0.9 &   37.1  & 37.7  &  %40.1 &  40.7 & {\color{red} $-62$\,700 }\\
%2016.1.01116.V &   4C\,01.28  & 4.4 & $-27.2$  & $-26.8$ & %$-26.1$ & $-25.6$ &  {\color{red}$-24$\,000 } \\
%\hline
%\end{tabular} 
%\end{table*}%
\begin{table*}
\caption{Flux and polarization properties of the polarization calibrators employed in Band 6 across the EHT campaign, as derived by \texttt{fluxscale} and \texttt{gaincal} in mode \texttt{XYf+QU} (and properly corrected for the $\pi$-radians ambiguities). 
A reference frequency of 220.987 GHz is assumed.}        
\label{table:polcal_b6}
\centering  
\small
\begin{tabular}{cccccccccc} 
\hline\hline                  
\noalign{\smallskip}
{Track} & { Source} & Flux (Jy)$^{a}$ & Spectral & { p(\%)} & \multicolumn{4}{c}{ EVPA (deg.)} & { RM } \\
            &              &  & Index$^{a}$ & & { SPW\,0} & { SPW\,1} & { SPW\,2} & { SPW\,3} & { (rad/m$^2$)} \\
\noalign{\smallskip}
\hline
\noalign{\smallskip}
D (Apr 5)  &     3C279 &  $8.9 \pm 0.9$  &  $-0.60 \pm 0.06$            & $13.23 \pm 0.04$  &  45.17 & 45.17 & 45.28 & 45.32 & $-10000 \pm 5000$ \\
B (Apr 6)  &     3C279 &      $8.9 \pm 0.9$  &  $-0.60 \pm 0.06$      & $13.04 \pm 0.04$ &  43.28 & 43.29 & 43.36 & 43.35 & $-5000 \pm 5000$ \\
A (Apr 10) &     3C279 &    $8.9 \pm 0.9$  &  $-0.60 \pm 0.06$   & $14.73 \pm 0.06$  &  40.18 & 40.19 & 40.18 & 40.20 & $-500 \pm 4000$ \\
E (Apr 11) &     3C279 &    $8.9 \pm 0.9$  &  $-0.60 \pm 0.06$       & $14.91 \pm 0.08$  &  40.13 & 40.14 & 40.01 & 40.08 & $6000 \pm 4000$ \\
&&&&&&&&&\\
C (Apr 7)  & B1921-293 & $3.1 \pm 0.3$   &  $-0.82 \pm 0.08$  & $5.97 \pm 0.07$ &  -48.89 & -49.02 & -49.49 & -49.59 & $44000 \pm 10000$ \\
\noalign{\medskip}
\hline    
\end{tabular} 
\tablenotetext{a}{The flux and spectral index for 3C279 are estimated after bootstrapping from Ganymede (observed in Track B), and then assumed constant on the other days. }
\end{table*}

%Previous calibration
%\begin{table*}
%\caption{Polarization properties of 3C\,279 in Band 6 across the EHT campaign. Taken from the %estimate of the X-Y phase tables by \texttt{gaincal} (and properly corrected for the %$\pi$-radians ambiguities)}              
%\label{table:polcal_b6_3c279}
%\centering  
%\small
%\begin{tabular}{|c|c|c|c|c|c|c|c|}
%\hline\hline                  
%\noalign{\smallskip}
%\bf{TRACK} & {\bf Flux$^{a}$ (Jy)} & {\bf Spectral} &  p (\%) & \multicolumn{4}{c|}{EVPA %(deg.)}\\
%           &                 &        Index     &        & spw\,0  & spw\,1  & spw\,2  & spw\,3 % \\
%\hline
%D (Apr 5)  &       8.9      &  $-0.60$            & 13.2  &  46.05 & 46.21  & 46.62  & 46.20 \\
%B (Apr 6)  &         8.9   &  $-0.60$      & 13.0 &  43.49 & 43.49  & 43.41  & 43.32 \\
%A (Apr 10) &      8.9      &  $-0.53$   & 14.7  &  40.11 & 40.06  & 39.83  & 39.89 \\
%E (Apr 11) &      8.9     &  $-0.52$       & 14.9  &  40.05 & 40.02  & 39.44  & 39.71 \\
%\noalign{\smallskip}
%\hline    
%\end{tabular} 
%\tablenotetext{a}{The flux and spectral index for 3C279 are estimated after bootstrapping from %Ganymede, and then assumed constant on the other days. Both assume a reference frequency of %220.987 GHz. }
%\end{table*}%

%________________________________________________________________
\subsection{Polarization Calibrators}
\label{polcals}
%________________________________________________________________

Based on a comparative study among all potential polarization calibrators observed in the VLBI campaign, the source 3C\,279 was established as the best calibrator. It is a strong mm source ($\sim$13 Jy in Band 3 and $\sim$9 Jy in Band 6) with a high fractional polarization (12-15\%) and was observed with a large parallactic-angle coverage. 
 3C\,279 was also observed on multiple (consecutive) days, allowing a check of the stability of the source (and/or  of the array) polarimetry across the whole campaign. 
During three nights (Apr 2, 3, 7) 3C279 was not included in the VLBI schedules, and alternative polarization calibrators had to be employed.  
 Despite the fact that their fractional polarizations are an order of magnitude lower than 3C\,279,  they allowed us to achieve a satisfactory calibration of the X-Y phases at the reference antenna.  
 Tables~\ref{table:polcal_b3} and \ref{table:polcal_b6} show the polarization properties of 3C\,279 and the 'alternative' polarization calibrators, in Bands 3 and 6, respectively.  

 %________________________________________________________________
\subsection{Special procedures}
\label{special-polcal}
\begin{figure*}[ht!]
%\centering
\includegraphics[width=10cm]{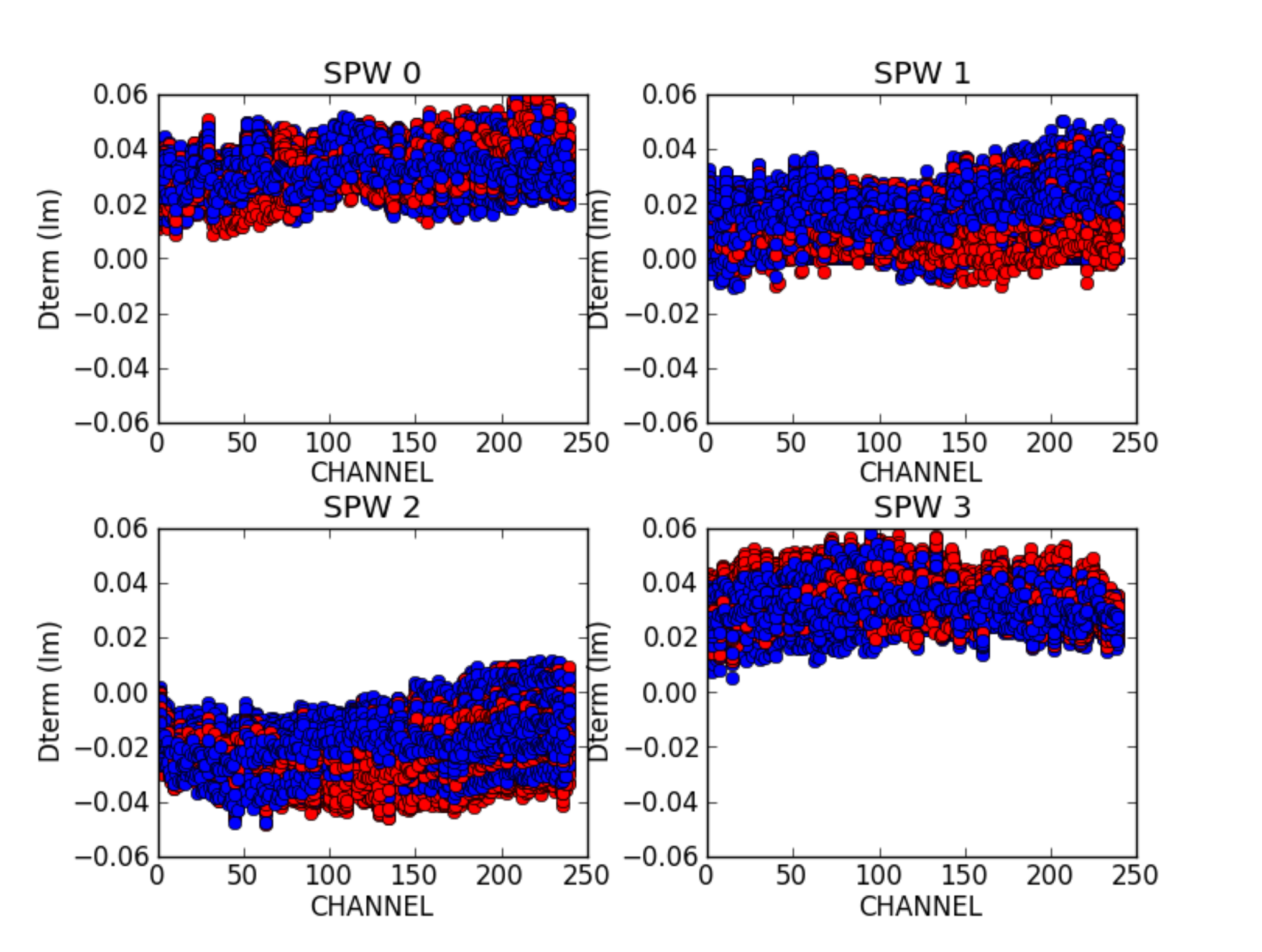} \hspace{-10mm}
\includegraphics[width=10cm]{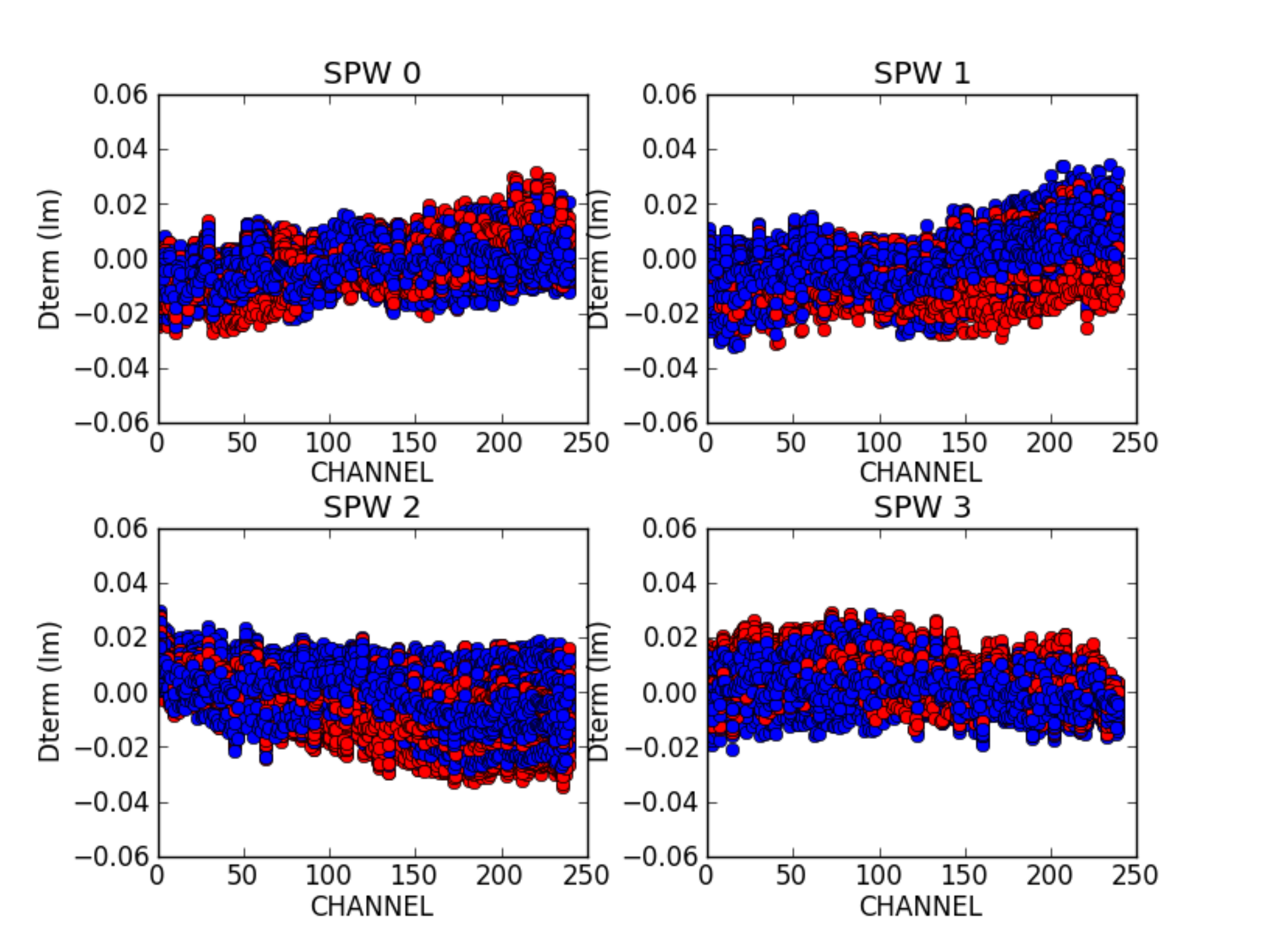}
\caption{Imaginary part of all the polarization D-terms estimated  in experiment 2016.1.00413.V. The left four panels do not account for the RM of the calibrator. The right four panels show the same data after accounting for the RM (i.e., calibrating with a different polarization model for each SPW). Red is for X; blue is for Y. Note the non-zero averages for all antennas when the RM is not taken into account.}
\label{fig:DT_IMAG}
\end{figure*}

\begin{figure*}[ht!]
%\centering
\includegraphics[width=10cm]{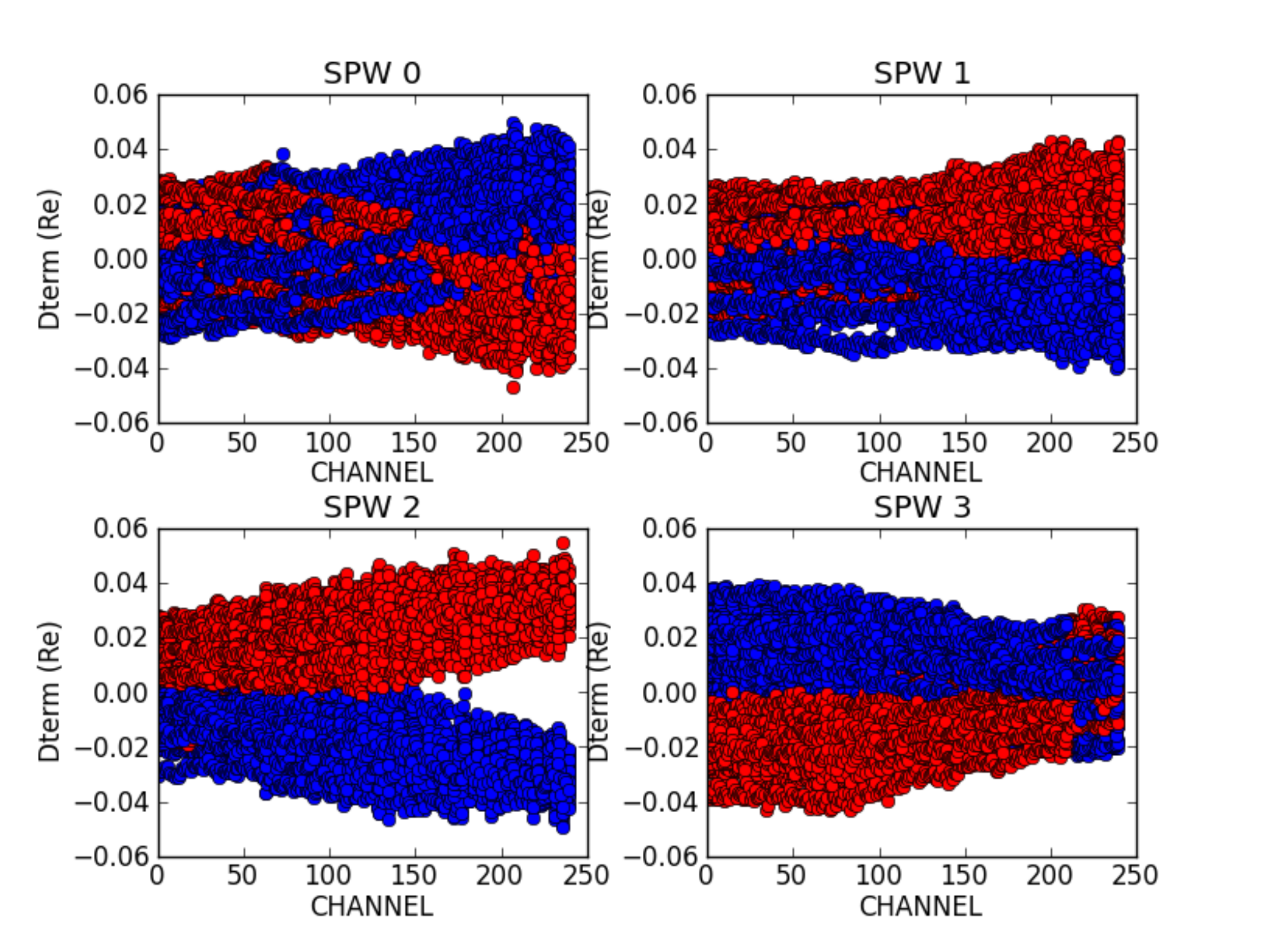} \hspace{-10mm}
\includegraphics[width=10cm]{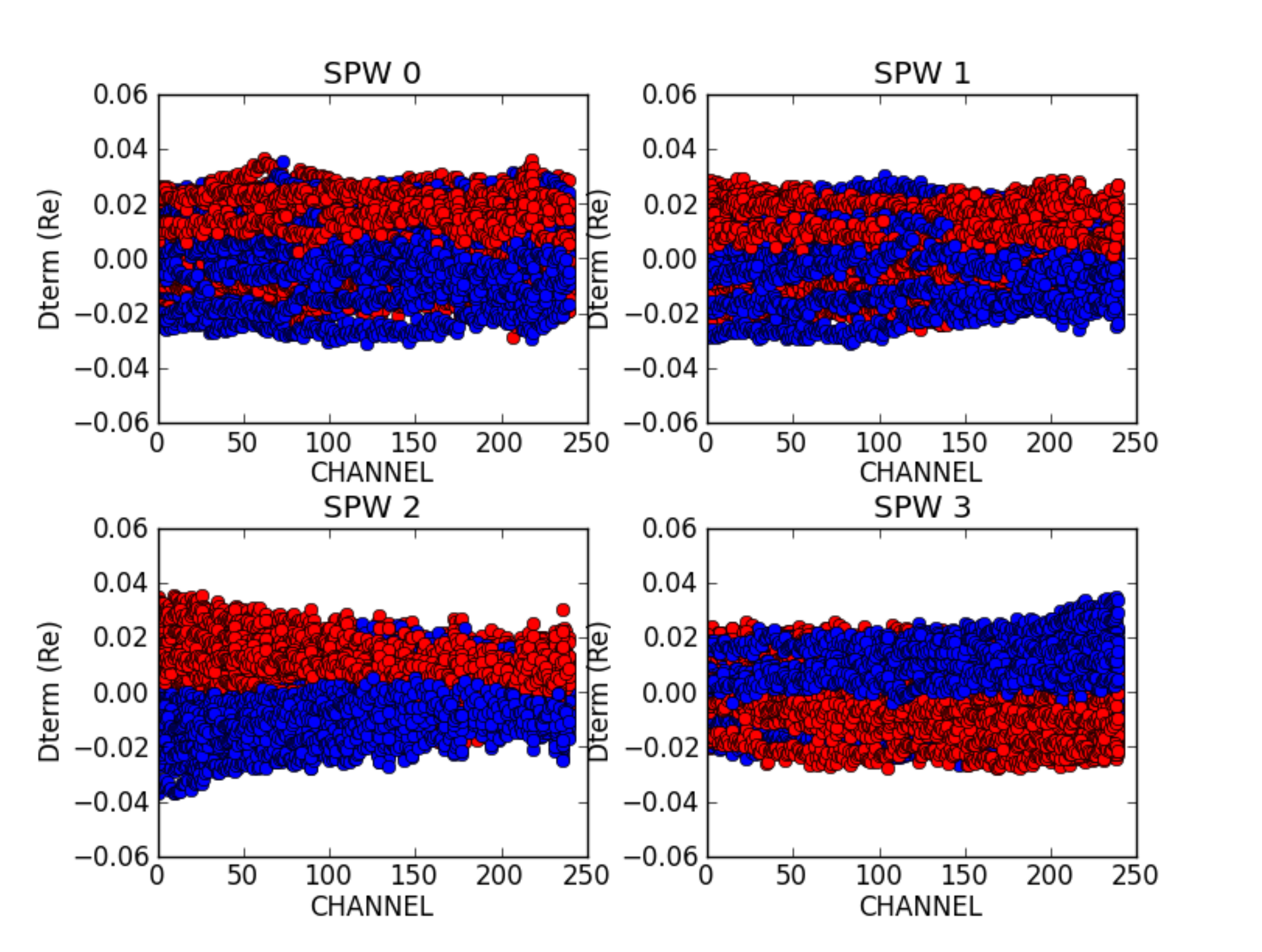}
\caption{{As in Fig.~\ref{fig:DT_IMAG}, but for the real part of the D terms.} Note the larger $X-Y$ symmetric shifts when the RM is not taken into account.}
\label{fig:DT_REAL}
\end{figure*}

%________________________________________________________________
Tables~\ref{table:polcal_b3} and \ref{table:polcal_b6}  show an appreciable  Faraday Rotation of the polarization calibrators used as alternatives to 3C279. 
A rotation of the electric vector position angle (EVPA) of several degrees across the SPWs (as seen in Band 3), yields  a rotation measure (RM) of order of $3 \times 10^4$~rad/m$^2$. Such a high rotation can potentially  bias the calibration of the leakage terms (D-terms) if  one single (i.e., frequency-averaged) polarization model is used in the \texttt{polcal} CASA task, as per the official QA2 procedure.
This effect is  shown in Figs. \ref{fig:DT_IMAG} and \ref{fig:DT_REAL}, which compare the D-terms  in one of the Band 3 experiments as estimated with a standard (QA2) procedure (i.e., one single model for source polarization in all SPWs),  versus independent models for each SPW. The systematics clearly seen in the former case are minimized in the latter.
Based on this evidence, each SPW was separately calibrated  in polarization, by accounting for the different EVPA of the polarization calibrator in each SPW (i.e., accounting for the RM in the calibrator), as well as for the spectral index. This is especially important for the GMVA (3mm) observations. 

%________________________________________________________________
%________________________________________________________________

\vspace{0.2cm}
A couple of final remarks are in order. 
The polarization calibration can be done using either \texttt{ALMAscans} or \texttt{APSscans} or both. 
During the April 2017 VLBI campaign  (good) polarization calibrators were not observed often enough in ALMA mode, therefore only \texttt{APSscans} were used (which ensured a good coverage of the parallactic angle).
Therefore,  the \texttt{ALMAscans} were not calibrated  in polarization. 
Finally, only data from  antennas that were phased
during the entire observing track are used for the polarization calibration\footnote{%
If the same antenna were used in the phased-array and as a comparison antenna over the course of the same observing track, it would have inconsistent X-Y phases that would bias the calibration.}.

%==============================================================================
\section{Calibration of phased-ALMA as a single VLBI station}
\label{vlbi}
%==============================================================================

The  QA2 calibration described in \S~\ref{calibration} and \S~\ref{polcal} provides the set of
calibration tables needed by  \texttt{PolConvert} to calibrate phased-ALMA as a single VLBI station. 

In summary, the QA2 process produces the following gain solution tables:

\begin{itemize}

\item \texttt{<label>.phase\_int.APP.XYsmooth} (hereafter denoted as $G_p$):  phase gains (per integration time).

\item \texttt{<label>.flux\_inf.APP} (hereafter denoted as $G_a$):   amplitude gains scaled to Jy units (per scan).

\item \texttt{<label>.bandpass\_zphs} (hereafter denoted as $B_0$):   bandpass (with zeroed phases).

\item \texttt{<label>.XY0.APP} (hereafter denoted as $XY_p^{r}$): 
cross-polarization phase at the TelCal phasing reference antenna.

\item \texttt{<label>.Gxyamp.APP} (hereafter denoted as $XY_a$): 
amplitude cross-polarization ratios for all antennas. 

\item \texttt{<label>.Df0.APP} (hereafter denoted as $D$): 
 D-terms at all antennas. 

\end{itemize}

In the following subsections, we describe how these calibration tables are applied by \texttt{PolConvert} to perform the polarization conversion (\S~\ref{polconvert}) and the flux calibration (\S~\ref{fluxcal}) of phased-ALMA. 

%________________________________________________________________
\subsection{Polarization Conversion}
\label{polconvert}

Details of this process are described in \citet{PCPaper}, and here we summarize the main concepts for completeness.

As described in \S~\ref{vlbicorr}, the DiFX software is "blind" to polarization and provides XL, XR, YL and YR correlation products.
The visibilities in mixed-polarization basis can be arranged in a matrix form as 

\begin{equation}
V_{+\odot} = \begin{pmatrix} V_{XR} & V_{XL} \\ V_{YR}  &  V_{YL} \end{pmatrix},
\label{MixMat1}
\end{equation}

\noindent where ALMA is assumed to be the first antenna in the baseline (linear polarization basis, denoted as $+$) and all the other stations observe in circular polarization basis (denoted as $\odot$). The matrix in pure circular basis (i.e., ALMA converted to circular) can be arranged as

\begin{equation}
V_{\odot\odot} = \begin{pmatrix} V_{RR} & V_{RL} \\ V_{LR}  &  V_{LL} \end{pmatrix}.
\label{MixMat2}
\end{equation}

The visibility matrix in circular-circular
polarization can be recovered directly from the matrix in mixed-polarization
by applying a simple matrix product:

\begin{equation}
V_{\odot\odot} = C_{\odot +} V_{+\odot},
\label{ConvMatrix_ideal}
\end{equation}

\noindent where (see Eq.~5 of \citealt{PCPaper})

$$C_{\odot +} = \frac{1}{\sqrt{2}}\begin{pmatrix} 1 & -i \\ 1  & i \end{pmatrix}$$ 

\noindent is the matrix  that  converts polarizations
from linear to circular ({\it C} stands for {\it Conversion}).

Equation~\ref{ConvMatrix_ideal} assumes that the visibilities are free  from
instrumental effects (absence of noise and perfectly calibrated signals). 
The observed visibility matrix is then related to the perfectly-calibrated visibility matrix by the equation (as computed by \texttt{PolConvert})  

\begin{equation}
V^{obs}_{\odot\odot} = C_{\odot +} \left(J^A\right)^{-1} V_{+\odot},
\label{ConvMatrix}
\end{equation}

\noindent where $J^A$ is the calibration Jones matrix of the phased array

{\tiny
\begin{equation}
J^A = \begin{pmatrix} \left< (B_0^{i})_X \, (G_p^{i})_X  \, G_a^{i} \right> & \left< (D^{i})_X \, (B_0^{i})_X \, (G_p^{i})_X \,  G_a^{i} \right> \\ \left< (D^{i})_Y \, (B_0^{i})_Y \, (G_p^{i})_Y \,  G_a^{i} \right>  &  \left< (B_0^{i})_Y \, (G_p^{i})_Y \,  G_a^{i} (XY_a^{i})\right> \, (XY_p^{r}) \end{pmatrix}.
\label{CalMatrix}
\end{equation}
}

$J^A$ includes all the calibration matrices (i.e., gain, bandpass, D-terms) of all the antennas in the phased array.
In particular, $G_a^{i}$ and $G_p^{i}$ are the amplitude and phase gains,  $B_0^{i}$ is the bandpass, and $D^{i}$ are the D-terms  for the $i$th ALMA antenna; 
 $XY_p^{r}$ is the phase offset between the X and Y signals of the reference
antenna (indicated with index $r$);  $\left<...\right>$ means averaging over phased antennas at each integration time. Note that the $G_a$ gains do not distinguish between X and Y (solutions are forced  to be the same for both polarizations using the "T'' mode - \S~\ref{ampcal}). 
 \texttt{PolConvert} interpolates individual gains   
(linearly, in amplitude-phase space) in both frequency and time directions to the values at the VLBI correlator (see Appendix~\ref{app:vlbicorr}; see also Appendix~\ref{glitch} for a discussion on possible issues related to this interpolation).

For the case where phased ALMA is the second station in the baseline : 

\begin{equation}
V^{obs}_{\odot\odot} = V_{\odot +} \left[(J^A)^{H} \right]^{-1} C_{\odot +}^{H},
\label{ConvMatrix2}
\end{equation}

\noindent where $H$ is the Hermitian operator. 
Note that the application of Eqs. \ref{ConvMatrix} and/or \ref{ConvMatrix2} automatically calibrates for the post-converted R-L delay and phase at ALMA. Thus, using phased ALMA as the reference antenna in the VLBI fringe-fitting will account for the absolute EVPA calibration and, furthermore, no fringe-fitting of the R-L delays will be needed.

%%--------------------------------------------------------------------------------------
\subsubsection{Post-conversion effects of residual cross-polarization gains}
\label{residual_crosspol}
%%--------------------------------------------------------------------------------------

In the estimate of the $XY_p^{r}$ solutions, CASA assumes that the amount of circular polarization in the calibrator is negligible. If this is not the case, Stokes $V$ appears as an imaginary component in the $XY$ and $YX$ correlation products, given in the sky frame. This component might be partially absorbed in the estimate of $XY_p^{r}$, hence introducing a phase offset between $X$ and $Y$ that cannot be corrected following standard QA2 procedures. Nevertheless, if a phase offset is introduced in $XY_p^{r}$, it will appear as a leakage-like effect in the polconverted VLBI visibilities, which can then be calibrated downstream in the VLBI data processing. The proof of this statement is straightforward. 

Given any gain Jones matrix in linear basis, $J_+$, it can be converted into circular basis as $J_\odot = C_{\odot +} J_+ C_{+ \odot}$. If $J_+$ has the form

\begin{equation}
J_+ = \left(\begin{array}{cc} 1 & 0 \\ 0 & \rho \end{array}\right),
\end{equation}

\noindent where $\rho$ is a residual complex gain between the phased sums in $X$ and $Y$, then 

\begin{equation}
J_\odot \propto \left(\begin{array}{cc} 1 & D \\ D & 1 \end{array}\right),
\end{equation}

\noindent which has the shape of a leakage matrix with equal D-terms for $R$ and $L$. In this equation, 

\begin{equation}
D = \frac{1- \rho}{1+\rho},
\label{DfacEq}
\end{equation}

\noindent which means that $\rho$ can be derived from $D$, so that it is possible to use the VLBI calibration in order to improve the alignment of the $X$ and $Y$ phases at ALMA, thus allowing us to detect Stokes $V$ with a higher accuracy.

According to the specifications devised by the APP, the $D$ factor in Eq. \ref{DfacEq} should be lower than 3\% in absolute value, which translates into a $\rho$ of less than $\sim$5\% in amplitude and $\sim$5~degrees in phase. These figures fall within the requirements of ordinary ALMA full-polarization observations (i.e., a 0.1~\% sensitivity in polarization or better, as per  the Call for Proposals\footnote{https://almascience.eso.org/proposing/call-for-proposals}).  Therefore, the requirement for a post-conversion polarization leakage below 3\% is easily met after the QA2 calibration is performed.

%________________________________________________________________
\subsection{Amplitude calibration}
\label{fluxcal}

\texttt{PolConvert} applies a modified version of Eqs. \ref{ConvMatrix} (or \ref{ConvMatrix2}) as following 

\begin{equation}
    V_{\odot\odot} = \frac{K}{\sqrt{K_{00}K_{11}}} V_{+ \odot}.
    \label{NormConvEq}
\end{equation}

\noindent where 
phased-ALMA  is assumed to be the first antenna in the baseline, 
 $K = C_{\odot +} \left(J^A\right)^{-1}$ is the total (calibration plus conversion)  matrix applied by \texttt{PolConvert} (Eq.~\ref{ConvMatrix}), 
 and $K_{00}$ and $K_{11}$ are the diagonal elements of $K$ (the ratio ensures that the amplitude correction applied in Eq. \ref{NormConvEq} is normalized).

 For compatibility with other VLBI stations, \texttt{PolConvert} stores the VLBI amplitude corrections $A$ for phased-ALMA as a combination of 
 $T_{sys}$ 
 (one value per intermediate frequency and integration time)
 and an instrumental gain given in degrees per flux unit or DPFU 
 (assumed to be stable over time and frequency)
  
\begin{equation}
 A = \sqrt{T_{sys}/\mathrm{DPFU}}= \sqrt{\frac{\left| K_{00} K_{11} \right|}{N}}
 \label{TsysEq0}
 \end{equation}

or equivalently 

\begin{equation}
T_{sys} = \frac{\left| K_{00} K_{11} \right|}{N} \mathrm{DPFU},
\label{TsysEq}
\end{equation}

where $N$ is the number of phased antennas at the given integration time. 
The $\sqrt{N}$ factor in Eq. \ref{TsysEq0} comes from the scaling between the amplitude calibration from the intra-ALMA cross-correlations and that of the summed signal, as we demonstrate in Appendix \ref{app:sqrtN}.

%-------------------------------------------------------------------------------------------------------------------------
\subsubsection{DPFU, T$_{\rm sys}$, and SEFD}
\label{SEFD}
%-------------------------------------------------------------------------------------------------------------------------

In principle, the DPFU of the phased array should scale with the number of phased antennas. 
Since the number of phased antennas  changes during the observations, the DPFU would also change. 
 In order  to keep the DPFU of phased-ALMA constant over a given epoch  (for calibration purposes), 
 we set the DPFU of phased-ALMA  to the antenna-wise {\it average} of DPFUs (instead of the antenna-wise {\it sum}). 
As a consequence, there is an $N$ factor that must be absorbed by the $T_{sys}$ in order to keep the same amplitude correction $A$,  according to Eq. \ref{TsysEq0}. 
In fact, we explicitly notice that the only meaningful calibration information for the VLBI fringes with ALMA is given by 
the combination of $T_\mathrm{sys}$ and DPFU (and not by their independent values), via the factor $\sqrt{K_{00}K_{11}/N}$ in Eq. \ref{TsysEq}. 
Using a different DPFU for phased-ALMA will thus result in a different set of $T_{sys}$ values computed by \texttt{PolConvert}, in such a way that the VLBI amplitude corrections remain unchanged.
  
To compute the antenna-wise DPFU average, DPFU$_i$ values for the individual antennas can be estimated 
using the following equation: 

\begin{equation}
    \mathrm{DPFU}_i = \frac{<A_{i,k}^{-1}>^2}{\left<(T_{sys})^{-1}_{i,k}\right>},
\end{equation}
 
\noindent where the sub-index $k$ runs over all scans where a $T_{sys}$ is measured at  the antenna $i$, $A_{i,k}$ is the amplitude correction of the antenna $i$ (see Eq. \ref{AmpCalVisEq} in Appendix \ref{app:sqrtN}) for scan  $k$, and $<...>$ is  the median operator (the median is more insensitive to outliers than the average). 
 In Cycle 4, there was a slightly different number of phased antennas on each day, therefore the average $<\mathrm{DPFU}>$ is also slightly different.
 We  therefore set the average DPFU $= 0.031$\,K/Jy as representative of the phased array during the whole campaign. 

\begin{figure*}[ht!]
\centering
\hspace{-2mm}{%
\includegraphics[%
  width=0.5\textwidth]
{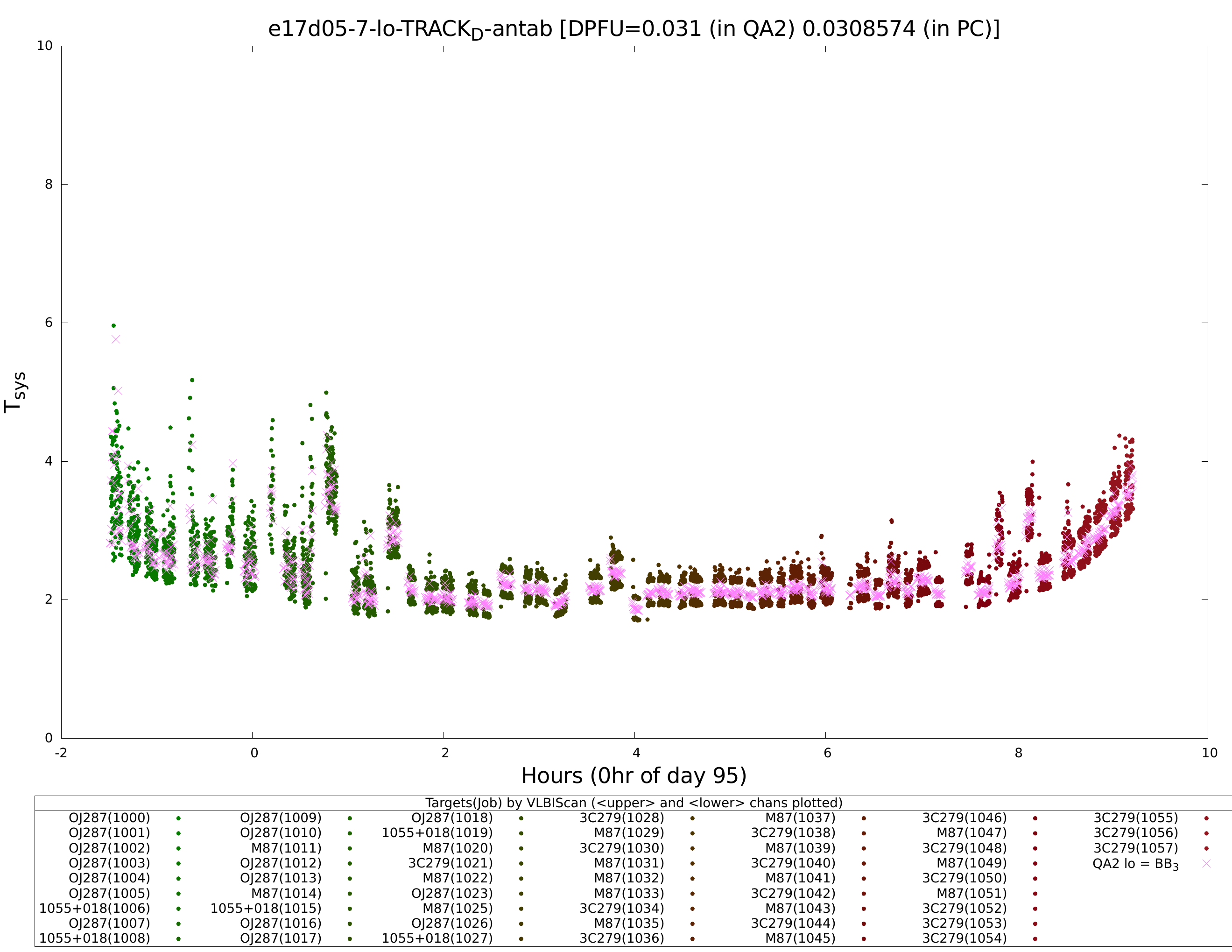}}%
\hspace{1mm}{%
\includegraphics[%
  width=0.5\textwidth]%
{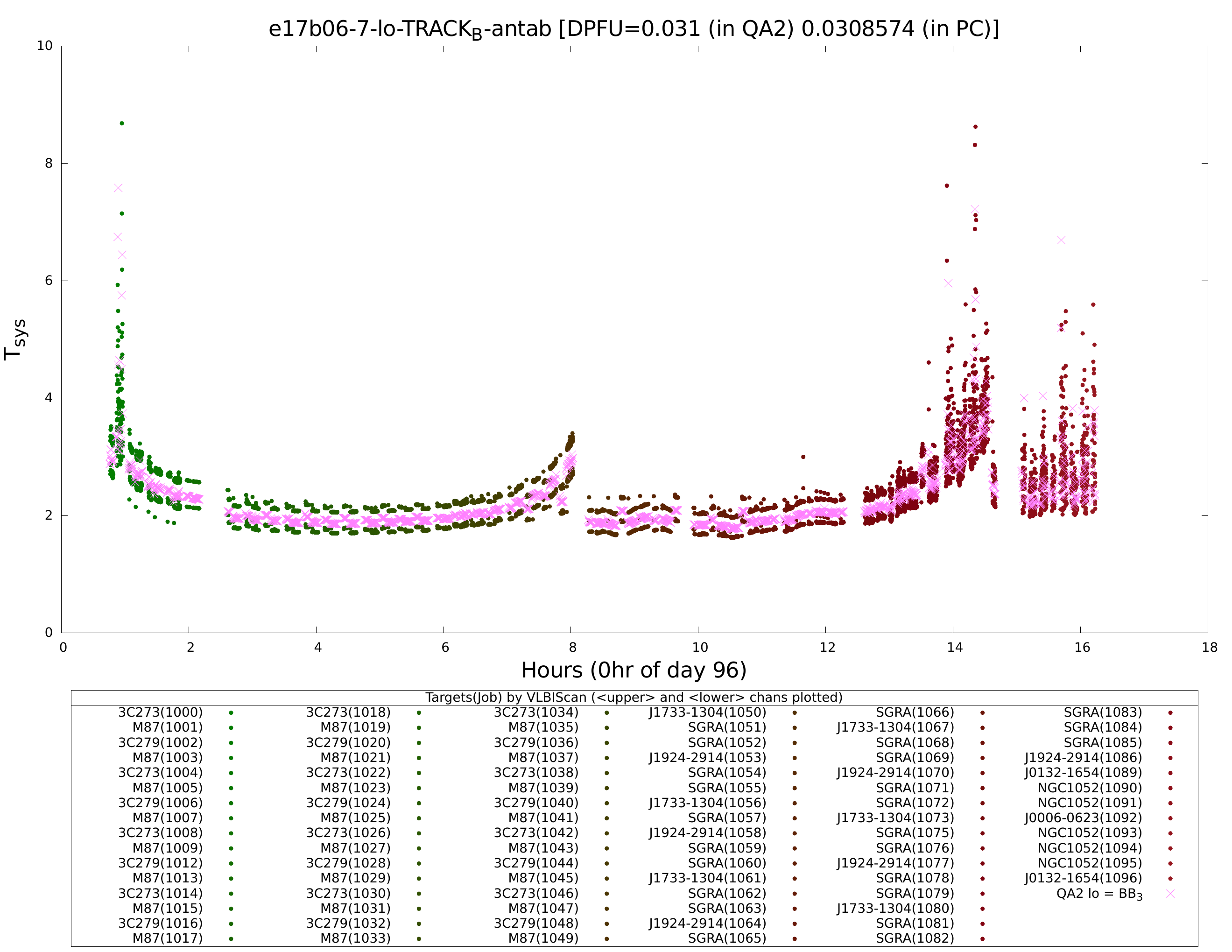}}

\hspace{-4mm}{%
\includegraphics[%
  width=0.5\textwidth]%
{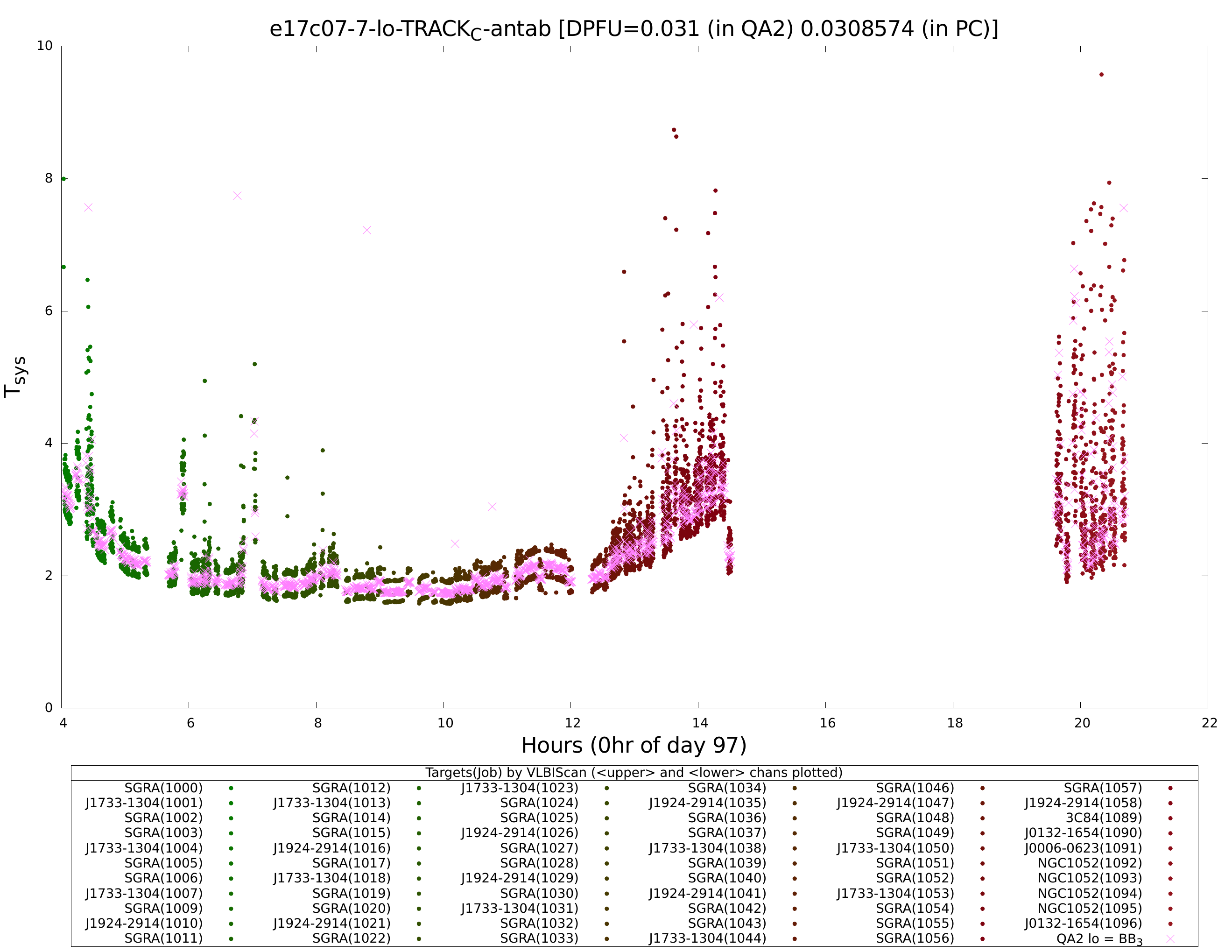}}%
\hspace{1mm}{%
\includegraphics[%
  width=0.5\textwidth]%
{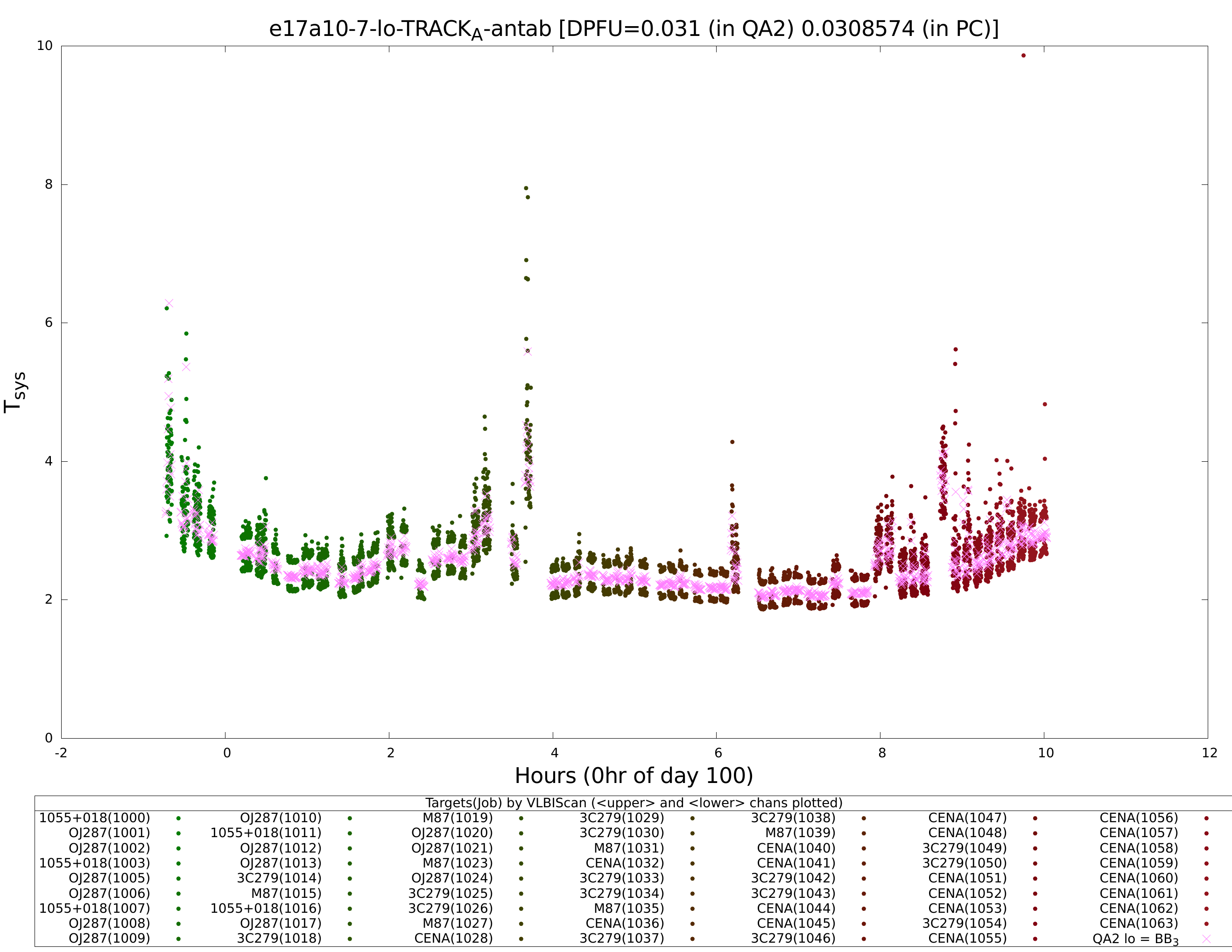}}

\hspace{0mm}{%
\includegraphics[%
  width=0.5\textwidth]%
{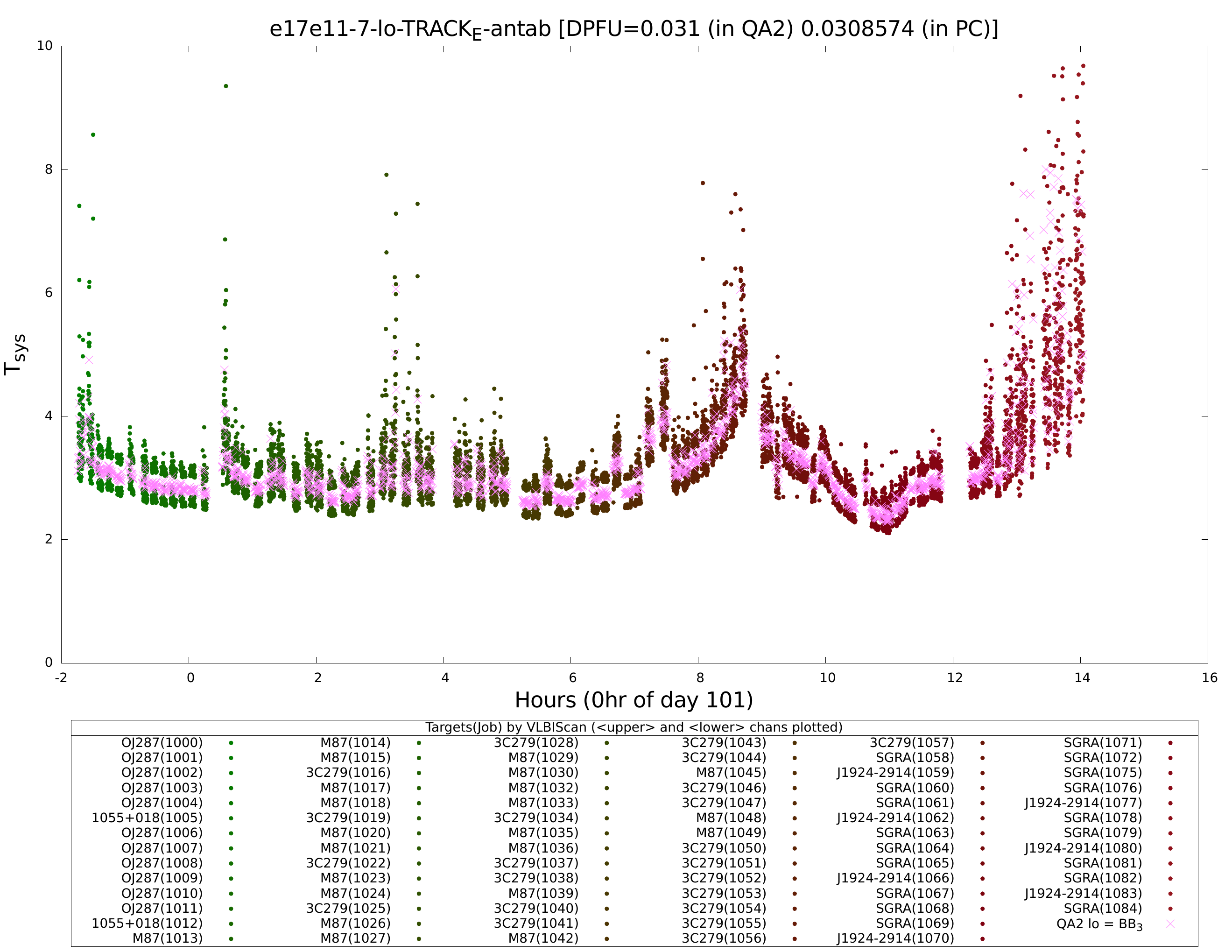}}
\caption{%
Comparison of QA2 (spectral-averaged) ANTAB tables with per-IF
ANTAB tables generated during polcoversion for the Band 6 experiments.  The five panels
are made for the EHT observing tracks on a per-VLBI scan basis (VLBI scans are typically
$\sim$4--8 min long).
For each scan, the
$T_\mathrm{sys}$ value is plotted from three sources:
(1) the time-averaged value provided by the QA2 deliverable (pink crosses),
(2) the time- and spectral-averaged value of the 16 lower IF
channels from \texttt{PolConvert} (colored dots), and
(3) that same average for the upper 16 IF channels (colored dots). 
Note that these last two points bracket the first, which is a bandpass effect.
The higher  \tsys\ values correspond to scans with lower phasing efficiency 
(e.g. at lower elevations,  when the targets are setting; see \S~\ref{pheff}). 
The green-through-red color gradient and the marker legend may be used to track down any issues. 
}
\label{fig:antab}
\end{figure*}

One could also define a {\it system-equivalent flux density} (SEFD) as the total system noise represented
in units of equivalent incident  flux density, which can be written as
\begin{equation}
   \mathrm{SEFD}= <T_{sys}> /\mathrm{DPFU}.
   \end{equation} 
Using Eq.~\ref{TsysEq} and DPFU $= 0.031$\,K/Jy, one can derive an effective {\it phased-array system temperature} for each scan (these are shown in figure~\ref{fig:antab} for  the Band 6 tracks). 
By taking the median of all \tsys\ values ($\sim$2.3~K), one  derives SEFD = 74~Jy in Band 6. 

One could also  compare this estimate with the  theoretically expected SEFD as provided by the ALMA observatory: 
\begin{equation}
 \mathrm{SEFD}= <T_{sys}> \ 2 k_B / \eta_A A_{\rm geom}, 
\end{equation} 
 where  \tsys\ is the  opacity corrected system temperature, $k_B$  is the BoltzmannÕs constant, $\eta_A$ is the aperture efficiency, and $A_{\rm geom}$ is the geometric collecting area of the telescope. 
For a diameter of 73~m for the area (equivalent to 37 12~m antennas),  $\eta_A = 0.68$ \citep{APPPaper}, and taking a representative value of 76~K for the opacity corrected \tsys\  (calculated for median elevations of the sources and typical atmospheric opacities in the 2017 observations), one derives SEFD = 74~Jy in Band 6, which matches perfectly the estimate from the QA2 gain calibration. 
We therefore take  SEFD  = 74~Jy as the representative sensitivity of phased-ALMA during the 2017 campaign  in Band~6. 
A similar analysis provides  SEFD  = 65~Jy in Band~3.

%----------------------------------------------------------------
%----------------------------------------------------------------

\subsubsection{ANTAB files}
\label{antab}

The amplitude calibration for phased-ALMA is computed 
via a linear interpolation of the ALMA antennas gains 
and  it  is stored in the "ANTAB" format (which is the standard file used in VLBI to store amplitude a-priori information, and readable by the AIPS task \texttt{ANTAB}).
In particular, 
the ANTAB files generated by \texttt{PolConvert} have $T_\mathrm{sys}$ entries every $\sim$0.4~sec (matching approximately the VLBI integrations) and 
per VLBI intermediate frequency (IF) band\footnote{The EHT data are recorded using 32 IFs, each 58\,MHz wide, covering a total bandwidth of 1.875\,GHz per sideband. The GMVA data are recorded using 4 IFs of 62.5\,MHz each, for a total bandwidth of 256~MHz (see Appendix~\ref{app:vlbicorr}).}. 
For assessment purposes, another version of ANTAB files is also generated directly from the antenna-wise average of all the  amplitude and phase gains coming from the QA2  
tables (using Eqs. \ref{CalMatrix} and \ref{TsysEq}) and provided in the deliverables to the VLBI correlation centers.  
The frequency and time sampling of these auxiliary ANTAB files is though much coarser than the values estimated by \texttt{PolConvert}, 
since they have $T_\mathrm{sys}$ entries only every sub-scan $\sim$18~sec  
and only one value per SPW (i.e., one value covering a total bandwidth of 1.875\,GHz).  
By inspecting these auxiliary tables at the correlation centers, calibration issues can be identified much earlier in the data processing pipeline (i.e. even before running \texttt{PolConvert}).
Figure~\ref{fig:antab}  shows the comparison plots 
between the QA2 (spectral-averaged) ANTAB tables with the per-IF
ANTAB tables generated by \texttt{PolConvert} for the Band 6 VLBI experiments. 
The plots are
generated as a consistency check of proper \texttt{PolConvert}
operation but do also provide a nice overview of ALMA
$T_\mathrm{sys}$ over the five days of EHT observations.

%----------------------------------------------------------------
\begin{figure*}[ht!]
%\centering
\hspace{-5mm}
\includegraphics[width=\textwidth]{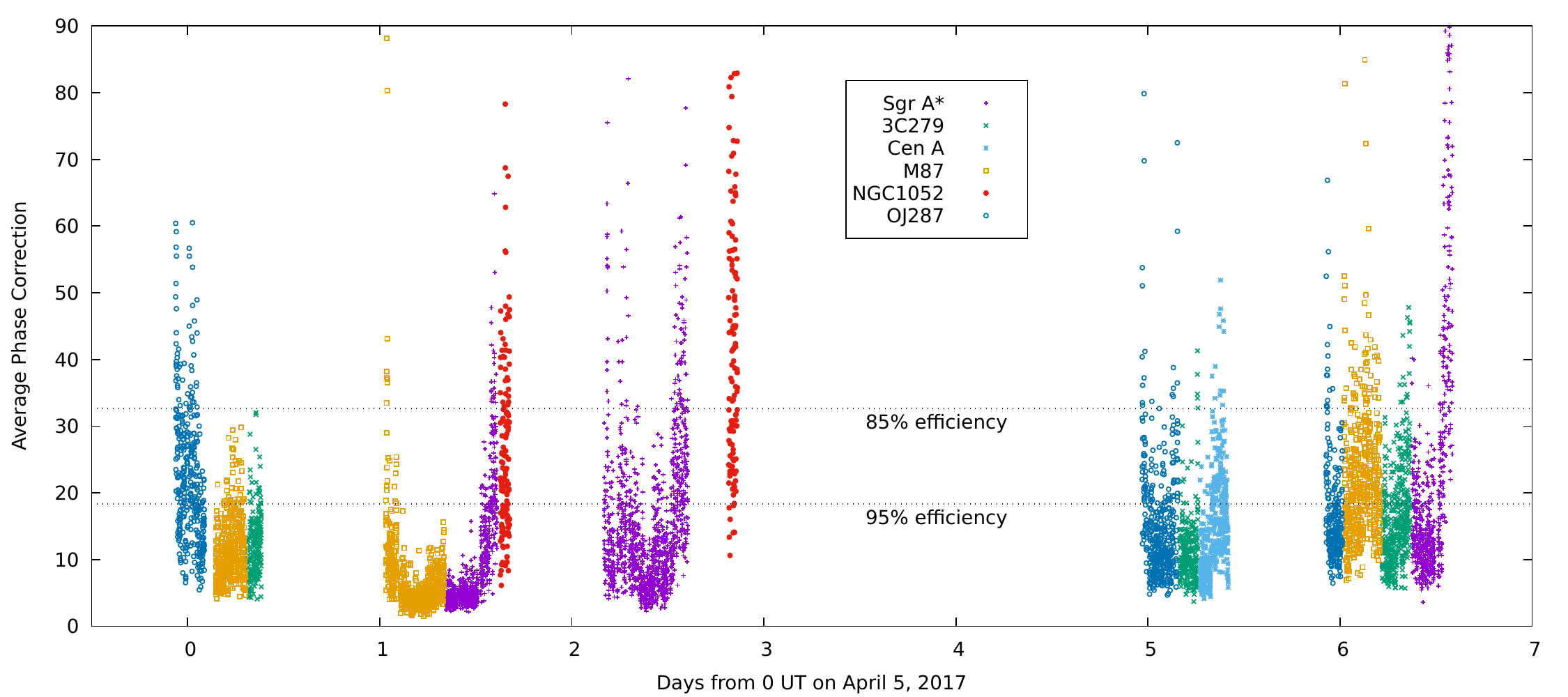} \hspace{-10mm}
\caption{%
Average phase corrections applied by the APS for all scans in the VLBI projects observed in Band 6 during the 2017 April EHT campaign.
The phasing errors (per TFB) calculated by TelCal are captured in the ASDM file metadata (\texttt{CalAppPhase.xml} tables); the correction made is the negative of the phase error.  
The points shown are averages over all 4 SPWs (and all spectral channels) and all the phased-antennas for every correlator sub-scan, colored by project and labelled by science target.  Guidelines at 85\% phasing efficiency (RMS phase errors of $32^{\circ}$)
and 95\% ($18^{\circ}$) are also shown.  
NGC1052 was observed in poor conditions (late morning, low elevation). 
The poor performance at the edges of the tracks
corresponds to low elevations.}
\label{fig:phrms}
\end{figure*}

%----------------------------------------------------------------
\subsubsection{Phasing efficiency}
\label{pheff}

An ideal phased array of $N$ elements   is
equivalent  to a single aperture with $N$ times the effective area of
one of the individual antennas. 
However, a real phased array will suffer from efficiency losses which translate in a decrease  of effective collecting area. 
These losses can be  characterized in terms of a "phasing
efficiency'', $\eta_{\rm p}$, where $\eta_{\rm p}$=1 corresponds to perfect efficiency. 
Following \cite{APPPaper},  the phasing efficiency  can be written as a function of the cross-correlation between the summed signal and that of a comparison antenna, 
divided by the averaged cross-correlations between the comparison antenna and the individual phased elements: 

\begin{equation}
\eta_p = \frac{\left<V_{sum}V_c\right>}{\sqrt{N}\left< V_i V_c\right>},
\end{equation}

\noindent where $N$ is the number of phased antennas\footnote{For an ideal phased array, the correlated amplitude is expected
to grow as the square root of the number of antennas contributing to the sum (see Eq.~\ref{AmpAPP_Eq}).}. 

The main source of efficiency losses in the APS is due to imperfections in the 
phasing solutions. These imperfections are the unavoidable consequence of residual delay errors, troposphere fluctuations, and cosmic source structure, but also the necessarily finite time  lag in the application of the phasing corrections (see \S~\ref{APS} and Fig.~\ref{casa-scans}). 
All of these factors increase the rms fluctuations in the phases of the antennas used to
form the phased sum, $\sigma_{\phi}$, 
which in turn induce a decrement in the correlated amplitude by a factor
$\epsilon \approx e^{\frac{-\sigma^{2}_{\phi}}{2}}$ \cite[e.g.,][]{Carilli1999,APPPaper}. 
In general, the atmosphere dominates $\epsilon$, hence efficiency losses will be larger during poor atmospheric conditions and at higher frequencies. 
In this respect, the phasing efficiency provides  a  metric for
comparing performance of the system under
various conditions (e.g., weather conditions, different observing frequencies, array configurations). 
In the following, by "phasing efficiency" we refer to the efficiency of the phase solver as reported by TelCal,
which does not include other losses \citep[see][for a list of all possible losses.]{APPPaper}.

The phases calculated by the APS and applied to the data are stored in the ASDM metadata (in the \texttt{CalAppPhase.xml} table). 
These phases can be used to understand the phasing efficiency. 
Figure~\ref{fig:phrms} displays per-sub-scan averages of the phasing corrections
applied to the Band 6 observations of the 2017 VLBI campaign.  
The average is over all 4 SPWs (and all spectral channels)\footnote{The performance is comparable in and across each sub-band.} and all the phased-antennas for every correlator sub-scan. 
When the efficiency is good, only small phase corrections ($\lesssim20^{\circ}$) are needed ($>$80\% of the time).  
When the
atmosphere is generating phase fluctuations faster than the APS
can correct them, larger phase corrections are generated and do not necessarily improve the result (i.e. the APS cannot keep up with the atmosphere).

\begin{figure*}[ht!]
%\centering
\hspace{-5mm}
\includegraphics[width=\textwidth]{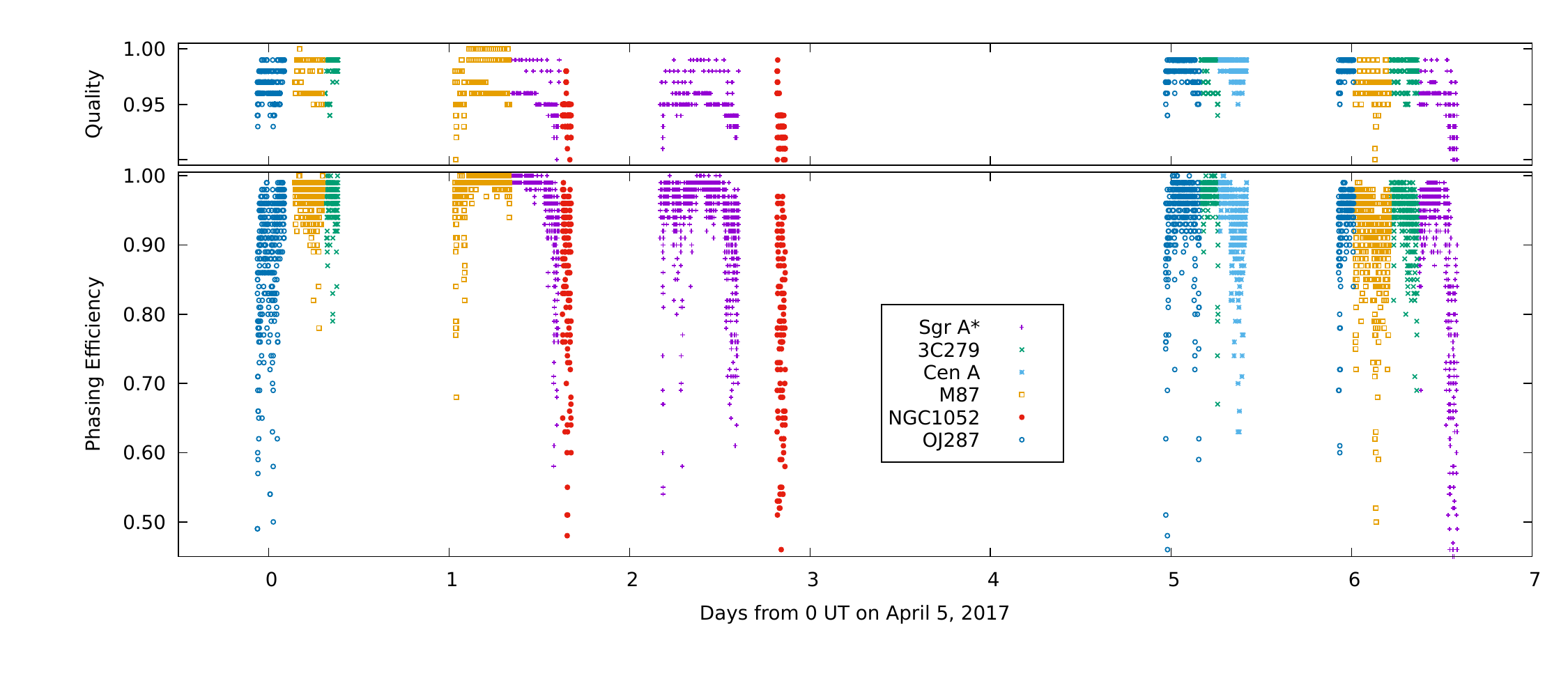} \hspace{-10mm}
\caption{%
Phasing efficiency and quality during phased-array operations in
Band 6 as part of the 2017 April EHT campaign. 
All scans are plotted and colored by science target for each project-track
combination.  
The phasing "quality''  is a "goodness-of-fit" parameter derived from the fitting process, which  is scaled so that unity corresponds to perfect phasing.
 The phasing efficiency
ranges from 0 (totally un-phased) to 1 (perfect phasing).
The large departures of efficiency below 0.8 correspond
to low target elevations or poor atmospheric conditions.}
\label{fig:pheff_av}
\end{figure*}

Fig.~\ref{fig:pheff_av} shows the corresponding phasing efficiency
calculated by TelCal and extracted from the same ASDM file metadata used to produce Figure~\ref{fig:phrms}. 
This efficiency is calculated by comparing the visibilities of the sum and reference antennas on baselines to one comparison antenna.    
The plot shows that in Cycle 4, for about 80\% of the time, the APS has met  90~\% phasing efficiency, which was the goal specified in the original operational requirements  \citep{APPPaper}.
When the phasing efficiency is less than 90\% the problem is ascribable to atmosphere and/or low target elevation.  
In fact, the VLBI tracks were
scheduled to cover each science target from elevation limit (15$^{\circ}$)
to elevation limit\footnote{ALMA does not normally observe at such low elevations in standard projects.}.  
Both Figures~\ref{fig:phrms} and \ref{fig:pheff_av}   show a clear systematic degradation of phasing efficiency associated with low elevations.

It is worth noting that while the  metadata contained in the \texttt{CalAppPhase.xml} table of the ASDM files are useful to understand,  quantify, or  display the phasing efficiency, 
they are not actually used to calibrate the data (e.g. 
to make an appropriate correction for the absolute VLBI correlation amplitudes).  
In fact,  the effects of phasing efficiency are automatically taken into account via the Jones matrix for the phased array (Eq. \ref{CalMatrix}).  The latter is computed at each integration time and for each frequency channel of the intra-ALMA visibilities, and provides  the full calibration of the summed VLBI signal \citep[see also Eqs. 13 and 14 of][]{PCPaper},
including the scaling factor  
that calibrates the VLBI amplitudes to Jy units (Eq.~\ref{NormConvEq}). 
 In other words, the effects of the phasing efficiency are already included in the amplitude- and phase-gains computed for the indvidual ALMA antennas  and are absorbed in the \tsys\ calculated by \texttt{PolConvert} (Eq.~\ref{TsysEq}). 
 As a complementary source of information for data quality assessment, 
the phasing efficiency can also be estimated from the gain calibration of the intra-ALMA visibilities as  
\begin{equation}
\eta_p = \frac{\left| \left< G^i_p G^i_a \right> \right|}{\left| \left< G^i_a \right> \right|}.
\end{equation}
where $G^i_p$ and $G^i_a$ are the phase and amplitude gains for the $i$-th phased antenna.

%==============================================================================
\section{Summary}
\label{summary} 
We have presented a detailed description of the special procedures for calibration and quality assurance  of ALMA observations as a phased-array in VLBI mode. 
These procedures were successfully applied  to the first science observations with phased-ALMA as part of Cycle-4, which were conducted in 2017 April in conjunction with the GMVA in  Band 3 (3mm) and with the EHT in Band~6 (1.3 mm), respectively.
The methodologies described here effectively turn the phased ALMA array into the world's most sensitive mm-VLBI station. 
The huge sensitivity of phased-ALMA is crucial  for the success of scientific experiments that require extremely high-angular resolution, 
including studies of black hole physics on event horizon scales 
and accretion and outflow processes around black holes in AGNs.

\begin{acknowledgements}
We are grateful to the MIT-Haystack and MPIfR correlation centers and  both the
EHTC Correlation and Calibration and Error Analysis working groups for their feedback.
This work was partially supported by 
the ERC Synergy Grant "BlackHoleCam: Imaging the Event Horizon of Black
Holes" (Grant 610058) and 
the Black Hole Initiative at Harvard University, which is supported by a grant from the John Templeton Foundation. 
The APP was supported by a Major Research Instrumentation award from the
National Science Foundation (award 1126433), an ALMA North American
Development Augmentation award, ALMA North America (NA) Cycle 3 and Cycle
4 Study awards, an ALMA NA Cycle-5 Development award, a Toray Science and
Technology Grant through the Toray Science Foundation of Japan, and a
Grant-in-Aid for Scientific Research on Innovative Areas (25120007),
provided by the Japan Society for the Promotion of Science and the
Ministry of Education, Culture, Sports, Science and Technology
(JSPS/MeXT).  Work at MIT Haystack was supported by NSF award AST-1440254.
ALMA is a partnership of ESO (representing its member states), NSF (USA) and NINS (Japan), together with NRC (Canada), MOST and ASIAA (Taiwan), and KASI (Republic of Korea), in cooperation with the Republic of Chile. The Joint ALMA Observatory is operated by ESO, AUI/NRAO and NAOJ.  The National Radio Astronomy Observatory is a facility of the National Science Foundation operated under cooperative agreement by Associated Universities, Inc. 
This paper makes use of the following ALMA data:\\
ADS/JAO.ALMA\#2016.1.00413.V, \\ ADS/JAO.ALMA\#2016.1.01116.V, \\
ADS/JAO.ALMA\#2016.1.01216.V, \\
ADS/JAO.ALMA\#2016.1.01114.V,\\
ADS/JAO.ALMA\#2016.1.01154.V,\\
ADS/JAO.ALMA\#2016.1.01176.V, \\
ADS/JAO.ALMA\#2016.1.01198.V, \\
ADS/JAO.ALMA\#2016.1.01290.V, \\
ADS/JAO.ALMA\#2016.1.01404.V. 
\end{acknowledgements}

%==============================================================================
\appendix
%==============================================================================

%==============================================================================
\section{Accuracy of the Absolute Flux-density Scale with Phased-ALMA}
\label{fluxscale-unc}

%------------------------------------------------------------------------------------------------------------------------------------------------
\subsection{Opacity corrections and global flux-density calibration for the entire VLBI campaign}
\label{tsys}
%------------------------------------------------------------------------------------------------------------------------------------------------

ALMA normally tracks the atmospheric opacity by measuring system temperatures at each antenna. %, taking one measurement every several scans. However,
The system temperatures are however not used in the phased-array data calibration. 
The reason for this choice is that 
the phased signal for VLBI is an {\it unweighted} sum of the signals from all the phased antennas, therefore 
all phased antennas should be treated equally in the gain calibration for \texttt{PolConvert}. 
In particular, 
in APS observations, antennas with bad or failed \tsys\  measurements cannot be removed from the analysis (since they have already been added to the phased sum); 
if \tsys\  were used, 
 opacity corrections would be applied only to a subset of phased antennas,  biasing the calibration by \texttt{PolConvert}. 
As explained in \S~\ref{ampcal}, the bulk of the opacity effects is removed with self-calibration, whereas $T_{\rm sys}$ (usually measured a few times per scheduling block) would correct for {\em second-order} opacity effects, 
related to the difference between the opacity correction in the observation of the primary flux calibrator and the (average) opacity in the observation of any given source.

In order to obtain a more accurate absolute flux-density calibration by accounting for these second-order opacity effects encoded in the $T_{\rm sys}$ measurements, 
the original ASDM files (containing the \tsys\ information) can be recovered and calibrated (in bandpass, flux, and phase) following ordinary QA2 procedures, but using the {\em same} primary flux-density calibrator used for the ALMA-VLBI QA2. The resulting visibilities (which will have the $T_{\rm sys}$ effects applied) may then be imaged with the same weighting and gridding as the ones used in the ALMA-VLBI QA2.

Here, we offer an alternative method to estimate the opacity correction 
which uses the antenna-wise average of valid $T_{sys}$ measurements (the same atmospheric opacity for all ALMA antennas is assumed)   to estimate {\it post-QA2} a global scaling factor, which is source-dependent and constant during the observing epoch, related to the difference between the opacity of a given source and the primary flux calibrator.
In particular, let $S^a_{QA2}$ be the flux-density estimate of source $a$, obtained from the \texttt{<label>.flux\_inf} QA2 table; $g^a(t)$ be the antenna-wise average of the amplitude gains; and  $T^a_{sys}(t)$ be the antenna-wise average of valid system temperatures (linearly interpolated in time for each source). Then, the opacity-corrected flux density for source $a$, $S^a_{\tau}$, will be given by

\begin{equation}
S^a_{\tau} = \left( \frac{\left< g^a(t) T^a_{sys}(t) \right>}{\left< g^{P}(t) T^P_{sys}(t) \right>}      \right)^2 S^a_{QA2}
\label{TauFluxEq}
\end{equation}

\noindent where the superindex $P$ stands for the primary flux-density calibrator and $\left<...\right>$ is the time average. 

Besides providing an opacity correction to the flux densities,  Equation~\ref{TauFluxEq} can also be used to derive a global absolute flux-density calibration for the whole VLBI campaign. 
If we assume the same SEFD for the ALMA system during the VLBI campaign (as assessed in \S~\ref{SEFD}), we can use Eq. \ref{TauFluxEq} to estimate the flux densities of all sources in all the VLBI tracks, by just using the gains $g^P(t)$ from the primary calibrator of one of the tracks. 
This is particularly useful to improve the  flux-density scale in tracks with no SSOs, as pointed out in \S~\ref{flux_calibrators}. 
In tables~\ref{tab:fluxes_b3} and \ref{tab:fluxes_b6}, we show the opacity-corrected flux densities for all the observed sources, 
 using Callisto from project 2016.1.00413.V  (in Band~3) and 
 Ganymede from Track B (in Band~6) as the primary flux calibrators, respectively.
 This is numerically equivalent to run the task \texttt{fluxscale} on all experiments combined, using the SSO in one of the experiments as primary calibrator  (and correcting for the $T_{sys}$ for each source). 
  The significance of this correction is given by the ratio $S^a_{\tau}/S^a_{QA2}$ (columns 6 and 7 in tables~\ref{tab:fluxes_b3} and \ref{tab:fluxes_b6}).  
  Although  in most cases this ratio is just a few percent,  in some cases it can be as high as 10-20\%. 
  These large corrections can be explained by a combination of significant air mass difference between the primary flux calibrator and   sources observed at low elevations (typical in VLBI observations)  and the flux variability of QSOs used as primary flux calibrator. 

%------------------------------------------------------------------------------------------------------------------------------------------------
\subsection{Comparison with ALMA calibrator `grid' source archive} 
\label{flux_comp_archive}
%------------------------------------------------------------------------------------------------------------------------------------------------

\begin{figure*}[ht!]
%\centering
\includegraphics[width=15cm]{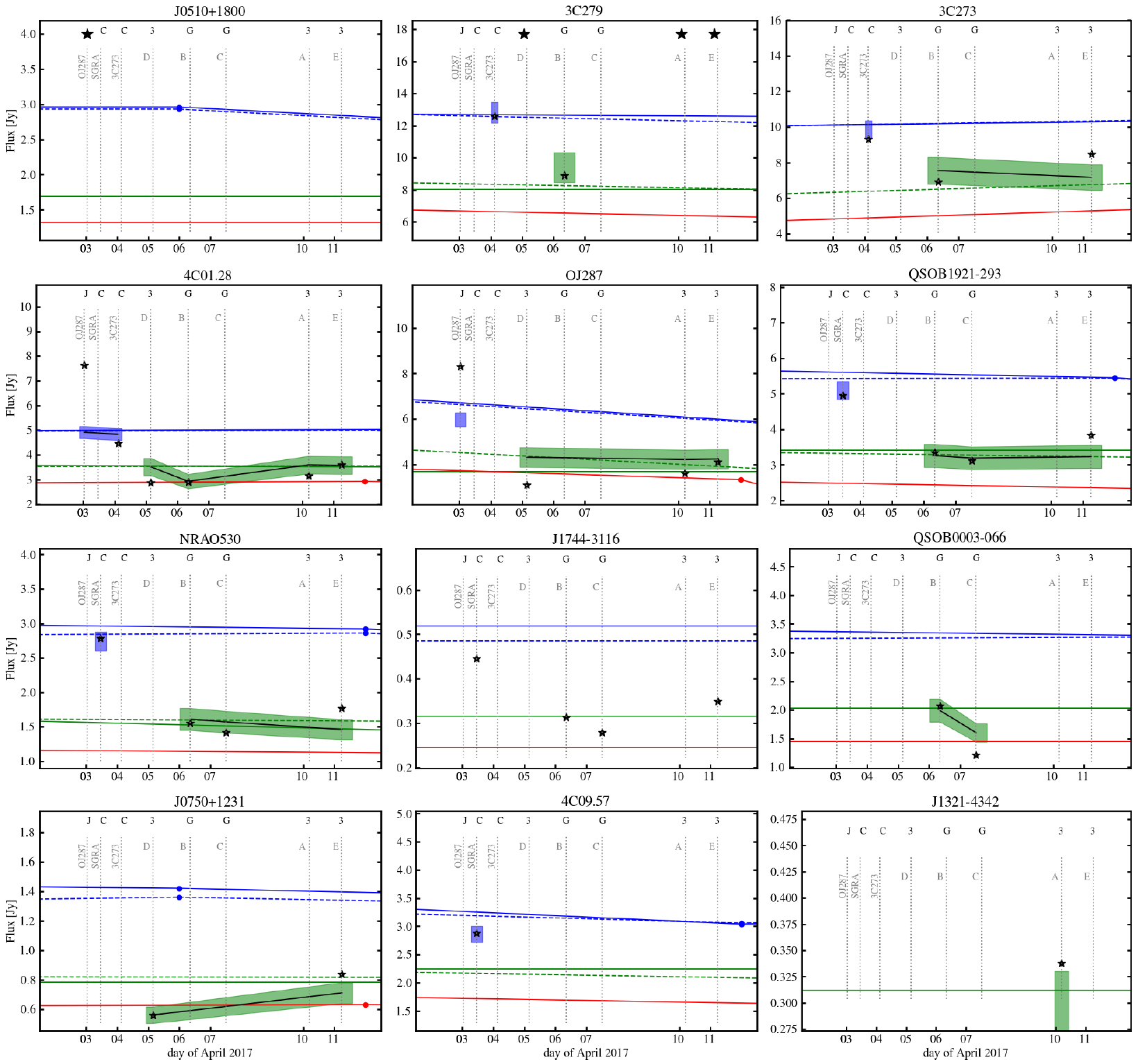} 
\caption{
Comparison between fluxes measured for all the sources observed in the ALMA-VLBI campaign and those retrieved from the ALMA calibrator source archive. 
Only sources with entries in the ALMA archive (close in time to the observations) are displayed. 
The measured flux values during the ALMA-VLBI observations (derived from  the CASA task \texttt{fluxscale}) are indicated as small stars. 
The solid black line shows the evolution of the flux estimate with \texttt{fluxscale}) after the T$_{\rm sys}$ correction (see Section~\ref{tsys} and Equation~\ref{TauFluxEq}), where the shaded regions indicate 5\% and 10\% uncertainty in Band\,3 (93\,GHz; blue shade) and Band\,6 (221\,GHz; green shade), respectively.
Three different bands are displayed from the ALMA archive: Band\,3 (blue),  Band\,6 (green), Band\,7 (red). The flux measurements (circles) are obtained from the ACA monitoring survey in Band\,3 and Band\,7 and their time evolution (lines) are obtained by interpolating these measurements. 
In particular, the solid and dashed blue lines show the Band\,3 flux time evolution at 91.46 and 103.49\,GHz, respectively. 
The red solid line shows the Band\,7 flux time evolution at 343.48\,GHz. 
The solid and dashed green lines show time evolution of the Band\,6 observed flux at 233\,GHz and the Band\,3 to 7 interpolated flux at 221\,GHz, respectively. 
The vertical dotted gray lines show the average time of observation at a given track (labeled at the top, to the left of corresponding line). At the top, the adopted flux calibrator source code is also labeled (C$\equiv$Callisto, G$\equiv$Ganymede, J$\equiv$J0510+1800,  3$\equiv$3C279; for the latter, the large black star indicates that the source is also its own flux calibrator). 
Note that for 4C~01.28 in Track~B, QSO B0003-066 in Tracks~B and C,  J0750+1231, and J1321-4342   only one \tsys\,measurement is available per track, making the \tsys\,correction less reliable (this  may explain some of the apparent discrepancies).  
}
\label{fig:fluxcomp_gs}
\end{figure*}

To assess the accuracy of the flux density calibration, the measured flux densities may be compared with values derived from independent flux monitoring done at ALMA. Specifically, every $\sim$10 days, the ACA observes bright reference sources (called Grid Sources or GSs) along with SSOs, with the goal of monitoring the flux evolution of the GSs, anchored to the modelled fluxes from the observed SSOs. The use of GSs as flux calibrators in regular observations is expected to provide a   5\% to 10\%  absolute flux calibration uncertainty. Some GSs were also observed during the ALMA-VLBI campaign either as ALMA calibrators or VLBI targets. As a reference, the list of observed sources considered as GSs at the time of writing are: 3C273 (J1229+0203); 3C279 (J1256-0547); 3C84 (J0319+4130); 4C01.28 (J1058+0133); 4C09.57 (J1751+0939); J0510+1800; J0750+1231; J1229+0203; OJ287 (J0854+2006); QSO~B0003-066 (J0006-0623); QSO~B1730-130 (J1733-1304); and QSO~B1921-293 (J1924-2914). This allows a direct comparison between fluxes measured during the ALMA-VLBI campaign and archival GS fluxes measured close in time. 
Figure~\ref{fig:fluxcomp_gs} shows such a comparison, which can be used to identify possible  systematic trends (e.g., variability of the flux calibrators). 
The expected flux density of GSs at a given time and frequency can be retrieved from the ALMA archive via the \texttt{getALMAflux()} function implemented in the \texttt{CASA analysis utils}. The flux values of the sources observed either in ALMA-mode or APS-mode are estimated in the uv-plane using the CASA task \texttt{fluxscale}, which adopts a point-source model. This assumption is correct since GSs are unresolved on ALMA baselines\footnote{The point-model assumption is correct for all the VLBI targets observed in Bands 3 and 6, except for Sgr~A* and M87, which have extended structure on ALMA baselines. 
A comparison between peak fluxes in the images produced in the QA2 process and flux values from \texttt{fluxscale}, shows consistent values generally to within 1\% also for these sources. The worst agreement is seen in Track E, with an angular resolution up to twice those of other tracks, resulting in a difference of up to 2\% and 10\%, for M87  and Sgr~A*, respectively. For Sgr~A*  in Track E, a more reliable  estimate of the flux density of the central compact component can be obtained from the image. }. 
Tables~\ref{tab:fluxes_b3} and \ref{tab:fluxes_b6} report the measured flux values (per SPW) for all sources observed in Band\,3 and Band\,6 (respectively) during the 2017 ALMA-VLBI campaign and, when available, the archival flux values for GSs.
The flux values per SPW are opacity-corrected using Equation~\ref{TauFluxEq}, whereas flux values  at the representative frequency are reported both with and without correction, respectively. 
In the following analysis, we consider only the opacity-corrected flux density values. 
The ratio between the measured and the archival flux values can then be used to identify possible systematics, and eventually  improve the flux calibration.
The tables also report the time difference between the VLBI observations and the archival entry. In the following, we consider only ACA measurements within 10 days from the VLBI observations for a meaningful comparison. 

%}

In Band~3, experiments 2016.1.00413.V and 2016.1.01216.V used a SSO, Callisto, as the absolute flux density calibrator. 
While for 2016.1.01216.V the flux measurements are generally consistent with 
the archival values (within the expected 5\% uncertainty),  sources in 2016.1.00413.V have flux values apparently lower (9\%-14\%), which may point to a $\sim$10\% systematic error. 
Experiment 2016.1.01116.V used J0510+1800 as the absolute flux density calibrator, and shows the most significant discrepancy seen in Band~3, where 
OJ287 appears to have a flux 15\% lower than expected. 

In Band~6, tracks B and C used a SSO, Ganymede, as a flux calibrator, and  the flux measurements are generally within 10\% of the archival values. 
 Exceptions are QSO~B0003-006 and 3C84 in Track C and J1058+0133 (4C01.28) in Track B, which appear to be 18\%, 12\%, and 17\% weaker than the archival predicted values. 
 While we cannot exclude intrinsic variability in these sources, we note that they have   only 1--2 \tsys\,measurements per track, making the \tsys\,correction to the flux scale less reliable. 
Another exception is 3C273 in Track B, which appear 15\% higher than the archival predicted value, possibly pointing to    variability. 
In tracks D, A, and E, 3C279 was used as flux calibrator. In order to provide a self-consistent calibration among the different tracks, the flux-density estimate for 3C279 in track B (which used Ganymede) was used in tracks D, A, and E (under the assumption that it remained constant across that week). ACA observed 3C279 in Band\,3 on Apr 1 and in Bands\,3 and 7 on Apr 13. These measurements can be used to predict a Band\,6 flux value at the representative frequency of 221 GHz, which appears to be lower than the measured value in Track B (8.91 vs. 8.21~Jy, corresponding to a $<$10\% difference). 
After using Eq. \ref{TauFluxEq}, the opacity-corrected  flux-density values for 3C279 (bootstrapped  from Track B) are 8.96, 9.39, 8.51, and 8.19 in Tracks D, B, A, and E, respectively. 
This indicates that the flux-density of 3C279   has apparently varied by 13\% during the EHT campaign (note the consistency between the ACA measurement and the bootstrapped value towards the end of the campaign).
As additional confirmation of the goodness of the flux calibration, GS sources in Tracks~D, A, and E do not show discrepancies larger than about 10\% with respect to the archived measurements (within 10 days from the VLBI campaign), with the exception of J0750+1231 (ALMA target) in track E, which is 14\% weaker than the archival value (note that also  this  source  had   only 1 \tsys\,measurement in the full track). 

\begin{figure*}[ht!]
%\centering
\includegraphics[width=15cm]{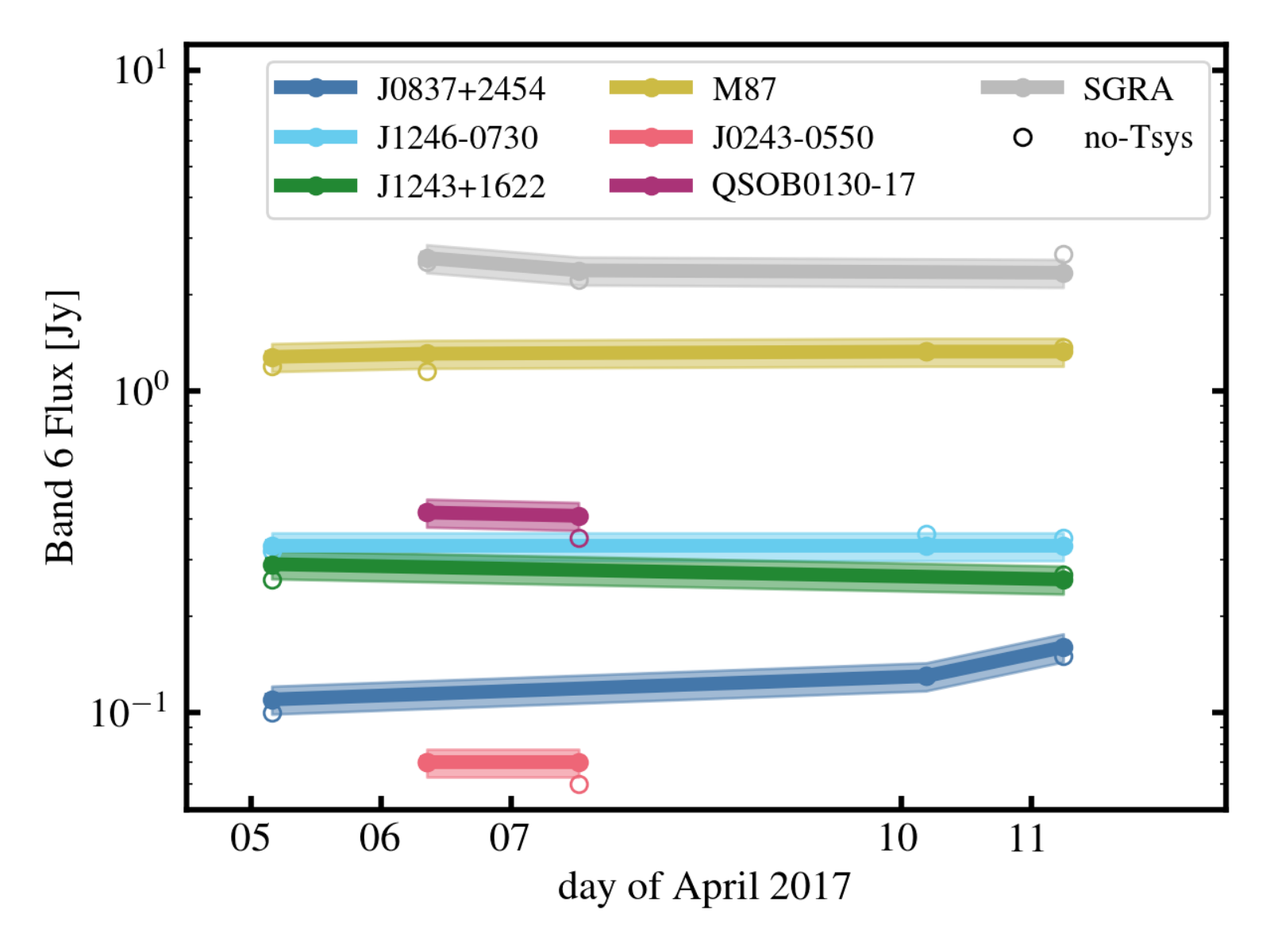} 
\caption{
Time evolution of the fluxes measured for the non-GS sources observed during the EHT campaign. 
The measured flux values are derived from  the CASA task \texttt{fluxscale} and are corrected using  T$_{\rm sys}$  measurements  following Equation~\ref{TauFluxEq}. 
%) are indicated as small stars. 
The thickness of the color bands indicates 10\% uncertainty. 
Note that only Sgr~A* and J0837+2454 show an appreciable variability in flux density over the course of the observing week.  
}
\label{fig:fluxcomp_nongs}
\end{figure*}

Besides the GS sources, Figure~\ref{fig:fluxcomp_nongs} shows also the time evolution of the fluxes measured for the non-GS sources observed in the EHT campaign, which appear to have constant flux density values over the course of the observing week (only Sgr~A* and J0837+2454 show an appreciable variability). 

It is worth noting that the flux-density values obtained  during QA2 from the  \texttt{fluxscale} task (Col. 7 in Tables~\ref{tab:fluxes_b3} and \ref{tab:fluxes_b6}) show larger (often $>10-20$\%) discrepancies with the archival predicted values  (Col. 8 in Tables~\ref{tab:fluxes_b3} and \ref{tab:fluxes_b6}), with respect to the corrected values using Eq. \ref{TauFluxEq}. 
In addition, there are also some apparent systematic trends, where Tracks B, C, and D exhibit a tendency toward lower fluxes compared to those measured in tracks A and E (between 10 and 20\%), Track C has generally lower values than B (of the order of 10\%), and Track~E  shows an apparent systematic increase in flux values with respect to Track~A  (also of the order of 10\%). 
These apparent systematic trends are reconducible to both the secondary opacity effects described in Appendix~\ref{tsys} and the flux variability of 3C279, and are fixed after correcting the flux estimates  with  Equation~\ref{TauFluxEq} and Ganymede in Track B as primary calibrator. 
The final flux-density values for SPW 0 to 3 are reported in columns 2 to 5 of Tables~\ref{tab:fluxes_b3} and \ref{tab:fluxes_b6}, respectively.  

\begin{longtable}{crrrrrrrrrc}
\caption{Comparison between fluxes measured for all the sources observed with ALMA in Band\,3 during the GMVA campaign and those retrieved from the ALMA calibrator ('grid') source archive.
The source flux values are listed separately for each project (labelled in each group's header row, along with the flux calibrator's name and respective adopted fluxes). 
Columns (Col.) 2-5 report the measured flux values for SPW=0,1,2,3 (corresponding approximately to 86, 88, 98, 100~GHz), respectively, as derived from the CASA's \texttt{fluxscale} task and corrected for \tsys following equation~\ref{TauFluxEq}. 
Col. 6 and 7  report flux values  at the representative frequency (93~GHz) as retrieved from  \texttt{fluxscale} with and without \tsys\ correction, respectively. 
Col. 8  reports the expected flux values from the ALMA archive at the representative frequency when an entry is found. 
Col. 9  reports the time difference in days between the VLBI observation and the archival entry. 
Col. 10  reports the time difference in days between the archived observations used to estimate the spectral index (this is relevant for Band~6 comparison). 
Col. 11  reports the ratio between the measured and the archive-predicted values. 
The quoted errors include the measurement error in quadrature with the systematic error in Band~3, corresponding to 5\% of the peak flux. } \label{tab:fluxes_b3} \\
\hline
%Track & RefCal & FluxCal & TransfFld & Flux & FluxArq \\
Source & Flux$_0$ & Flux$_1$ & Flux$_2$ & Flux$_3$ & Flux & FluxFS & FluxArq & Age & $\Delta$t & Ratio\\
\hline
\multicolumn{11}{c}{2016.1.00413.V}\\
\multicolumn{11}{c}{(Callisto) }\\
\hline
4C09.57 & 2.91 & 2.90 & 2.80 & 2.78 & 2.86 & 2.88 & 3.41 & 5 & 0 & 0.84$\pm$0.04 \\
QSOB1730-130 & 2.85 & 2.82 & 2.65 & 2.62 & 2.74 & 2.79 & 3.01 & 5 & 0 & 0.91$\pm$0.05 \\
QSOB1921-293 & 5.27 & 5.24 & 4.98 & 4.93 & 5.09 & 4.95 & 5.72 & 5 & 0 & 0.89$\pm$0.04 \\
J1744-3116 & \ldots & \ldots & \ldots & \ldots & \ldots & 0.45 & 0.52 & -33 & 2 & \ldots \\
SgrA & 2.11 & 2.08 & 2.18 & 2.21 & 2.14 & 2.04 & \ldots & \ldots & \ldots & \ldots\\
\hline
\multicolumn{11}{c}{2016.1.01116.V}\\  \multicolumn{11}{c}{(J0510+1800, F=3.01\,Jy, $\alpha$=-0.44)}\\
\hline
OJ287 & 6.25 & 6.19 & 5.83 & 5.71 & 5.98 & 8.32 & 7.00 & 2 & 0 & 0.85$\pm$0.04 \\
4C01.28 & 5.05 & 4.99 & 4.78 & 4.70 & 4.91 & 7.62 & 5.07 & 2 & 0 & 0.97$\pm$0.05 \\
J0830+2410 & \ldots & \ldots & \ldots & \ldots & \ldots & 0.97 & 0.62 & 145 & 0 & \ldots \\
\hline
\multicolumn{11}{c}{2016.1.01216.V}\\
\multicolumn{11}{c}{(Callisto) }\\
\hline
3C273 & 10.11 & 10.08 & 9.71 & 9.63 & 9.86 & 9.33 & 10.35 & 3 & 0 & 0.95$\pm$0.05 \\
3C279 & 13.21 & 13.12 & 12.61 & 12.50 & 12.81 & 12.61 & 12.98 & 3 & 0 & 0.99$\pm$0.05 \\
J1058+0133 & 5.01 & 4.96 & 4.71 & 4.66 & 4.83 & 4.48 & 5.07 & 3 & 0 & 0.95$\pm$0.05 \\
J1224+0330 & \ldots & \ldots & \ldots & \ldots & \ldots & 0.72 & 0.76 & 147 & 0 & \ldots \\
\hline
\end{longtable} 
\begin{longtable}{crrrrrrrrrc}
\caption{
Comparison between fluxes measured for all the sources observed with ALMA in Band\,6 during the EHT campaign and those retrieved from the ALMA calibrator source archive.
The source flux values are listed separately for each project (labelled in each group's header row, along with the flux calibrator's name and respective adopted fluxes). 
Columns (Col.) 2-5 report the measured flux values for SPW=0,1,2,3 (corresponding approximately to 213, 215, 227, 229 GHz), respectively, as derived from the CASA's \texttt{fluxscale} task and corrected for \tsys following equation~\ref{TauFluxEq}. 
Col. 6 and 7  report flux values  at the representative frequency (221~GHz) as retrieved from  \texttt{fluxscale} with and without \tsys\ correction, respectively. 
Col. 8  reports the expected flux values from the ALMA archive at the representative frequency when an entry is found. 
Col. 9  reports the time difference in days between the VLBI observation and the archival entry. 
Col. 10  reports the time difference in days between the archived observations used to estimate the spectral index (this is relevant for Band~6 comparison). 
Col. 11  reports the ratio between the measured and the archive-predicted values. 
The quoted errors include the measurement error in quadrature with the systematic error in Band~6, corresponding to 10\% of the peak flux.
} \label{tab:fluxes_b6} \\
\hline
%Track & RefCal & FluxCal & TransfFld & Flux & FluxArq \\
TransfFld & Flux$_0$ & Flux$_1$ & Flux$_2$ & Flux$_3$ & Flux & FluxFS & FluxArq & Age & $\Delta$t & Ratio\\
\hline
\multicolumn{11}{c}{Track D}\\  
%\multicolumn{11}{c}{(3C279, F=8.91\,Jy, $\alpha$=-0.6)}\\  
3C279 & 9.20& 9.15 & 8.83 &  8.81 & 8.96 & 8.91 & 8.21 & 4 & 0 & 1.09$\pm$0.11 \\
\hline
J0750+1231 & --- & --- & 0.56 & 0.56 & --- & 0.50 & 0.83 & -1 & 0 & --- \\
OJ287 & 4.45 & 4.47 & 4.22 & 4.19 & 4.32 & 3.09 & 4.78 & 4 & 1 & 0.90$\pm$0.09 \\
4C01.28 & 3.57 & 3.61 & 3.43 & 3.42 & 3.51 & 2.87 & 3.51 & 4 & 1 & 1.00$\pm$0.10 \\
J0837+2454 & 0.12 & 0.12 & 0.11 & 0.11 & 0.11 & 0.10 & 0.17 & -78 & 0 & 0.63$\pm$0.06$^*$ \\
J1246-0730 & 0.34 & 0.34 & 0.32 & 0.32 & 0.33 & 0.32 & 0.35 & 170 & 0 & 0.95$\pm$0.10$^*$ \\
J1243+1622 & 0.30 & 0.29 & 0.28 & 0.28 & 0.29 & 0.26 & 0.26 & 563 & 0 & 1.12$\pm$0.11$^*$ \\
M87 & 1.33 & 1.32 & 1.24 & 1.23 & 1.28 & 1.20 & \ldots & \ldots & \ldots & \ldots\\
\hline
\multicolumn{11}{c}{Track B}\\
\multicolumn{11}{c}{(Ganymede) }\\
\hline
3C273 & 7.74 & 7.70 & 7.42 & 7.35 & 7.55 & 6.93 & 6.59 & 5 & 0 & 1.15$\pm$0.11 \\
3C279 & 9.56 & 9.51 & 9.18 & 9.16 & 9.39 & 8.91 & 8.21 & 5 & 0 & 1.14$\pm$0.11 \\
J1058+0133 & 2.99 & 2.99 & 2.91 & 2.86 & 2.92 & 2.90 & 3.51 & 5 & 1 & 0.83$\pm$0.08$^*$ \\
QSOB1730-130 & 1.65 & 1.67 & 1.57 & 1.55 & 1.61 & 1.55 & 1.59 & -6 & 1 & 1.01$\pm$0.10 \\
QSOB1921-293 & 3.34 & 3.31 & 3.18 & 3.16 & 3.26 & 3.35 & 3.27 & -6 & 1 & 1.00$\pm$0.10 \\
QSOB0003-066 & 2.04 & 2.02 & 1.95 & 1.94 & 1.99 & 2.07 & 1.96 & 7 & 0 & 1.02$\pm$0.10$^*$ \\
J1744-3116 & \ldots & \ldots & \ldots & \ldots & \ldots & 0.31 & 0.26 & -30 & 2 & \ldots \\
ngc1052 & 0.44 & 0.44 & 0.41 & 0.41 & 0.43 & 0.43 & 0.49 & 117 & 0 & 0.88$\pm$0.09$^*$ \\
J0243-0550 & 0.07 & 0.07 & 0.07 & 0.07 & 0.07 & 0.07 & 0.11 & 149 & 0 & 0.66$\pm$0.07$^*$ \\
J1225+1253 & \ldots & \ldots & \ldots & \ldots & \ldots & 0.11 & 0.06 & 564 & 0 & \ldots \\
J1243+1622 & \ldots & \ldots & \ldots & \ldots & \ldots & 0.25 & 0.26 & 564 & 0 & \ldots \\
M87 & 1.36 & 1.35 & 1.27 & 1.26 & 1.31 & 1.16 & \ldots & \ldots & \ldots & \ldots\\
QSOB0130-17 & 0.43 & 0.42 & 0.41 & 0.40 & 0.42 & 0.42 & \ldots & \ldots & \ldots & \ldots\\
SgrA & 2.63 & 2.63 & 2.53 & 2.63 & 2.60 & 2.53 & \ldots & \ldots & \ldots & \ldots\\
\hline
\multicolumn{11}{c}{Track C}\\
\multicolumn{11}{c}{(Ganymede) }\\
\hline
QSOB1730-130 & 1.62 & 1.60 & 1.54 & 1.53 & 1.57 & 1.41 & 1.59 & -5 & 1 & 0.99$\pm$0.10 \\
QSOB1921-293 & 3.27 & 3.25 & 3.11 & 3.09 & 3.19 & 3.12 & 3.27 & -5 & 1 & 0.97$\pm$0.10 \\
QSOB0003-066 & 1.67 & 1.68 & 1.57 & 1.51 & 1.60 & 1.21 & 1.94 & -6 & 0 & 0.82$\pm$0.08$^*$ \\
3c84 & 10.23 & 10.17 & 9.48 & 9.07 & 9.76 & 7.31 & 11.09 & -27 & 2 & 0.88$\pm$0.09$^*$ \\
J1744-3116 & \ldots & \ldots & \ldots & \ldots & \ldots & 0.28 & 0.26 & -29 & 2 & \ldots \\
ngc1052 & \ldots & \ldots & \ldots & \ldots & \ldots & 0.39 & 0.49 & 118 & 0 & \ldots \\
J0243-0550 & 0.07 & 0.07 & 0.07 & 0.07 & 0.07 & 0.06 & 0.11 & 150 & 0 & 0.66$\pm$0.07$^*$ \\
QSOB0130-17 & 0.42 & 0.42 & 0.40 & 0.39 & 0.41 & 0.35 & \ldots & \ldots & \ldots & \ldots\\
SgrA & 2.42 & 2.41 & 2.32 & 2.41 & 2.38 & 2.22 & \ldots & \ldots & \ldots & \ldots\\
\hline
\multicolumn{11}{c}{Track A}\\  
%\multicolumn{11}{c}{(3C279, F=8.91\,Jy, $\alpha$=-0.6)}\\ 
3C279 & 8.74 & 8.69 & 8.39 &  8.38 & 8.51 & 8.91 & 8.00 & -3 & 0 & 1.06$\pm$0.11 \\
\hline
OJ287 & 4.31 & 4.33 & 4.12 & 4.09 & 4.22 & 3.60 & 4.06 & -3 & 1 & 1.04$\pm$0.10 \\
4C01.28 & 3.66 & 3.67 & 3.51 & 3.50 & 3.59 & 3.15 & 3.55 & -3 & 1 & 1.01$\pm$0.10 \\
J0837+2454 & 0.14 & 0.14 & 0.13 & 0.13 & 0.13 & 0.13 & 0.17 & -73 & 0 & 0.75$\pm$0.07$^*$ \\
J1246-0730 & 0.34 & 0.34 & 0.32 & 0.32 & 0.33 & 0.36 & 0.35 & 175 & 0 & 0.95$\pm$0.10$^*$ \\
J1321-4342 & 0.31 & 0.30 & 0.30 & 0.30 & 0.30 & 0.34 & 0.27 & -334 & 0 & 1.09$\pm$0.11$^*$ \\
M87 & 1.38 & 1.37 & 1.29 & 1.27 & 1.33 & 1.33 & \ldots & \ldots & \ldots & \ldots\\
CenA & 5.70 & 5.67 & 5.62 & 5.62 & 5.66 & 5.89 & \ldots & \ldots & \ldots & \ldots\\
\hline
\multicolumn{11}{c}{Track E}\\  
3C279 & 8.38 & 8.33 & 8.04 &  8.03 & 8.19 & 8.91 & 8.00 & -2 & 0 & 1.02$\pm$0.11 \\
\hline
QSOB1730-130 & 1.51 & 1.50 & 1.44 & 1.43 & 1.46 & 1.77 & 1.59 & -1 & 1 & 0.92$\pm$0.09$^*$ \\
QSOB1921-293 & 3.30 & 3.29 & 3.17 & 3.14 & 3.23 & 3.84 & 3.27 & -1 & 1 & 0.99$\pm$0.10 \\
OJ287 & 4.35 & 4.35 & 4.17 & 4.15 & 4.24 & 4.13 & 4.06 & -2 & 1 & 1.04$\pm$0.10$^*$ \\
4C01.28 & 3.63 & 3.62 & 3.50 & 3.48 & 3.57 & 3.59 & 3.55 & -2 & 1 & 1.01$\pm$0.10 \\
3C273 & 7.36 & 7.32 & 7.07 & 7.04 & 7.17 & 8.47 & 6.78 & -2 & 0 & 1.06$\pm$0.11$^*$   \\
J0750+1231 & 0.73 & 0.73 & 0.69 & 0.69 & 0.71 & 0.84 & 0.83 & 5 & 0 & 0.86$\pm$0.09$^*$ \\
J1744-3116 & \ldots & \ldots & \ldots & \ldots & \ldots & 0.35 & 0.26 & -25 & 2 & \ldots \\
J0837+2454 & 0.16 & 0.16 & 0.15 & 0.15 & 0.16 & 0.15 & 0.17 & -72 & 0 & 0.92$\pm$0.09 \\
J1246-0730 & 0.34 & 0.34 & 0.32 & 0.32 & 0.33 & 0.35 & 0.35 & 176 & 0 & 0.95$\pm$0.10 \\
J1243+1622 & 0.27 & 0.27 & 0.26 & 0.26 & 0.26 & 0.27 & 0.26 & 569 & 0 & 1.00$\pm$0.10 \\
M87 & 1.39 & 1.38 & 1.29 & 1.28 & 1.33 & 1.37 & \ldots & \ldots & \ldots & \ldots\\
SgrA & 2.39 & 2.38 & 2.26 & 2.36 & 2.34 & 2.67 & \ldots & \ldots & \ldots & \ldots\\
\hline
\multicolumn{11}{c}{* These sources have   only 1--2 \tsys\,measurements per track, making the \tsys\,correction to the flux scale less reliable}\\
\multicolumn{11}{c}{(this  may explain some of the apparent discrepancies with the ALMA archive predicted values).}
\end{longtable}

%==============================================================================
\section{Amplitude calibration scaling factor for the summed signal.}
\label{app:sqrtN}
%++++++++++++++++++++++++++++++++++++++++++++++++++++
%
The amplitude corrections, $A_i$, of each ALMA antenna, $i$, are derived from the intra-ALMA cross-correlations, so that, at any given time,

\begin{equation}
S_{ij} = A_i A_j \rho_{ij},
\label{AmpCalVisEq}
\end{equation}

\noindent where $\rho_{ij}$ is the correlation coefficient for the baseline between ALMA antennas $i$ and $j$, and $S_{ij}$ is the correlated flux density. The value of the correlation coefficient is

$$ \rho_{ij} = \frac{\left<s_i s_j \right>}{\sqrt{\left<n_i^2\right>\left<n_j^2\right>}}, $$

\noindent where $\left< ... \right>$ indicates average over the correlator integration time, $s_i$ is the source signal recorded at antenna $i$ and $n_i$ is the total signal at that antenna ($s_i$ is assumed to be much smaller than $n_i$). Both $s_i$ and $n_i$ can be given in units of Jy, so that they contain the combined effects of $T_{\rm sys}$ and DPFU. We can then write

\begin{equation}
A_i =  \sqrt{\left<n_i^2\right>} = \sqrt{\frac{T_{sys}^i}{\mathrm{DPFU^i}}},
\label{Rad_Eq}
\end{equation}

 \noindent where $\mathrm{DPFU_i}$ is the instrumental gain (in degrees per Jy) of antenna $i$. Any extra amplitude correction factor for $A_i$ (e.g., related to the signal digital re-quantization in the VLBI backend) can also be included in this equation, though we can rescale the DPFU$^i$ to absorb its effects, and keep Eq. \ref{Rad_Eq} simple. 

In the case of the phased sum, by denoting  with $s_p$ and $n_p$ the source and noise signals, respectively, the cross-correlation to another VLBI station, $l$, is

\begin{equation}
\rho_{lp} = \frac{ \left< s_l \sum_k{s_k} \right>}{\sqrt{\left<n_l^2\right>\sum_k{\left<n_k^2\right>}}},
\label{AVLBI_Eq}
\end{equation}

\noindent where $s_p = \sum_k{s_k}$ and $\left< n_p^2 \right> = \sum_k{\left< n_k^2 \right>}$. We assume that the system temperature dominates the signal amplitude,  i.e. $\sum_{ij}{\left< s_i s_j\right> }$ is much smaller than $\sum_k{\left< n_k^2 \right>}$. If all ALMA antennas have similar DPFUs and $T_{sys}$ (so that $\left< n_i^2 \right> \sim \left< n_j^2 \right> \sim \left<A \right>^2$), then

\begin{equation}
\sqrt{\sum_k{\left<n_k^2\right>}} = \sqrt{N} \left< A \right>,
\label{AvgA_Eq}
\end{equation}

\noindent where $N$ is the number of phased antennas. Substituing Eqs. \ref{AvgA_Eq} into Eq. \ref{AVLBI_Eq}, and comparing to Eq. \ref{Rad_Eq}, we get

\begin{equation}
\left< s_l s_p \right> = \left< s_l \sum_k{s_k} \right> = \rho_{lp} \left( \sqrt{N} \left< A \right> \right)  \sqrt{\frac{T_{sys}^l}{\mathrm{DPFU}}} \eta_{lp} 
\label{AmpAPP_Eq} 
\end{equation}

In essence, the ALMA contribution to the amplitude calibration of the VLBI fringes (i.e., $\sqrt{N} \left< A \right>$) is related to the amplitude calibration of the APP visibilities (i.e., $\left< A \right>$) by the factor $\sqrt{N}$.

%==============================================================================
\section{Effects of phase gain noise on  \texttt{PolConvert}ed ALMA auto-correlations}
\label{glitch}

The gains estimated by the \texttt{gaincal} task during the QA2 process have a random noise contribution. 
The thermal noise in the phase gains stored in the \texttt{<label>.phase\_int.APP} tables is likely the highest of the whole set of QA2 tables, since these gains are derived from the shortest integration times ($\sim$4-sec). 
If the gain noise is independent for $X$ and $Y$, the true calibration matrix for the phased ALMA will be such that (see Eq.~\ref{CalMatrix})

\begin{equation}
J^A_{00} = \left< (B_0^{i})_X \, (G_p^{i})_X  \, G_a^{i} \right>(1+N^a_X)\,e^{N^p_X},
\label{noiseEq}
\end{equation} 

\noindent where $N^a_X$ and $N^p_X$ are time-variable random values around zero, related to the noise in gain amplitude and phase (respectively) of the $X$ polarization. There is a similar expression for $J^A_{11}$ and the $Y$ polarization. 
Given this noise, the  \texttt{PolConvert}ed visibilities are affected by a time-variable leakage-like term (D-term matrix) that depends on the ratio 
$\rho(t)$ of gain noises,  i.e. 

\begin{equation}
\rho(t) = \frac{1+N^a_X}{1+N^a_Y}\,e^{N^p_X-N^p_Y}.
\label{rhoEq}
\end{equation}

\noindent The form of the post-conversion leakage matrix is

\begin{equation}
D(t) = \begin{pmatrix} 1 & d_R \\ d_L  &  1 \end{pmatrix}~~\textrm{with}~~d_R = d_L = \frac{1-\rho(t)}{1+\rho(t)}
\label{DtermEq}
\end{equation}

\noindent (see \S~\ref{residual_crosspol} for a derivation).

\begin{figure*}[th!]
\centering
\includegraphics[width=15cm]{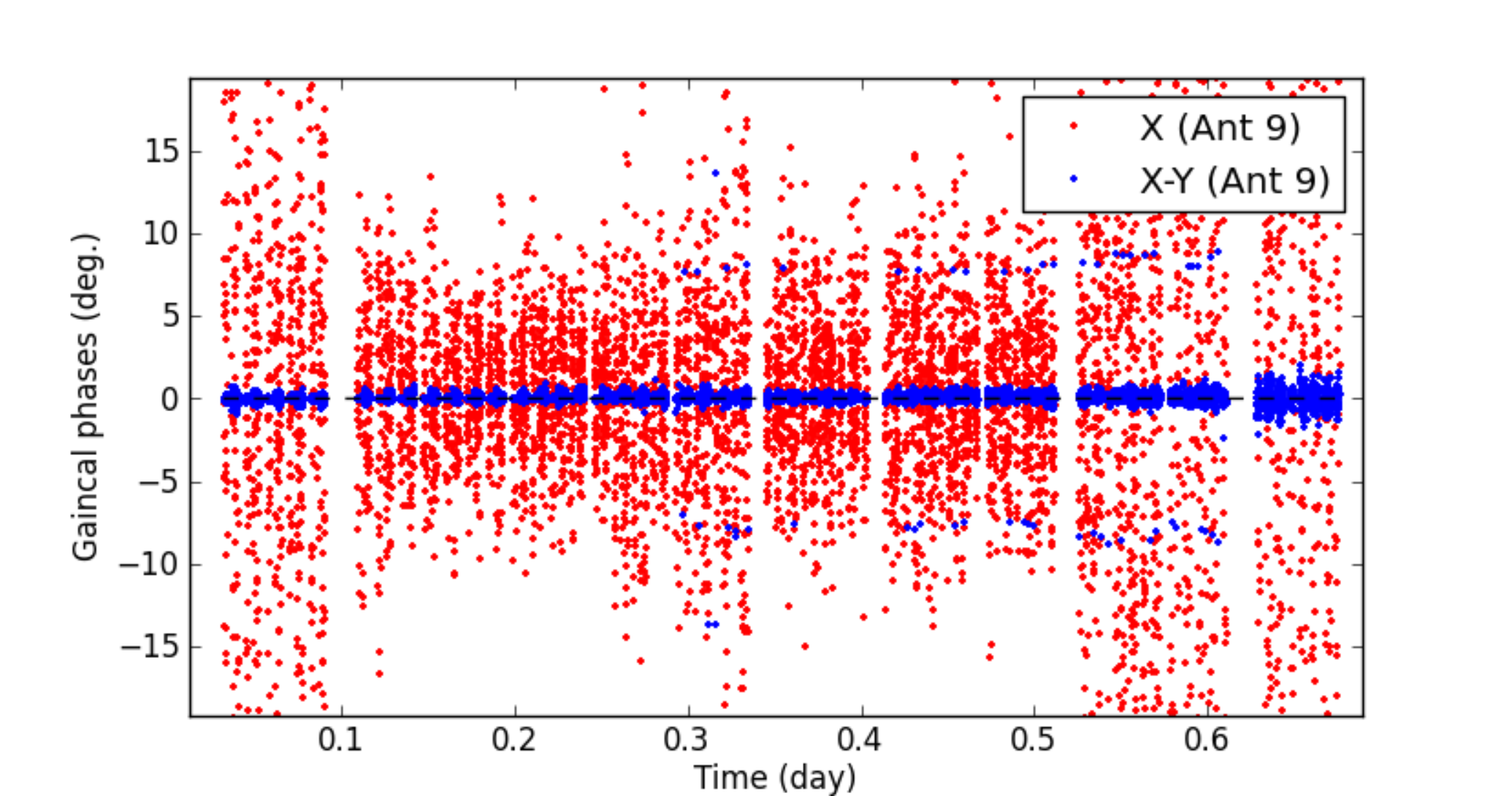}
\caption{Difference between the $X$ and $Y$ phase gains of an ALMA antenna during track B (blue) and actual phase gains in $X$ (red). Outliers around $-50$ deg. (initial integration times at the start of each scan) are not shown.}
\label{XYTimeFig}
\end{figure*}

\begin{figure*}[th!]
\centering
\includegraphics[width=15cm]{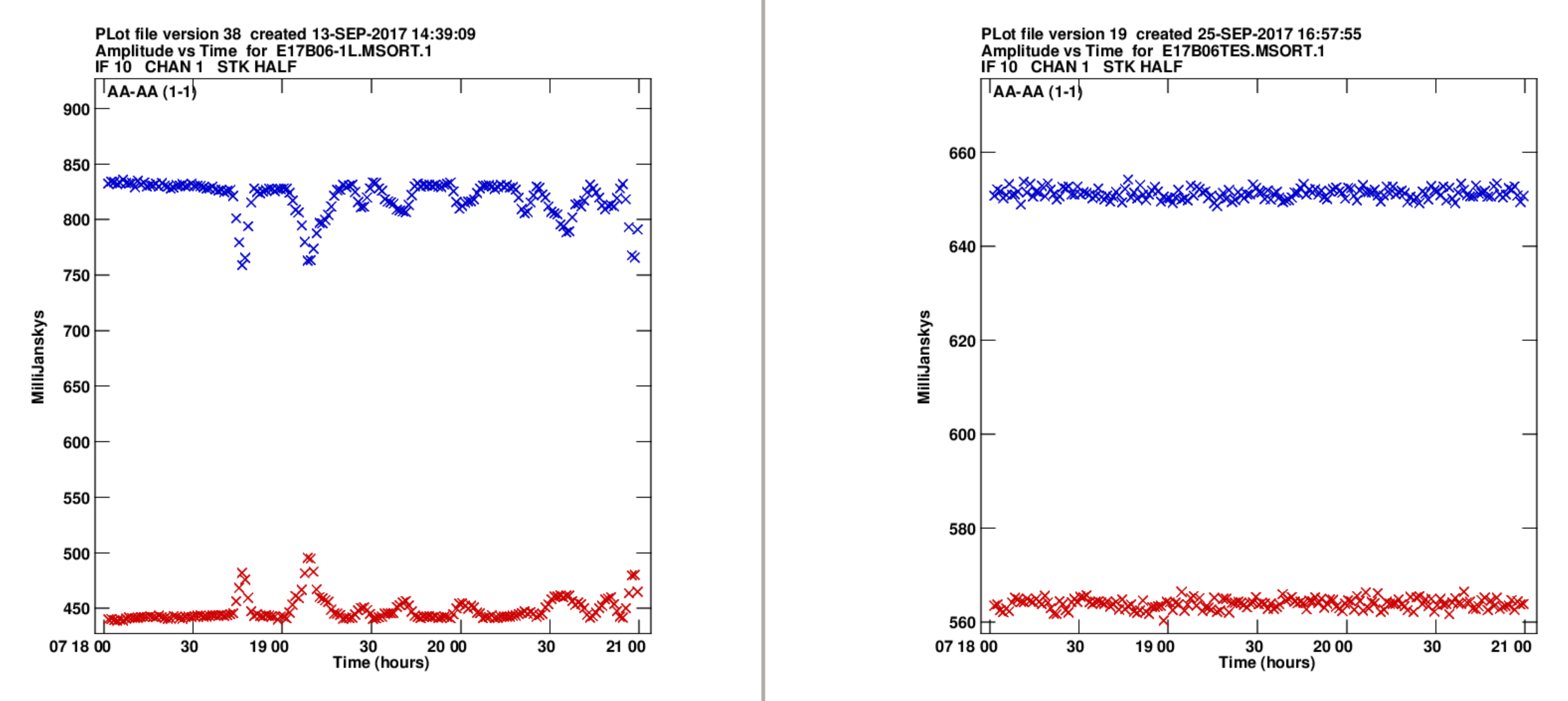}
\caption{Left:  \texttt{PolConvert}ed ALMA auto-correlations of a calibrator scan, obtained  using the G-mode gains (i.e., independent phase gains for each polarization). Note the anti-correlated variability. 
Right: the same scan, but converted using the T-mode gains. Note the different amplitude levels.}
\label{AutoCorrFig}
\end{figure*}

Since the tropospheric effects (which introduce the fastest changes in the phase gains) shall be independent of polarization, we would expect the cross-polarization difference in phase gains to be stable (or slowly varying in time). We show in Fig. \ref{XYTimeFig} the angle of the X-Y phase-gain difference, $arg\left[\left< G^i_a \, (G^i_p)_X\right>/\left< G^i_a \, (G^i_p)_Y\right>\right]$, for track B. Fast-changing differences of 2--3 degrees are seen, which, according to Eq. \ref{DtermEq}, would result in running D-terms of $\sim 3\%$ (i.e., within the specifications of the APP). These values are however computed from the QA2 tables without interpolation into the VLBI correlation accumlation times. If the interpolated gains are applied to the VLBI fringes, cases of random amplitude variability can be seen  in the  \texttt{PolConvert}ed auto-correlation coefficients, with opposite signs for R and L (e.g., Fig. \ref{AutoCorrFig}, left). 
This variability can be caused by a D-term matrix like that of Eq. \ref{DtermEq}, and would be a consequence of higher  noise in some of the phase gains  in the \texttt{<label>.phase\_int.APP} table at specific  integration times. One way to check whether these "glitches" are indeed due to phase-gain noise is to impose the same phase gains to both polarizations, so that $N^p_X = N^p_Y$ in Eq. \ref{rhoEq}. 
This can be done by selecting the option "T'' in the  \texttt{PolConvert} keyword "\texttt{gainmode}'', which replaces the phase gains, $(G^i_p)_X$ and $(G^i_p)_Y$, of each antenna by their geometric average (the so-called  "T-mode" gain), i.e.

\begin{equation}
(G^i_p)_T = \sqrt{(G^i_p)_X \times (G^i_p)_Y}
\label{TmodeEq}
\end{equation}

\begin{figure*}[th!]
\centering
\includegraphics[width=15cm]{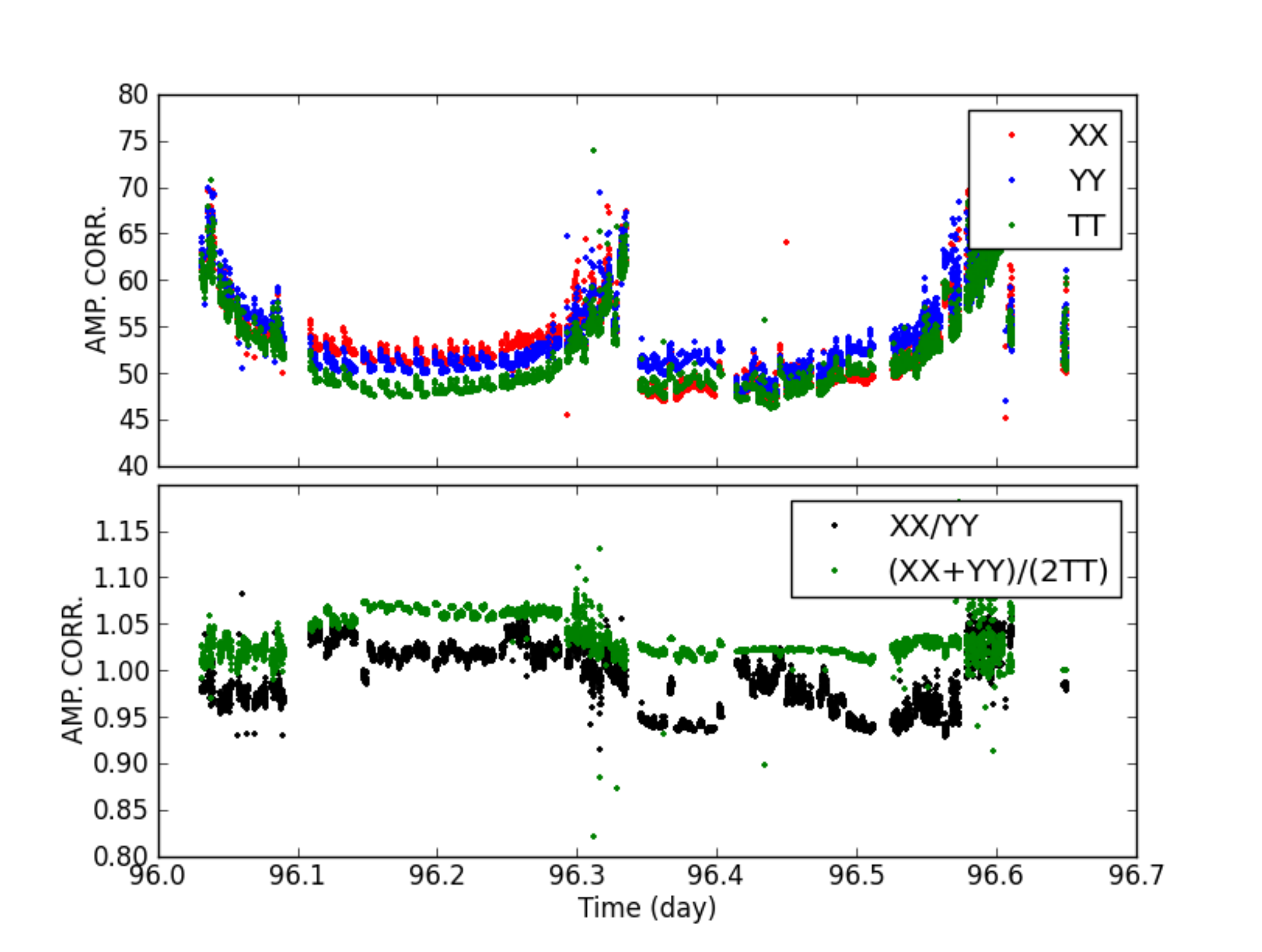}
\vspace{-5mm}
\caption{Top: total amplitude corrections for the $X$ (red) and $Y$ (blue) ALMA streams, derived from the QA2 tables in G mode (i.e., $\left<G^i_a\,(G^i_p)_X\right>$ for $X$ and similar for $Y$). The amplitude corrections for both streams resulting from the T-mode calibration is shown in green. 
Bottom: ratio of amplitude corrections between $X$ and $Y$ (black), which indicates a different  phasing efficiency in $X$ and $Y$. 
The difference of amplitude correction between T-mode and the average of corrections between $X$ and $Y$ for G-mode is also shown (green).}
\label{XYTimeFig2}
\end{figure*}

Using these T-mode gains,  \texttt{PolConvert} produces auto-correlations which appear free from these leakage-like glitches (Fig. \ref{AutoCorrFig}, right). However, we also see that the amplitude ratio changes, which indicates that some ellipticity has been introduced in the  \texttt{PolConvert}ed signals. This effect can be understood if the TelCal's phasing efficiency is slightly different between the two polarizations. A polarization-dependent phasing efficiency could be explained either by mechanical effects (i.e., antenna deformation with elevation) or different $T_{sys}$ (on average) for each polarizer. We can check this hypothesis by plotting the overall amplitude correction for the phased $X$ and $Y$ streams using the "G'' mode phase gains (i.e., $(G^i_p)_X$ and $(G^i_p)_Y$). The results are shown in Fig. \ref{XYTimeFig2}. Indeed, there is a varying difference in the amplitude corrections for X and Y (Fig. \ref{XYTimeFig2} bottom; black points), which are due to the different phasing efficiencies in each polarization. This translates into the need of using independent phase gains for $X$ and $Y$ at each integration time. Otherwise, the different phasing efficiencies will not be properly accounted in the conversion process. 

In light of these results, a better approach to the calibration would be to apply the "T'' mode gain, $(G^i_p)_T$, modified by a smoothed version of the X-Y phase differences, $(G^i_p)_{diff}$, i.e.

\begin{equation}
(G^i_p)_{diff} = \overline{\sqrt{(G^i_p)_X / (G^i_p)_Y}},
\label{TDiffEq}
\end{equation}

\noindent where the over-bar represents a smoothing operator (i.e., a running average or a running median) in phase space, over time windows of a given length. With this smoothed difference (free of the rapid phase noises), we can recover the efficiency-corrected phase gains as

\begin{equation}
(G^i_p)_X = (G^i_p)_T \times (G^i_p)_{diff} ~~\textrm{and}~~ (G^i_p)_Y = (G^i_p)_T / (G^i_p)_{diff}.
\label{TSmoothEq}
\end{equation}

In the QA2 process, these "smoothed" X-Y phase gains are stored in the \texttt{<label>.phase\_int.APP.XYsmooth} tables (\S~\ref{cross-phase} and \S~\ref{vlbi}).

%==============================================================================

\section{Additional details of the VLBI correlation process}
\label{app:vlbicorr}

Correlation of VLBI data from ALMA is performed after the
observations at either of two specialized VLBI correlators: 
one at MIT-Haystack Observatory (Westford, MA, USA) and the other  
at the MPIfR (Bonn, Germany). 
The designated
correlators both currently use the DiFX software correlator
\citep{Deller2011}, which produces a set of SWIN files (one per VLBI scan) and metadata.
The archive produced during the QA2 stage is sent to the VLBI correlators and contains all the calibration information needed to run   \texttt{PolConvert} (see \S~\ref{vlbi}), plus extra information useful for assessment. 
The QA2-related calibration tables are then applied to the SWIN  files, using  \texttt{PolConvert},
and new versions of the SWIN files are generated, performing a proper polarization conversion for the phased-ALMA station.  

The Haystack and MPIfR correlators have adopted a similar work flow and 
 have a similar hardware architecture. 
 Specifically, a set of Mark6 playback units  provide the baseband data streams to the the processing cores (which do the correlation) and the output is reported to a single "manager" process that writes one binary output file per scan.  
 Two utilities, \texttt{difx2mark4} and \texttt{difx2fits}, are used  to create output files in \texttt{FITS-IDI} format, which is read in input by either AIPS or CASA, or in \texttt{Mark4} format, which is read by the HOPS.  
 AIPS, CASA, and HOPS are the software packages currently used for (mm-)VLBI data processing \citep{EHT2019III}. 
In the event, the EHT went through 3 DiFX correlations
and 4 polconversions to arrive at the final data products for the 2017 VLBI projects.

We note that the frequency and time resolution used by the ALMA correlator (Tables ~\ref{table:freq_b3} and \ref{table:freq_b6}) differ from those used by the VLBI correlators.
The choices at the ALMA correlator are driven by the APS system choices discussed in
\S~\ref{APS}. 
The choices at the VLBI correlators are constrained by the recordings
made at the other stations and the numerology of the "zoom band" configuration used
in DiFX \citep[see][]{APPPaper}.  
\texttt{PolConvert} interpolates the ALMA gains to match (in frequency
and time) the spectral and time resolution of the VLBI fringes.  
For the case of the GMVA, the DDC mode of the VLBA is sufficient to align a pair of 128 MHz channels with 4 of ALMA's 62.5 MHz channels (with a 0.5-sec acquisition period). 
For the EHT, the frequency channelization with the ALMA setup is slightly more complicated. 
The VLBI digital back ends \texttt{R2DBE} record a single 2.048 GHz channel which covers the 1.875~GHZ span of the ALMA SPW. 
Each 1.875~GHz SPW is broken up into 32 spectral IFs of 58.59375~MHz. 
Since the latter is not at all a normal spacing for VLBI, the current DiFX correlation setup  is for a 58~MHz IF, 
which results in a small ($< 1\%$) loss in SNR relative to the full possible bandwidth.
Data were then correlated with a frequency resolution of 0.5~MHz (116 individual frequency channels in each of the 58~MHz-wide IFs) and an accumulation period of 0.4~sec.
A full description of the correlation and calibration procedures of EHT data is provided in \citet{EHT2019III}.


\begin{thebibliography}{99}

\bibitem[Asada \& Nakamura(2012)]{AsadaNakamura2012} Asada, K., \& Nakamura, M.\ 2012, \apjl, 745, L28 

\bibitem[Boccardi et al.(2017)]{Boccardi2017} Boccardi, B., Krichbaum, T.~P., Ros, E., \& Zensus, J.~A.\ 2017, \aapr, 25, 4 

\bibitem[Briggs(1995)]{Briggs1995} Briggs, D.~S.\ 1995, Bulletin of the American Astronomical Society, 27, 112.02 

\bibitem[Brogui{\`e}re et al.(2011)]{Broguiere2011} Brogui{\`e}re, D., Lucas, R., Pardo, J., \& Roche, J.-C.\ 2011, Astronomical Data Analysis Software and Systems XX, 442, 277 

\bibitem[Carilli \& Holdaway(1999)]{Carilli1999} Carilli, C.~L., \& Holdaway, M.~A.\ 1999, Radio Science, 34, 817 

\bibitem[Deller et al.(2011)]{Deller2011} Deller, A.~T., Brisken, W.~F., Phillips, C.~J., et al.\ 2011, \pasp, 123, 275 

\bibitem[Doeleman et al.(2008)]{Doeleman2008} Doeleman, S.~S., Weintroub, J., Rogers, A.~E.~E., et al.\ 2008, \nat, 455, 78 

\bibitem[Doeleman et al.(2012)]{Doeleman2012} Doeleman, S.~S., Fish, V.~L., Schenck, D.~E., et al.\ 2012, Science, 338, 355 

\bibitem[Fish et al.(2013)]{Fish2013} Fish, V., Alef, W., Anderson, J., et al.\ 2013, arXiv:1309.3519 

%#################################################
\bibitem[EHT Collaboration et al. (2019a)]{EHT2019I}  The Event Horizon Telescope Collaboration \ 2019a, \apjl, in press 

\bibitem[EHT Collaboration et al. (2019b)]{EHT2019II}  The Event Horizon Telescope Collaboration \ 2019b, \apjl, in press 

\bibitem[EHT Collaboration  et al. (2019c)]{EHT2019III}  The Event Horizon Telescope Collaboration \ 2019c, \apjl, in press 
%#################################################

\bibitem[Goddi et al.(2006)]{Goddi2006} Goddi, C., Moscadelli, L., Torrelles, J.~M., Uscanga, L., \& Cesaroni, R.\ 2006, \aap, 447, L9 

\bibitem[Goddi et al.(2017)]{Goddi2017} Goddi, C., Falcke, H., Kramer, M., et al.\ 2017, International Journal of Modern Physics D, 26, 1730001-239 

\bibitem[Hamaker et al.(1996)]{Hamaker1996} Hamaker, J.~P., Bregman, J.~D., \& Sault, R.~J.\ 1996, \aaps, 117, 137 

\bibitem[Issaoun et al.(2017)]{Issaoun2017} Issaoun, S., Goddi, C., Matthews, L.~D., et al.\ 2017, \aap, 606, A126 

\bibitem[Jorstad et al.(2001)]{Jorstad2001} Jorstad, S.~G., Marscher, A.~P., Mattox, J.~R., et al.\ 2001, \apjs, 134, 181 

\bibitem[Kellermann et al.(1998)]{Kellermann1998} Kellermann, K.~I., Vermeulen, R.~C., Zensus, J.~A., \& Cohen, M.~H.\ 1998, \aj, 115, 1295 

\bibitem[Krichbaum et al.(1998)]{Krichbaum1998} 
  Krichbaum, T.~P., Graham, D.~A., Witzel, A., et al.\ 1998, \aap, 335, L106 
  

\bibitem[Mart{\'{\i}}-Vidal et al.(2016)]{PCPaper} Mart{\'{\i}}-Vidal, I., Roy, A., Conway, J., \& Zensus, A.~J.\ 2016, \aap, 587, A143 

\bibitem[Matthews et al.(2018)]{APPPaper} Matthews, L.~D., Crew, G.~B., Doeleman, S.~S., et al.\ 2018, \pasp, 130, 015002 

\bibitem[Nagai et al.(2016)]{Nagai2016} Nagai, H., Nakanishi, K., Paladino, R., et al.\ 2016, \apj, 824, 132 

\bibitem[Pearson \& Readhead(1988)]{Pearson1988} Pearson, T.~J., \& Readhead, A.~C.~S.\ 1988, \apj, 328, 114 


\bibitem[Rudolf et al.(2007)]{Rudolf2007} Rudolf, H., Carter, M., \& Baryshev, A.\ 2007, IEEE Transactions on Antennas and Propagation, 55, 2966 

\bibitem[Sault et al.(1996)]{Sault1996} Sault, R.~J., Hamaker, J.~P., \& Bregman, J.~D.\ 1996, \aaps, 117, 149 

\bibitem[Thompson et al.(2017)]{ThompsonMoranSwenson2017} Thompson, A.~R., Moran, J.~M., \& Swenson, G.~W., Jr.\ 2017, Interferometry and Synthesis in Radio Astronomy, by A.~Richard Thompson, James M.~Moran, and George W.~Swenson, Jr.~3rd ed.~Springer, 2017 

\bibitem[Tilanus et al.(2014)]{Tilanus2014} Tilanus, R.~P.~J., Krichbaum, T.~P., Zensus, J.~A., et al. 2014, arXiv:1406.4650 

\end{thebibliography}
\end{document}